\documentclass[12pt]{article}
\usepackage{amsmath}
\usepackage{graphicx}
\usepackage{enumerate}

\usepackage{url} % not crucial - just used below for the URL 

\usepackage{amsmath,amssymb, amsfonts}
\usepackage{afterpage}

\usepackage{booktabs}
\usepackage{color,soul}
\usepackage{caption}
\usepackage{enumitem}
\usepackage{anyfontsize}
\usepackage[T1]{fontenc}
\usepackage{graphicx}
\usepackage{overpic}
 \usepackage{subfigure}
\usepackage[leftcaption]{sidecap}
\usepackage{booktabs}
\usepackage{floatrow}
\usepackage{microtype}

\usepackage{algorithm,algorithmic}
\usepackage{url}
\usepackage{mathrsfs}
\usepackage{MnSymbol, multirow}
\usepackage[authoryear]{natbib}
\usepackage{babel}
\usepackage[dvipsnames]{xcolor}

\usepackage[hidelinks]{hyperref}
\hypersetup{colorlinks=true, citecolor=blue, linkcolor=BrickRed}

\usepackage{setspace}
\usepackage{subfigure}
\usepackage{times}

\newtheorem{theorem}{Theorem}

\newtheorem{definition}{Definition}
\newtheorem{lemma}{Lemma}

\newtheorem{remark}{Remark}

\newtheorem{innercustomgeneric}{\customgenericname}
\providecommand{\customgenericname}{}
\newcommand{\newcustomtheorem}[2]{%
	\newenvironment{#1}[1]
	{%
		\renewcommand\customgenericname{#2}%
		\renewcommand\theinnercustomgeneric{##1}%
		\innercustomgeneric
	}
	{\endinnercustomgeneric}
}

\newcustomtheorem{myassumption}{Assumption}

\newcommand{\cH}{\mathcal{H}}
\newcommand{\cum}{{\textbf{cum}}}
\newcommand{\floor}[1]{\lfloor #1 \rfloor}
\newcommand{\cM}{\mathcal{M}}
\newcommand{\cT}{\mathcal{T}}
\newcommand{\Exp}{\text{Exp}}
\newcommand{\Log}{\text{Log}}
\newcommand{\ve}{\varepsilon}
\newcommand{\bi}{\textbf{i}}
\newcommand{\cP}{\mathcal{P}}

\def\T{{{\scriptscriptstyle \top}}}
\def\rlog{\textrm{Log}}

\newcommand{\vertiii}[1]{{\left\vert\kern-0.25ex\left\vert\kern-0.25ex\left\vert #1 
		\right\vert\kern-0.25ex\right\vert\kern-0.25ex\right\vert}}

\usepackage{xr}
\makeatletter

\newcommand*{\addFileDependency}[1]{% argument=file name and extension
\typeout{(#1)}% latexmk will find this if $recorder=0
\@addtofilelist{#1}
\IfFileExists{#1}{}{\typeout{No file #1.}}
}\makeatother

\newcommand*{\myexternaldocument}[1]{%
\externaldocument[s-]{#1}%
\addFileDependency{#1.tex}%
\addFileDependency{#1.aux}%
}
\myexternaldocument{supp}
\newlength\myindent
\newcommand\bindent{%
  \begingroup
  \setlength{\itemindent}{\myindent}
  \addtolength{\algorithmicindent}{\myindent}
}
\newcommand\eindent{\endgroup}

\newcommand{\blind}{1}

\usepackage{pdfpages}

\begin{document}

\def\spacingset#1{\renewcommand{\baselinestretch}%
{#1}\small\normalsize} \spacingset{1}

%%%%%%%%%%%%%%%%%%%%%%%%%%%%%%%%%%%%%%%%%%%%%%%%%%%%%%%%%%%%%%%%%%%%%%%%%%%%%%

\if1\blind
{
  \title{\bf Stationarity of Manifold Time Series }

  \author{Junhao Zhu\thanks{JZ is  partially supported by CANSSI (Canadian Statistical Sciences Institute), Data Science Institute and Medicine by Design, University of Toronto}\hspace{.2cm}, Dehan Kong\thanks{
    DK and ZZ acknowledge financial support from a 
    \textit{Catalyst Grant from Data Science Institute and Medicine by Design, University of Toronto. }}\hspace{.2cm}, Zhaolei Zhang\footnotemark[2]\hspace{.2cm},\\
 University of Toronto,\\
    and  \\
    Zhenhua Lin\thanks{ZL research is partially supported by the NUS startup grant A-0004816-00-00} \\
    National University of Singapore}
    \date{}
  \maketitle
} \fi

\if0\blind
{
  \bigskip
  \bigskip
  \bigskip
  \begin{center}
    {\LARGE\bf Stationarity of Manifold Time Series }
\end{center}
  \medskip
} \fi

\bigskip
\begin{abstract}
In modern interdisciplinary  research, manifold time series data have been garnering more attention. A critical question in analyzing such data is ``stationarity'', which reflects the underlying dynamic behavior and is crucial across various fields like cell biology, neuroscience and empirical finance.  Yet, there has been an absence of a formal definition of stationarity that is tailored to manifold time series. This work bridges this gap by proposing the first  definitions of first-order and second-order stationarity for manifold time series. Additionally, we develop novel statistical procedures to test the stationarity of manifold time series and study their asymptotic properties. Our methods account for the curved nature of manifolds, leading to a more intricate analysis than that in Euclidean space. The effectiveness of our methods is evaluated through numerical simulations and their practical merits are demonstrated through analyzing a cell-type proportion time series dataset from a paper recently published in Cell. The first-order stationarity test result aligns with the biological findings of this paper, while the second-order stationarity test provides numerical support for a critical assumption made therein. 
\end{abstract}

\noindent%
\vfill

\newpage
\spacingset{1.9} % DON'T change the spacing!

	\section{Introduction}
	\label{sec:introduction}
	Recent advances of scientific research   introduce various  complex data; a notable category among these is manifold time series, which refer to temporal data with values  residing on manifolds. Central to the exploration of these datasets is a crucial question: is a manifold time series ``stationary''?  This inquiry is vital for a thorough understanding of the data’s dynamic nature and its implications in the broader context of the study. 
	
	For example, in cell biology, the pioneering study by \cite{schiebinger2019optimal} introduced Waddington Optimal Transport (WOT) for investigating cellular developmental paths and transitions between cell types by tracking changes in cell-type proportions over time. These proportions, represented on a unit sphere  \citep[e.g.,][]{scealy2010}, form a spherical time series. Stationarity in this context reflects dynamic equilibrium in cellular development, such as stable populations in stem cell differentiation. The relevance of  ``stationarity'' in manifold time series (here, the unit sphere) to WOT emerges in two key ways. First, WOT seeks to capture the evolving trend in a spherical time series of non-stationary cell-type proportions, yet it lacks a formal method to distinguish genuine non-stationarity from random fluctuations. Secondly, WOT implicitly presumes the constancy of randomness from cellular proliferation and apoptosis or sequencing platform technical noises over time, without thorough statistical justification. These aspects relate to first- and second-order stationarity in manifold time series. % Statistically, these two considerations touch upon first-order and second-order stationarity in manifold time series analysis, respectively. 
 
%For example, in cell biology, a pioneering study by \cite{schiebinger2019optimal} published in Cell developed a  method known as Waddington Optimal Transport (WOT) for investigating cellular developmental paths and transitions between cell types by tracking changes in cell-type proportions over time. Due to the inherent constraints, these proportions are aptly represented within a unit sphere \citep[e.g.,][]{scealy2010}, leading to the concept of spherical time series for cell-type proportions. In this scenario, stationarity in the time series corresponds to a state of dynamic equilibrium in the cellular development and transition processes, such as the stable cellular populations seen in adult stem cell differentiation. The relevance of WOT to the notion of ``stationarity'' in manifold time series (here, the unit sphere) emerges in two key ways. First, WOT seeks to capture the evolving trend in a spherical time series of non-stationary cell-type proportions, yet it does not establish a formal method to determine whether temporal variations are due to genuine non-stationarity or mere random fluctuations within the cell population. Secondly, WOT implicitly presumes the constancy of randomness from cellular proliferation and apoptosis or sequencing platform technical noises over time, without thorough statistical justification. Statistically, these two considerations touch upon first-order and second-order stationarity in manifold time series analysis, respectively. 

As another example, in neuroscience, there is a growing interest in  modelling time series with values in the manifold of symmetric positive definite (SPD)  matrices to study dynamic resting state  functional connectivity and to reveal the fundamental mechanisms underlying brain networks \citep{yang2020current}.
% \citep{andrew2014,  valsasina2019characterizing,yang2020current}
 Typically, one interesting question is to determine the ``stationarity'' of the SPD-matrices-valued manifold time series. Scientists are interested in whether the observed temporal fluctuation in  functional connectivity values reflects  a reliable ``non-stationarity'', or  merely attributes to noise and statistical uncertainty. 
The above examples show that determining/testing ``stationarity'' of manifold  time series is pivotal for advancing our knowledge in these complex biological fields. The concept of ``stationarity'' in manifold time series is not  restricted  to biological studies. In empirical finance, an important question is to determine whether the correlation matrices of returns, as a time series residing in a sub-manifold of SPD matrices, undergoes some systematic shift over time \citep{corrisktest}. 
Several promising results  from spherical or general non-Euclidean time series analysis were proposed.   
\cite{fisher1994} and \cite{ZHU2024105389} mainly focus on  estimation of  auto-regressive models in  sphere-valued time series.   \cite{dubey2020frechet, wang2023online} and \cite{jiang2024two} investigated change-point detection in non-Euclidean data, assuming time series are segmented into blocks with constant mean and variance. However, their methods did not address more general forms of weak stationarity or account for continuous underlying dynamics in the time series.
 %investigated change-point detection in non-Euclidean data, though their approaches did not consider a more general form of weak stationarity. These seminal pioneering studies addressed change-point scenarios where the time series is segmented into blocks with constant mean and variance within each block. However, they did not consider cases where changes in the time series are driven by some underlying continuous  dynamics. 
 \cite{vandelft2024} explored testing for strong stationarity in time-varying metric measure spaces, where each data point in the time series is a metric space instead of a point within a given manifold. 
A visible gap remains: none of the existing works  have formally defined the concept of first and second-order stationarity for manifold time series. The existing weak stationarity definition and testing methods \citep{2013zhou,aue2020,anne2021} are only applicable to data in Euclidean or Hilbert spaces.

To bridge this gap, we propose the first definition of the first-order and second-order stationarity of manifold time series, and develop corresponding testing procedures to determine whether a manifold time series exhibits either first-order or second-order stationarity, based on  our extension of locally stationary time series to manifolds.  
The notion of local stationarity, originally formulated for  time series in Euclidean space, assumes  a data-generating scheme varying smoothly within local time intervals \citep[e.g.,][]{priestley1988non,dal1997,zhouwu2009}. 
In our work,  local stationarity allows a proper definition of the second-order stationarity of a manifold time series which may not be first-order stationary, and
connects the manifold calculus with tools of asymptotic statistics to facilitate derivation of asymptotic properties of our test statistics. Local stationarity is a reasonable assumption in our real data application to cell biology, as demonstrated in Figure \ref{fig:fle} and related works of  cell biology  \citep[e.g.,][]{lahnemann2020eleven}.
% From an application standpoint, many observed manifold time series is locally similar while evolving over time.
% In single cell data science, there is an acute need for statistically sound methods to infer cell state trajectories within cell types and continuous transitions between cell types, which is observed in tissue generation \citep{lahnemann2020eleven}. The study by \cite{schiebinger2019optimal} exemplifies this through the visualization of time-varying cell populations originated  from mouse embryonic fibroblasts (MEFs), as presented in Figure \ref{fle}, indicating temporal evolution and structural similarity between consecutive time points and suggesting local stationarity in cell-type proportions. 

% Similarly in neuroscience, \cite{hutchison2013dynamic} illustrated dynamic variations in brain functional connectivity through a time series of SPD matrices by sliding-window analysis.

% For instance, in the study by \cite{schiebinger2019optimal}, the visualization of time-varying cell populations from mouse embryonic fibroblasts (MEFs) using force-directed layout embedding illustrates this concept. Each dot in their visualization represents an individual cell, showing that cellular populations evolve over time yet maintain similar structures at consecutive time points. This hints that the cell-type proportions, considered as a spherical time series, exhibit a locally stationary structure.

Our main contributions are summarized as follows:
\begin{enumerate}
    \item We propose the first definition of  the first-order stationarity and second-order stationarity for manifold time series. Our definition incorporates the stationarity of multivariate time series in Euclidean space as a special case. As demonstrated in the above examples, these concepts are crucial for addressing  practical scientific inquiries.
    \item We develop procedures to test the  first-order stationarity of manifold time series based on the Cumulative Sum (CUSUM) \citep{cusum1954} of residuals in the  tangent space at the sample intrinsic mean. The tangent space at the sample intrinsic  mean is not identical to the tangent space at the population intrinsic mean due to the  curved nature of manifolds as shown in Figure \ref{tpmu}(a). This property makes the CUSUM statistic in manifolds more complicated than the counterpart in Euclidean space.  We show  that the asymptotic null distribution of the $L^2$-norm of the CUSUM of residuals induced by the Riemannian metric converges to the sup-norm of a process in the tangent space at the population intrinsic mean, with the  form $U(t)-\cH(t)^{-1}\circ \cH(1)\circ U(1)$, where $U(\cdot)$ is a
centered Gaussian process with an  unknown covariance operator  and  $\cH(\cdot)$ is  an unknown invertible linear operator induced by the curvatures.
     We propose a test that leverages techniques of Gaussian multiplier bootstrap to mimic  $U(\cdot)$ and estimates the operator-valued function $\cH(t)$. We establish the consistency of our method and provide the local alternative distribution to show that our method has local power with a rate of $O(T^{-1/2})$, {where $T$ is the length of the time series.}
    \item Third, we develop a second-order stationarity test for the manifold time series, and establish asymptotic properties for the test statistic.  {One of the major challenges lies in determining the asymptotic distribution of the test statistics since the curved nature of manifolds introduces an additional $O_p(T^{-1/2})$  term  to the test statistic compared to Euclidean space. } % The asymptotic analysis of the test statistics in the manifold case is more complicated than linear spaces 
    % since the estimation error of the intrinsic mean and the curved nature of manifolds contribute to the test statistics. } 
    % \lin{emphasize the challenges of this generalization?}
    Surprisingly, under certain regularity conditions, the null distribution of the  test statistic for manifold time series is invariant to manifold curvatures, asymptotically converges to a Gaussian distribution  and aligns with its counterparts in Euclidean space. 
    {In contrast, under the alternative hypothesis, although the test statistic still asymptotically follows a Gaussian distribution, it exhibits a difference in  variance from its Euclidean counterpart.} 
    % \item Fourth, we provide numerical  analysis  on simulation data to demonstrate the consistency and power of methods. We also apply our testing procedures to test the first and second-order stationarity of cellular developmental time series \citep{schiebinger2019optimal}. \kong{This point may be removed as the simulations and data applications seem to be standard.}
\end{enumerate}

%\lin{highlight some technical challenges, if any?}

We structure the rest of the paper,  as follows. In Section \ref{bg}, we introduce background of    Riemannian manifolds and Euclidean time series. Section \ref{sec:station}  defines the first- and second-order stationarity for  manifold time series. In Section \ref{test}, we develop statistical tests for the stationarity of manifold time series, and establish the corresponding asymptotic properties of the test statistics under null and alternative hypotheses. Simulations and real data application are presented in Section \ref{sim} and  Section \ref{realdata1}, respectively.  In Section \ref{discuss}, we end with a brief discussion.

% (first order bias , second order ...)

\section{Background}\label{bg}
Before introducing the definition of stationarity and the methods of stationarity test within the context of manifold time series, we briefly review the concepts of stationarity in Euclidean space $\mathbb{R}^D$, some background of Riemannian manifolds, and the intrinsic mean.

\subsection{Stationarity and locally stationary}\label{bgsta}
The notion of   stationarity is important as it guarantees the consistency and validity  
of most of  modelling and testing in time series data analysis \citep{shumway2000time}. The definition of stationarity is given as follows:
%\begin{definition}
\begin{itemize}
    \item A collection of    $\mathbb{R}^D$-valued random vectors  $\{X_i\}_{i=1}^T$ is first-order stationary if 
$\mathbb{E}(X_i)\equiv \mu$ for some constant $\mu\in \mathbb{R}^D$. It is  second-order stationary if the auto-covariance matrix $\mathbb{E}\{(X_i-\mathbb{E}X_i)(X_j-\mathbb{E}X_j)^\T \}$  only depends on the lag $|i-j|$. If a time series is both first and second-order stationary, then it is stationary.
\end{itemize}
%\end{definition}

In real-world time series data, the stationarity may not always hold. Instead, the  local stationarity was introduced \citep{dal1997,zhouwu2009}. It offers a way to relax the stationarity assumption, enabling flexible modeling of changes in mean and dependency structures. \cite{zhouwu2009} defines the local stationarity of time series in Euclidean space $\mathbb{R}^D$ as follows: 
\begin{itemize}
    \item A collection of $\mathbb{R}^D$-valued random vectors  $\{X_i\}_{i=1}^T$ is a locally stationary time series if there exists an unknown measurable filter function $G$ such that 
$X_i = G(i/T,\mathcal{F}_i)$, where  $\mathcal{F}_i=(\cdots,\epsilon_0,\cdots,\varepsilon_{i-1}, \epsilon_{i})$, $\{\varepsilon_i\}_{i \in \mathbb{Z}}$ are i.i.d. random variables, and $G$ satisfies some smooth conditions.
% \end{definition}
\end{itemize}
{The above definition includes many time series models, such as  time-varying linear processes 
and  time-varying GARCH models \citep{bollerslev1986generalized} satisfying some regularity conditions \citep{wu2011,2013zhou}. If the filter $G$ is  further independent of $t$, then the time series is stationary.}

\subsection{Riemannian manifold}\label{subsec:manifold}
Below we briefly introduce some basic concepts of Riemannian manifolds {that are essential to our development, with slight emphasis on geometric intuition rather than mathematical rigour}. We refer readers to a self-contained note by \cite{lin2022aos} for more details and to the textbook by \cite{do1992riemannian} for a more comprehensive treatment.

A  topological space $\mathcal{M}$ is called a differential 
 manifold of dimension $D$ if it admits a maximal differentiable atlas that consists of coordinate systems $(U_\alpha,\textbf{x}_\alpha)$ for $\alpha\in J$, such that $\bigcup_{\alpha\in J}U_\alpha=\mathcal M$ and $\textbf x_\alpha \circ\textbf x_\beta^{-1}$ is differentiable whenever $U_\alpha\cap U_\beta\neq\emptyset$, where  $J$ is an index set and each
 $\textbf{x}_\alpha:U_\alpha\rightarrow \mathbb R$ is a coordinate map.
% A topological space $\mathcal{M}$ is called a differential 
% manifold of dimension $d$ if there exists a family of injections $\textbf{x}_\alpha: U_\alpha\subset \mathbb{R}^D \to \mathcal{M}$ of open subsets $U_\alpha$ of $\mathbb{R}^D$ to $\mathcal{M}$ such that  (remove)
% \begin{enumerate}
%     \item $\mathcal{M} =\bigcup_\alpha \textbf{x}_\alpha(U_\alpha)$
%     \item Whenever  $\alpha_1,\alpha_2$ are a pair satisfying $ \textbf{x}_{\alpha_1}(U_{\alpha_1})\cap \textbf{x}_{\alpha_2}(U_{\alpha_2}) = W\neq \emptyset$, the sets $\textbf{x}_{\alpha_1}^{-1}(W)$ and $\textbf{x}_{\alpha_2}^{-1}(W)$ are open subsets in $\mathbb{R}^D$ and the function $\textbf{x}_{\alpha_2}\circ \textbf{x}_{\alpha_1}^{-1}$ is differentiable.
%     \item The family of injections $\{(U_\alpha,\textbf{x}_\alpha)\}$ is  maximal relative to the above two conditions.
% \end{enumerate}
% The mapping $(U_\alpha, \textbf{x}_\alpha)$ with $p\in \textbf{x}_\alpha(U_\alpha)$ is called a coordinate system of $\mathcal{M}$ at $p$.
% The family of injections $\{(U_\alpha,\textbf{x}_\alpha)\}$   satisfying (a) and (b) above is called a differentiable structure on $\mathcal{M}$. A function $f:\mathcal{M}\to\mathbb{R}$ is differentiable at $p\in\mathcal{M}$ if there   exists a coordinate system  $(U_\alpha, \textbf{x}_\alpha)$ at $p$ such that $f\circ \textbf{x}_\alpha^{-1}$ is differentiable at $\textbf{x}_\alpha^{-1}(p)$.
A curve $c:(-\epsilon,\epsilon) \to\mathcal{M}$ is differentiable at $p$ if $p=c\left(0\right)$ and there exists a coordinate system  $(U_\alpha, \textbf{x}_\alpha)$ such that $p\in U_\alpha$ and $\textbf{x}_\alpha\circ c$ is differentiable at $0$. The tangent vector to the curve $c$ at $t=0$ is a linear functional $c^\prime(0)$ such that 
for any function $f$ differentiable at $p$  we have $c^\prime(0) f={d(f\circ c)(0)}/{dt}$.
% $c^\prime(0) f=\frac{d(f\circ c)}{dt}\big|_{t=0}$. 
The tangent space  at $p$ is the linear space of all tangent vectors at $p$, denoted by $\mathcal{T}_p\mathcal{M}$. 
The aggregation of all tangent spaces $\bigcup_{p\in\mathcal{M}} \cT_p\cM$ is called the tangent bundle of $\cM$, denoted by $\cT\cM$.

A differentiable manifold $\mathcal{M}$ is a Riemannian manifold if it is additionally equipped with a Riemannian metric  which defines a smoothly varying inner product $\langle\cdot,\cdot\rangle_p: \mathcal{T}_p\mathcal{M}\times \mathcal{T}_p\mathcal{M}\to \mathbb{R}$ for each point $p$ in $\mathcal{M}$. The Riemannian metric also induces a norm $\|\cdot\|_p$ on each $\cT_p\cM$, and induces a distance function on $\mathcal{M}$, denoted by $d_\mathcal{M}(\cdot,\cdot)$, so that $\mathcal{M}$ endowed with $d_\mathcal{M}(\cdot,\cdot)$ is a metric space. In addition, the Riemannian metric uniquely determines an affine connection called Levi-Civita connection $\nabla: \mathcal{T}_p\mathcal{M}\times \mathcal{T}_p\mathcal{M}\to \mathcal{T}_p\mathcal{M}$, which allows us to  connect nearby tangent spaces and to define the directional derivatives of tangent vectors. An important geometric characteristic of manifold is the curvature.
Formally, the curvature on a Riemannian manifold $\cM$ is defined as a tensor, given by 
$\mathcal{R}_\cM(U,V)  = \nabla_V\nabla_U-\nabla_U\nabla_V+\nabla_{[U,V]}$, where $U,V$ are two vector fields on $\cM$ and $[U,V]=UV-VU$. 
Given a point $p\in\cM$ and a two-dimensional subspace of $\cT_p\cM$ spanned by two linearly independent tangent vectors $u,v\in\cT_p\cM$, the sectional curvature
is defined  as $\kappa(u,v,p) = \langle \mathcal{R}_\cM(u,v)u,v\rangle_p/(\|u\|_p^2\|v\|_p^2-\langle u,v\rangle^2_p) $. 
If $\kappa(u,v,p)\leq 0 ~(\geq 0) $ for any $(u,v,p)\in \cT_p\cM\times\cT_p\cM\times \cM$, then we say $\mathcal{M}$ is a non-positively-curved (non-negatively-curved) manifold. For Euclidean space $\mathcal{M}=\mathbb{R}^D$, one can show that $\mathcal{R}_\cM\equiv0$  and $\kappa(u,v,p)\equiv 0$.  Intuitively, the deviation of the curvature tensor or the sectional curvature from $0$ quantifies how a manifold bends or curves.

Let $c(t)$ be a  differentiable curve with $c(0)=p$, and $v$ be a tangent vector in $\mathcal{T}_p\mathcal{M}$. The parallel transport of $v$ along $c(t)$ is a vector field  $V(t)$ defined on $\mathcal{T}_{c(t)}\mathcal{M}$ such that $V(0)=v$ and $\nabla_{c^\prime(t)}V(t)=0$.  Denote the parallel transport of $v\in\mathcal{T}_{c(s)}\mathcal{M}$ to $\mathcal{T}_{c(t)}\mathcal{M}$ along $c$
by $\mathcal{P}^{c(t)}_{c(s)}(v)$. The collection $\{E_1(t),\cdots, E_D(t):0\leq t\leq 1\}$, denoted by $\mathbf{E}$, is called a parallel orthonormal frame on $\mathcal{T}_{\mu(t)}\mathcal{M}$, if it satisfies the following conditions:
\begin{enumerate}
    \item $E_k(t) = \mathcal{P}_{c(0)}^{c(t)} E_k(0)\in \mathcal{T}_{c(t)}\mathcal{M}$, for any $t\in[0,1]$ and $k\in \{1,\cdots,d\}$.
    \item $\langle E_k(t),E_l(t) \rangle_{c(t)}=\delta_{kl}$ for any $t\in[0,1]$ and $k,l\in \{1,\cdots,d\}$, where $\delta_{kl}$ equals to $1$ if $k=l$, and $0$ if $k\neq l$.
\end{enumerate}
We write $\mathbf{E}(t)=\{E_1(t),\cdots, E_D(t)\}$. 

A differentiable  curve  $\gamma$ is a geodesic if $\nabla_{\gamma^\prime (t)}\gamma^\prime(t)= 0$. The concept of geodesic generalizes the straight line in Euclidean space. 
 For any  $p\in \mathcal{M}$ and $v\in\mathcal{M}$, there exists  a unique geodesic such that $\gamma_v(0)=p$ and $\gamma_v^\prime (0)=v$, which gives rise to the Riemannian exponential map $\text{Exp}_p(v)=\gamma_v(1)$. There is a neighborhood $\mathcal E_p\subset \cT_p\cM$ such that $\text{Exp}_p$ is bijective on $\mathcal E_p$. Therefore, restricting $\text{Exp}_p$ to $\mathcal E_p$, we can define its inverse. This inverse is called the Riemannian logarithmic map at $p$, denoted by $\rlog_p$, satisfying$\rlog_p(\text{Exp}_pv) =v$ for $v\in\mathcal E_p$.

Let $f_X(\cdot) = d^2_\mathcal{M}(\cdot,X)/2$, and denote $\partial_p f_X\in\mathcal{T}_p\mathcal{M}$ the Riemannian gradient of $f_X$ at $p$, that is, 
for any tangent vector $u\in\mathcal{T}_p\mathcal{M}$, $u(f_X)(p) = \langle \partial_p f_X,u \rangle_p$. We also  let  $\mathcal{S}_p\mathcal{M}$ denote  the space of {self-adjoint operators} on $\mathcal{T}_p\mathcal{M}$ and  $H(p,X)$ denote 
 the Riemannian Hessian operator of the function $f_X(\cdot)$ at $p$, i.e., the operator in $\mathcal{S}_p\mathcal{M}$ such that for any tangent vectors $u,v\in\mathcal{T}_p\mathcal{M}$, 
$\langle H(p,X)u,v\rangle_p = \langle \nabla_u\partial_pf_X , v\rangle_p= \langle \nabla_v\partial_pf_X , u\rangle_p=\langle H(p,X)v,u\rangle_p.$

\subsection{Intrinsic mean}
{In curved Riemannian manifolds, the concepts of algebraic addition and the usual mean/average do not apply. The notion of the intrinsic mean, proposed by \cite{fmean1948}, serves as a well established generalization of the traditional mean in the literature. For a random element $X$  in a metric space $\mathcal{M}$ with a distance function $d$, we say $\mu$ is the intrinsic mean (or Fr\'{e}chet mean) of $X$ if 
\begin{equation}\label{eq:intrinsic-mean}
\mu=\arg\min_{p\in\mathcal{M}}\mathbb{E}d_\mathcal{M}^2(X,p).
\end{equation}
Unlike the arithmetic mean, which is well-defined for data in Euclidean spaces, the intrinsic mean extends the idea of finding a central point to spaces where the notion of averaging as simple arithmetic might not make sense. In particular, the intrinsic mean mimics the Euclidean mean in the sense that it minimizes the average squared distance to $X$. The intrinsic mean is a popular tool to model metric-space (including Riemannian manifolds as a special case) valued data in different contexts, such as regression for non-Euclidean data \citep{alex2019,lin2022aos}, change-point detection \citep{jiang2024two,dubey2020frechet} in metric space, and generalized principal component analysis for manifold-valued data \citep{xavier2018aos}. 
 To ensure the unique exisistence of $\mu$ in Eq.\eqref{eq:intrinsic-mean}, 
we assume one of the following conditions:
\begin{enumerate}[label=(\textbf{M\arabic*})]
    \item\label{M1} $\mathcal{M}$ is a simply connected and complete manifold, with bounded non-positive sectional curvatures.
    \item\label{M2}  $\mathcal{M}$ is a simply connected and complete subset of a complete Riemannian manifold with  positive sectional curvatures upper bounded by $\kappa>0$, and satisfies a bounded  diameter condition: $\sup_{p,q\in\mathcal{M}}d_\cM(p,q)<  \pi/{\kappa}^{1/2}$.
\end{enumerate}

\section{Stationarity on Riemannian Manifolds}\label{sec:station}
%\subsection{Definition of stationarity for manifold time series}\label{def:stationary}
In this section, we introduce the definition of    stationarity and local stationarity of manifold time series.
Let $\mathcal{M}$ be a Riemannian manifold of dimension $D$  satisfying conditions \ref{M1} or \ref{M2},  and $\mu(t):[0,1]\to \mathcal{M} $  be a smooth curve on $\mathcal{M}$, associated with a parallel orthonormal frame  $\mathbf{E}=\{E_1(t),\cdots,E_D(t):0\leq t\leq1\}$. 
For any $e=(e^1,\cdots,e^D)\in\mathbb{R}^D$,    $e^\T\mathbf{E}(t)$ denotes the vector in $\mathcal{T}_{\mu(t)}\mathcal{M}$  with coordinate-representations $(e^1,\cdots,e^D)$ under the basis  $\{E_1(t),\cdots,E_D(t)\}$, i.e., $e^\T \mathbf{E}(t)=\sum_{j=1}^D e^j E_j(t) $.

%\zhu{The  first-order stationarity  of manifold time series is defined as follows
\begin{definition}[first-order stationarity]
A  manifold time series $\{X_i\}_{i=1}^T$ on $\mathcal{M}$  is  first-order stationary if  there exists $\mu\in\cM$ such that 
$\mu = \arg\min_{p\in\cM}\mathbb{E}d_\cM^2(X_i,p)$ holds for all $i=1,\cdots,T$, i.e., when its intrinsic mean stays constant.
\end{definition}
Before defining the second-order stationarity for manifold time series, we need to introduce the notion of local stationarity. Traditionally, the second-order stationarity in Euclidean space is defined for first-order stationary   time series. However, it is common in practice that a time series is trend-stationary, i.e., it is second-order stationary after subtracting a deterministic trend. In order to incorporate this wider sense of second-order stationary in manifold time series, we first introduce the local stationarity.

\begin{definition}[local stationarity] \label{locstadef}
    A manifold   time series $\{X_i\}_{i=1}^T$ on $\mathcal{M}$ is  locally stationary  with the mean function $\mu(t)$ if there exists a parallel orthonormal frame  $\mathbf{E}=\{E_1(t),\cdots,E_D(t):0\leq t\leq1\}$ and  an  $\mathbb{R}^D$-valued  processes $\{e_i\}_{i=1}^T$ such that, with $t_i=i/T$,
    \begin{itemize}
        \item  $e_i = G_\mathbf{E}(t_i,\mathcal{F}_i)$ for some unknown measurable filter function, where  $\mathcal{F}_i=(\cdots,\varepsilon_0,\cdots,\varepsilon_{i-1}, \varepsilon_{i})$ and $\{\varepsilon_i\}_{i \in \mathbb{Z}}$ are i.i.d random variables, 
     \item  $\rlog_{\mu(t_i)}X_i =e_i^\T \mathbf{E}(t_{i})$ with $\mu(t_i)=\arg\min_{p\in \cM}
     \mathbb{E}d^2_\cM(X_i,p)$.
     \end{itemize}
\end{definition}
Local stationarity in manifold time series describes a data generating mechanism that varies continuously over time, where in a short time interval, the statistical characteristics for the time series, such as the intrinsic mean of the time series, do not significantly change. In addition, $\rlog_{\mu(t_i)}X_i =e_i^\T \mathbf{E}(t_{i})$ implies $X_i=\mathrm{Exp}_{\mu(t_i)}\{e_i^\T \mathbf E(t_i)\}$, ensuring that the observations $X_1,X_2,\ldots, X_T$ sampled from the data generating mechanism fall onto the manifold $\mathcal M$. In contrast, analyses of the manifold  time series while ignoring the manifold structure (e.g., via embedding the manifold into a Euclidean space and performing the analyses therein) may not preserve this important property. 

\begin{remark}
Throughout this manuscript, our definition of local stationarity follows the framework of \cite{zhouwu2009}. We also recognize an alternative definition for Euclidean and functional time series discussed in \cite{dal1997, lsfts2018}, which differs from that of \cite{zhouwu2009} by a factor of $O_p(1/T)$ under certain regularity conditions. Our theoretical results can be extended to accommodate this alternative with minimal adjustments.
\end{remark}

To introduce the concept of second-order stationarity, we note that for a locally stationary manifold   time series $\{X_i\}_{i=1}^T$ as in the above definition, we have $\mathbb{E}[e_i]=0$. For a fixed orthonormal frame $\mathbf{E}$,
let $\mathcal{C}_{ij} = \mathbb{E}(e_i e_j^\T)$ be the covariance matrix of coordinate-representation for $\rlog_\mu(i/T)$ and $\rlog_\mu(j/T)$ under  $\mathbf{E}$. %We then introduce the definition of first-order  stationarity and second-order stationarity of manifold time series:
\begin{definition}[second-order stationarity]
A locally stationary manifold time series $\{X_i\}_{i=1}^T$ on $\mathcal{M}$ with mean function $\mu(t)$ is second-order stationary if $\mathcal{C}_{ij}$ depends on $i,j$ only through $|i-j|$.
\end{definition}
 If a manifold time series is both first- and second-order stationary, then we say it is stationary.
Our definition of stationarity extends the traditional notion from Euclidean space to general Riemannian manifolds. When $\mathcal{M}$ is the Euclidean space endowed with the canonical inner product, 
then our definition of both first- and second-order stationarity is identical to the classical definition as given in Section \ref{bgsta}. The definition is also invariant to the choice of the parallel orthonormal frames, i.e., if $\mathbf{E}$ and $\mathbf{E}^\prime$ are two parallel orthonormal frames along $\mu(t)$ and $\{X_i\}_{i=1}^T$ is first-order and/or second-order stationary under $\mathbf{E}$, then it is also first-order and/or second-order stationary  under $\mathbf{E}^\prime$. In fact, the concept of second-order stationarity in Euclidean space also (implicitly) depends on parallel orthonormal frames; see Remark \ref{rmk1}  for elaboration.

The above three definitions provide tools to characterize dynamic states of different levels for manifold time series. For example, for the aforementioned cell developmental data, first-order stationarity of cell-type composition time series indicates that cell-type transitions reach an equilibrium state, while second-order stationarity suggests that the randomness in cell-type transitions, caused by  noise in sampling procedures  or cellular birth and death, remains constant over time. In addition, the proposed local stationarity can serve as a valuable tool for modeling multi-resolution and continuous cellular developmental processes, particularly observed  in tissue generation \citep{lahnemann2020eleven}.

\begin{remark}\label{rmk1}
	One may notice that the definition of second-order stationarity in Riemannian manifold is defined through the parallel orthonormal frame, while in Euclidean space, the definition of stationarity appears to be free of orthonormal frames. However, we show that even for Euclidean space, the  second-order stationarity is implicitly defined on the parallel orthonormal frame, and the second-order stationary may not hold if the basis along the mean is no longer a parallel orthonormal frame. For example, let $\{(Z_{i,1},Z_{i,2})\}_{i=1}^T$ be an i.i.d sequence of standard Gaussian random vectors in $\mathbb{R}^2$,
	and $X_i = (i/T + 0.5\cdot Z_{i,1}+0.5\cdot Z_{i,2},~{i}/{T}+ 0.3\cdot Z_{i,1}+2\cdot Z_{i,2})$. Then $\{X_i\}_{i=1}^T$ is  a second-order stationary time series with a linear trend. Let $E_1=(1,0)$ and $E_2= (0,1)$ be the canonical  orthonormal basis in $\mathbb{R}^2$, and $E_1(t) = \cos(t)E_1 +\sin(t)E_2$ and $E_2(t) = -\sin(t)E_1 +\cos(t)E_2$  be a set of
	time-varying  orthonormal basis for $0\leq t\leq 1$, which is no longer parallel. The  coordinate representation of the   detrend time series  $\{X_i-{i}/{T}\}_{i=1}^T$
	under the frame $\{E_1({i}/{T}),E_2({i}/{T})\}$,  is not stationary because the autocovariance matrix of the coordinate representation $\{e_i\}_{i=1}^T$  depends on $i$.
\end{remark}

\section{Tests of Stationarity}\label{test}
The real-world examples in the introduction  highlight the considerable scientific importance of assessing the stationarity in manifold time series. In this section, we introduce detailed statistical testing procedures for both first- and second-order stationarity in manifold time series.

\subsection{First-order stationarity test}
Let $\mathcal{M}$ be a Riemannian manifold of dimension $D$, and $\{X_i\}_{i=1}^T$ be a locally stationary time series with mean function $\mu(t)$ 
satisfying Definition \ref{locstadef}. Let $\mathbf{E}=\{E_1(t),\cdots,E_D(t):0\leq t\leq1\}$ be a fixed  parallel orthonormal frame on $\mu(t)$ and     $\{e_i\}_{i=1}^T$ be the coordinate-representation of $\rlog_{\mu(i/T)}X_i\in\mathcal{T}_{\mu(i/T)}\mathcal{M}$ under the basis $\{E_1({i}/{T}),\cdots,E_D({i}/{T}) \}$, for $i=1,\cdots,T$.
 We consider the following null and alternative:
$$\boldsymbol{H_0}:~\mu(t)\equiv \mu~\text{for some constant }\mu\in\mathcal{M},~\text{versus}~~\boldsymbol{H_1}:~\mu(t)~\text{is a non-constant smooth curve.}$$ We employ a CUSUM statistic to construct a test for these hypotheses. First, we estimate $\mu$ by the  empirical intrinsic mean  $\hat\mu  = \arg\min_{p\in\mathcal{M}}{T}^{-1}\sum_{i=1}^T d^2_\mathcal{M}(p,X_i).$
Then, with $v_i = \Log_{\hat{\mu}} X_i$ and $S_j= \sum_{i=1}^j v_i$, we introduce the test statistic  $$Q_T = \max_{1\leq j\leq T}\|T^{-1/2}S_j\|_{\hat\mu}.$$ Under $\boldsymbol{H_1}$, one would expect the CUSUM statistic  $Q_T$ to be larger compared to its value when $\boldsymbol{H_0}$ is valid.

To develop a test based on the CUSUM statistic  $Q_T$, we study the asymptotic property of $Q_T$, starting with introducing some technical definitions and  regularity conditions.
As the manifold time series may contain complex dependency structures,   we first introduce an additional quantity to quantify the temporal dependency; similar dependency measures can also be found in \cite{wu2005nonlinear} and \cite{2013zhou}.
\begin{definition}\label{def:delta}
    Let $\{X_i\}_{i=1}^T$ be a locally stationary time series as in Definition \ref{locstadef}, and  $\{\varepsilon^\prime\}_{i\in\mathbb{Z}}$ an i.i.d copy of $\{\varepsilon\}_{i\in\mathbb{Z}}$.
    Assume that $\max_{1\leq i\leq T}\mathbb{E}\|e_i\|_p^p <\infty $ for some positive  $p$, where $\|\cdot\|_p$ is the $L_p$-norm in Euclidean space. Then for any integer $k>0$, the $k$-th physical dependence measure is 
    \begin{equation}
        \delta_p(k,G_\mathbf{E}) : =\sup_{0\leq t\leq 1}(\mathbf{E}\|G_\mathbf{E}(t,\mathcal{F}_k)-G_\mathbf{E}(t, (\mathcal{F}_{-1},\varepsilon_0^\prime,\varepsilon_1,\cdots,\varepsilon_k))\|^p_p)^{1/p}. 
    \end{equation}
    If $k\leq0$, we take $\delta_p(k,G_\mathbf{E}):=0$ conventionally.
\end{definition}
We also assume the following regularity conditions for establishing the asymptotic distributions of the test statistic.
\begin{enumerate}[label=(\textbf{A\arabic*})]
\item \label{LipHess} The Hessian tensor $H(p,X)$ is  $L_H$-Lipschitz continuous in $p$ given $X$, and $L_H$-Lipschitz continuous in $X$ almost surely for any fixed $p$, where $L_H<\infty$ is uniformly bounded.
		\item\label{G:Lip} There exists some finite constant $C$ such that 
   $\mathbb{E}\|G_\mathbf{E}(t,\mathcal{F}_0)-G_\mathbf{E}(s,\mathcal{F}_0)\|_2\leq C|s-t| $ ,
  and 
$$\mathbb{E}d_\mathcal{M}\left(\Exp_{\mu(t)}\{G_\mathbf{E}(t,\mathcal{F}_0)^\T \mathbf{E}(t)\}, \Exp_{\mu(s)}\{G_\mathbf{E}(s,\mathcal{F}_0)^\T \mathbf{E}(s)\} \right)\leq C|t-s|, ~~~\forall s,t\in[0,1].$$
%\lin{Can Lipschitz continuity of $\mu$ imply this condition? }\zhu{This can not be implied by the Lip-cont of $\mu$ and $C_\mathbf{E}$, especially in a negatively-curved space, the distance of two points on manifold may explode even though we restrict the noise scale in tangent space  (if we can assume compact support,  then I think Lip-cont of $\mu$ and $C_\mathbf{E}$ can work. ) }
    \item\label{depend} $\delta_4(k,G) = O(\alpha^k)$ for some $\alpha\in [0,1)$, where $\delta_4(k,G)$ is defined in Definition \ref{def:delta}.
     \item\label{Sigma} Let $\Sigma_\mathbf{E}(t) = \sum_{k\in\mathbb{Z}} \mathbb{E}\{G_\mathbf{E}(t,\mathcal{F}_0)G_\mathbf{E}(t,\mathcal{F}_k)^\T\}$ for $t\in[0,1]$, where $\mathbb{Z}$ is the set of all integers. We assume the smallest eigenvalue of $\Sigma_\mathbf{E}(t)$ is  bounded away from $0$  uniformly over $t\in[0,1]$. 
      \item\label{G:subexp} $\sup\limits_{0\leq t\leq 1}\mathbb{P}(\|G_\mathbf{E}(t,\mathcal{F}_0)\|_2\geq M)\leq \exp(-C_1M)$ for some constant $C_1<\infty$ and any $M>0$, i.e., $G_\mathbf{E}(t,\mathcal{F}_0)$ is uniformly sub-exponential. 
\end{enumerate}
The above assumptions, whose Euclidean counterparts are common in the literature, are further discussed in Remark \ref{rmk2}. A concrete example satisfying the above conditions is provided in Remark \ref{example}. The following lemma plays an important role in the investigation of the asymptotic properties of $Q_T$, and $\mathcal S_\mu\mathcal M$ is defined in Section \ref{subsec:manifold}.
\begin{remark}\label{rmk2}
    The assumption \ref{LipHess} holds when the support of  data    is a  bounded subset of $\mathcal{M}$, and  can be replaced with sub-Gaussian conditions, for example, $\max_{1\leq i\leq T}\mathbb{P}(d_\cM(X_i,\mu)>M)\leq \exp(-CM^2)$ for some positive  constant $C<\infty$ and any $M>0$.  Euclidean counterparts of Assumptions \ref{G:Lip}-\ref{Sigma} are common in the literature of stationarity test, such as \cite{2013zhou}. The condition \ref{G:subexp} is required to control the variation induced by the curved nature of manifolds. Stronger conditions were used in previous works of non-Euclidean data analysis. For example,   \cite{alex2019,dubey2020frechet} assumed bounded support of data. %, and   \cite{AOS2095} assumed a sub-Gaussian condition.  }  
\end{remark}

\begin{remark}\label{example}
	We give an  example satisfying Assumptions \ref{LipHess}-\ref{G:subexp}.  Let $\mathcal{M}$ be  the space of $3\times 3$ SPD matrices  with the affine-invariant metric \citep{moakher2005differential}. Let $\mu(t)$ be a geodesic  such that $\mu(0)=I_3$ and $\mu(1)=1.5 I_3$. 
	Let  $\{E_{j,k}(0)\}_{1\leq j\leq k\leq 3}\subset \text{Sym}_3$ be a  set of $3\times 3$ symmetric matrices  with $1$ at the  $(j,k)$ and $(k,j)$ entries   and $0$ at the remaining entries. One can show that  $\{E_{j,k}(0)\}_{1\leq j\leq k\leq 3}$ is  an orthogonal basis of $\mathcal{T}_{\mu(0)}\text{Sym}_3^+$, with $\|E_{j,k}(0)\|_{\mu(0)} = 1 $ for $j=k$, and $\|E_{j,k}(0)\|_{\mu(0)} = \sqrt{2} $ for $j\neq k$. Let $\{E_{j,k}(t):1\leq j\leq k\leq 3,~0\leq t\leq1 \}$ be the parallel orthogonal  frame along $\mu(t)$ with initial value $E_{j,k}(0)$. For simplicity in notations, we also let $t_i=i/T$. A time-varying auto-regressive processes satisfying our conditions are given as follows: 
	\begin{equation*}
		\rlog_{\mu(t_i)}  X_{i+1} = \left(0.05+0.25 t_i \right) \cP_{\mu(t_{i})}^{\mu (t_i)} \rlog_{\mu(t_{i+1})} X_{i} + \{(t_i-0.5)^2+0.2 \}{\ve_i}, 
	\end{equation*}
	where $\ve_i = \sum_{1\leq j\leq k\leq 3}Z_{i,j,k} E_{j,k}(t_i)$, and the collection of $Z_{i,j,k}$ are independent Gaussian random variables such that $Z_{i,j,k}\sim \mathcal{N}(0,1)$ if $j=k$ and $Z_{i,j,k}\sim \mathcal{N}(0,1/4)$. Here,   $\cP_{\mu( s)}^{\mu(t)}$ is the parallel transport map from $\mu( s)$ to $\mu( t)$ along $\mu$.
\end{remark}

% \begin{remark}
%     The assumption \ref{LipHess} holds when the support of  data    is a  bounded subset of $\mathcal{M}$, and  can be replaced with sub-Gaussian conditions, for example, $\max_{1\leq i\leq T}\mathbb{P}(d_\cM(X_i,\mu)>M)\leq \exp(-CM^2)$ for some positive  constant $C<\infty$ and any $M>0$.  Euclidean counterparts of Assumptions \ref{G:Lip}-\ref{Sigma} are common in the literature of stationarity test, such as \cite{2013zhou,zhou2023}. The condition \ref{G:subexp} is required to control the variation induced by the curved nature of manifolds. {Stronger conditions were used in previous works of non-Euclidean data analysis. For example,   \cite{alex2019,dubey2020frechet} assumed bounded support of data, and   \cite{AOS2095} assumed sub-Gaussian condition.  }  
% \end{remark}

\begin{lemma}\label{lem1}
Let $H_i = H(\mu,X_i)$.
If  Assumptions \ref{LipHess}-\ref{G:subexp} hold and   $\mu(t)\equiv \mu$ for some constant $\mu\in\mathcal{M}$, 
then $d(\hat{\mu}, \mu)=O_p(T^{-1/2})$. In addition, there uniquely exists 
$\cH (t):[0,1]\to \mathcal{S}_\mu\mathcal{M} $, an $\mathcal{S}_\mu\mathcal{M}$-valued function,   such that
$\sup_{1\leq k\leq T }\|\cH({k}/{T})-{T}^{-1}\sum_{i=1}^k H_i\|_\mu = O_p(T^{-1/2})$.  
\end{lemma}
We are ready to present the asymptotic null distribution of $Q_T$ in the following theorem. 
\begin{theorem}\label{thm1}
If  Assumptions \ref{LipHess}-\ref{G:subexp} hold and that  $\mu(t)\equiv \mu$ for some constant $\mu\in\mathcal{M}$, then 
\begin{equation}\label{gp}
    Q_T \overset{\mathcal{D}}{\to }\sup_{0\leq t\leq 1} \|U(t)-\cH(t)\circ \cH^{-1}(1)\circ U(1)\|_{\mu},
\end{equation}
where $\cH$ is introduced in Lemma \ref{lem1} and $U(t) =u(t)^\T \mathbf{E}(0)$  with $u(t)$ being a centered Gaussian process
with covariance function 
$\Sigma_u(t,s) =\int_0^{\min(t,s)}\Sigma_\mathbf{E}(\xi)d\xi.$
\end{theorem}
Theorem \ref{thm1} states that the null distribution of the test statistic $Q_T$ converges to the distribution of the sup-norm of a centered Gaussian process defined on $\mathcal{T}_\mu\mathcal{M}$.
 \begin{figure}[t]
    \centering
    \includegraphics[width=0.95\textwidth]{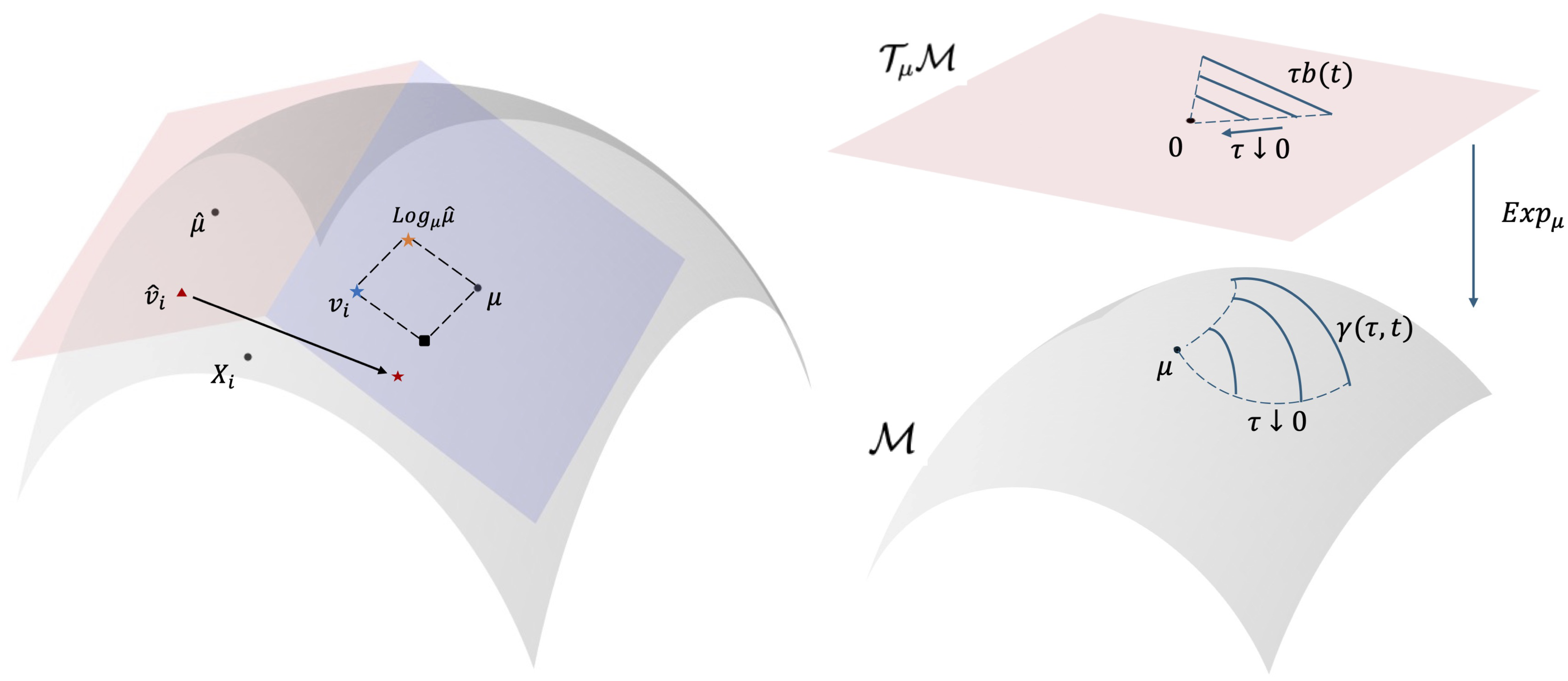}
    \vspace{-0.1cm}
    \caption{Left Panel:
    Illustration on how a curved manifold  differs from Euclidean space and affects the CUSUM statistics. Assume  $\mu$ is the population intrinsic mean,  $\hat{\mu}$ is the sample intrinsic mean, and $X_i$ is a data point in $\cM$. Let  $v_i=\Log_\mu X_i$ and $\hat{v}_i =\Log_{\hat{\mu}} X_i $.
    % Points in $\cM$ are represented by black dots, points in $\cT_\mu\cM$ are represented by stars, and points in  $\cT_{\hat\mu}\cM$ are represented by triangles. 
    The red star  {\color{BrickRed}$\filledstar$} represents $\cP_{\hat\mu}^\mu v_i$, and the square $\blacksquare$
    represents $v_i-\Log_\mu\hat\mu$. In Euclidean space, $\cP_{\hat\mu}^\mu \hat{v}_i  = X_i-\hat\mu$, $v_i=X_i-\mu$, $\Log_\mu\hat\mu=\hat\mu-\mu$, and thus $\cP_{\hat\mu}^\mu v_i = v_i-\Log_\mu\hat\mu$, or equivalently $X_i-\hat\mu= (X_i-\mu)-(\hat{\mu}-\mu)$. However, in a curved manifold, as shown in the figure,  $\cP_{\hat\mu}^\mu v_i$ ({\color{BrickRed}$\filledstar$}) deviates from $v_i-\Log_\mu\hat\mu$ ($\blacksquare$); this deviation contributes to the CUSUM statistics, which is unknown and  need to be estimated from data. Right Panel: Illustration of the local alternative. We consider a perturbation $\tau(T)\cdot  b(t)$  on the tangent space $\cT_\mu\cM$. Let $\gamma(s,t) = \Exp_\mu \big(s \cdot b(t)\big)$. As $T\to \infty$, $\tau(T)$ converges to $0$, and  the mean function $\mu_T(t) =\gamma(\tau(T),t)$ converges to $\mu$. 
    % \kong{$\tau$ and $\tau(T)$ mean the same thing? May need to unify or explain to avoid confusing?}
    % Intuitively, $\cP_{\hat\mu}^\mu\hat{e}_i$, which is represented by the red star in this figure, and $\Log_\mu\hat{\mu}$, represented by the blue dot,  serve as counterparts of $X_i-\hat{\mu}$ and $\hat{\mu}-\mu$ in curved manifold, respectively. However,  $\cP_{\hat\mu}^\mu\hat{e}_i+\Log_\mu\hat{\mu}$, represented by the blue triangle, is not identical to $e_i$, i.e., the identity $ X_i-\hat{\mu}=(X_i-\mu)-(\hat\mu-\mu)$ in Euclidean space cannot be extended to curved manifolds. In manifold,  the deviation of $\cP_{\hat\mu}^\mu\hat{e}_i$  (i.e., $X_i-\hat{\mu}$ in Euclidean space) from  $\Log_\mu X_i-\Log_\mu\hat{\mu}$  (i.e., $(X_i-\mu)-(\hat\mu-\mu)$ in Euclidean space) contributes to the CUSUM statistics,  which is unknown 
    % and  need to be estimated from data. \lin{The figure seems messy. Perhaps use the same type of symbols for the same type of objects, e.g., dots for points on the manifold and star for tangent vectors of $T_\mu\cM$ and triangle for tangent vectors from $T_{\hat \mu}\cM$, and make this clear in the figure caption?}
% \kong{It may be better to use different symbols for {\color{BrickRed}$\filledstar$} and {\color{violet}$\filledstar$} because readers or some referees may print in black and white, so they can not see the color difference. }
} 
     \label{tpmu}
\end{figure}
In Euclidean space and Hilbert space, the operator valued function $\cH(t)$ is given by $\cH(t)=t \circ \text{Id}$. In this case, we have $\cH(t)\circ \cH^{-1}(1)=t \circ \text{Id}$ and $Q_T$ weakly converges to $\sup_{0\leq t\leq 1}\|U(t)-tU(1)\|_2$, which is identical to the convergence of the asymptotic distribution of $T^{-1/2}\max_{1\leq k\leq T} \|\sum_{1\leq j \leq k} X_j - T^{-1} \sum_{1\leq l \leq T} X_l \|$ as given in  \cite{2013zhou}. 
 However, for a first-order stationary time series in a general  Riemannian manifold with non-vanishing  curvatures, $\cH(t)\circ \cH^{-1}(1)\neq t\circ \text{Id}$,  
and the test proposed by \cite{2013zhou} is no longer valid since it does not include the additional term $\cH(t)$ induced by the curvature. Intuitively, the difference between   $\cH(t)\circ \cH^{-1}(1)$ and $t\circ \text{Id}$  is induced by the deviation shown in   Figure \ref{tpmu}, i.e., the deviations of $\cP^\mu_{\hat{\mu}}\Log_{\hat\mu}X_i$ from $\Log_\mu X_i-\Log_\mu\hat\mu$. 
% \kong{a bit more specific about deviation of what.}

The limiting process established by Theorem \ref{thm1} 
includes two components, specifically, a Gaussian random process $U(t)$ with a complicated covariance function and a deterministic operator-valued function $\cH(t)$. To perform a valid test under null hypothesis, we propose to approximate the deterministic function $\cH(t)$ by a CUSUM statistic  and bootstrap  the random process $U(t)$ by adapting the Gaussian multiplier bootstrap in \cite{2013zhou}. Specifically, for $t=k/T$ with some positive integer $k$,  we take $\widehat\cH(t)=T^{-1}\sum_{i=1}^k H(\hat\mu,X_i)$ as an estimate of $\cH(t)$.
% Otherwise, $k/T<t<(k+1)/T$    for some $k$, and we take  \lin{$\widehat\cH(t)=T^{-1}\sum_{i=1}^k H(\hat\mu,X_i)+(t-k/T)  H(\hat\mu,X_{k+1})$.} 
Roughly speaking, $\widehat\cH(\cdot)$ can be viewed as a plug-in estimate of $\cH(\cdot)$ by substituting $\mu$ with $\hat\mu$.

We bootstrap the Gaussian process $U(t)$  by a moving-block multiplier bootstrap procedure, as follows. Let $n$ be a fixed block size.
For each bootstrap sample, generate i.i.d standard Gaussian random variables $\{R_k\}_{k=n}^{T-n+1}$. For $t=k/T$ with  $k\in\{1,\ldots,T\}$, define 
$U_\star(t)=\sum_{j=1}^k\{n(T-n+1)\}^{-1/2}R_j\sum_{i=j}^{j+n-1} \rlog_{\hat\mu}X_i $. Via resampling from $U_\star$, we can obtain an estimate of the null distribution of $Q_T$; see 
Algorithm \ref{boot1}, where a test procedure is  provided in Step 5. The following theorem establishes the consistency of the Gaussian multiplier bootstrap method with curvature term adjustment  under the null, showing that the proposed test procedure is asymptotically valid.

\begin{theorem}\label{thm2}
    Suppose that the assumptions in Theorem \ref{thm1} hold  and the block-size $n:=n(T)$ satisfies 
    $\lim_{T\to\infty}n(T) =\infty$, and $\lim_{T\to\infty}T^{-1}n(T) =0$. Under $\boldsymbol{H}_0$, conditioning on $\{X_i\}_{i=1}^T$, we then have
    \begin{equation}
        Q_T^{(b)} \overset{\mathcal{D}}{\to} \sup_{0\leq t\leq 1} \|U(t)-\cH(t)\circ \cH^{-1}(1)\circ U(1)\|_{\mu}.
\end{equation} 
\end{theorem}

\begin{algorithm}[htbp]
\caption{Curvature Adjusted  Multiplier Bootstrap (CAMB)
}\label{boot1}
\begin{algorithmic}
\STATE\textbf{Input:} Manifold time series $\{X_i\}_{i=1}^T$,  bootstrap sample size $B$, and the significant level $\alpha $.
\STATE \textbf{1.} Estimate empirical intrinsic mean $\hat\mu =\arg\min_{p\in\mathcal{M}}T^{-1}\sum_i d_\mathcal{M}^2(p,X_i)$. Estimate the Riemannian Hessian tensor $\hat{H}_i$ by the plug-in estimator $\hat{H}_i = H(\hat{\mu},X_i)$, and the tensor-valued process $\hat{\cH}_j ={T}^{-1}\sum_{i=1}^j \hat{H}_i,~j=1,\cdots,T$.
\STATE \textbf{2.} Compute the residuals $v_i= \rlog_{\hat\mu}X_i$, and determine the moving-block size $n$ 
% \kong{maybe it is slightly better to use other notation for moving-block size as $n$ usually refers to sample size unless $n$ is commonly used in time series literature. How about $ h$?}
by the minimum-volatility method \citep{politis2012subsampling}.
\STATE \textbf{3. } Compute the CUSUM $S_j = \sum_{i=1}^j v_i $ for $1\leq j\leq T$, the test statistic $Q_T$,  and 
the moving-block local sum $S_{j,n} =\sum_{i=j}^{j+n-1}v_i$ for $1\leq j\leq T-n+1$.
\STATE \textbf{4. } Generate bootstrap samples of $Q_T$:
\bindent
\FOR{$b=1,\cdots,B$} 
\STATE\textbf{i.} Generate $T-n+1$  i.i.d
standard normal random variables $\{R^{(b)}_j\}_{j=1}^{T-n+1}$.
\STATE \textbf{ii.}  $V^{(b)}_{k,n} = \sum_{j=1}^k\{{n(T-n+1)}\}^{-1/2}S_{j,n} R^{(b)}_j,~\text{for}~  k=n,\cdots, T-n+1$.
\STATE\textbf{iii.} $Q^{(b)}_{T}= \max_{n\leq k\leq T-n+1}\|V^{(b)}_{k,n}- \hat\cH_k \circ \hat\cH_T^{-1}\circ  V^{(b)}_{T-n+1,n}\|_{\hat\mu} $.
\ENDFOR
\eindent
\STATE \textbf{5. } 
Obtain the bootstrap  $p\text{-value}=({B}^{-1})\sum_{b=1}^B I\{Q_T^{(b)}\geq Q_T\}$,  and reject $H_0$ if $p\text{-value}\leq \alpha$.
\end{algorithmic}
\end{algorithm}

%\begin{remark}
%    In Euclidean space, test statistics can be constructed using  $L_1$, $L_2$ or $L_\infty$ -norms of the demeaned version of data, while in our paper the test is based on  the $L_2$-norm naturally induced by  the Riemannian metric. \lin{The reason is that the consistency is only guaranteed by this canonical Riemannian norm in curved manifolds.}
%\end{remark}
Next we study the asymptotic local power of the proposed test, where we utilize tools of parametrized surfaces in manifolds \citep{do1992riemannian}. Let $\mu\in\mathcal{M}$ be a constant, and $b(t):[0,1]\to \cT_\mu\cM$ be a smooth curve,  $\gamma(s,t) =\Exp_\mu\{s\cdot b(t)\},~0\leq s,t\leq 1 $ be a parametrized surface near $\mu$, and $\{E_j(s,t), j =1,\cdots,d,~0\leq s, t\leq 1 \}$ a collection of vector fields such that
\begin{itemize}
    \item For fixed $s$,  $\{E_j(s,t),~0\leq t\leq1~ ,j=1,\cdots,d\}$ is a parallel orthonormal frame along $\gamma$;
    \item $E_j(s,t):[0,1]\times [0,1]\to \cT\cM$ is smooth   on $[0,1]\times [0,1]$ for all $j=1,\cdots,d$.
\end{itemize} 
We consider the following local alternative hypothesis under the locally stationary scheme:
\begin{equation}\label{localalter}
    \mu_T(t) = \gamma\big(\tau(T),t\big), \text{ with }\tau(T)~\text{being a non-negative sequence}\text{ s.t. }\lim_{T\to\infty}{\tau(T)}=0,
\end{equation}
for a locally stationary time series $\{X_i\}_{i=1}^T$ as in Definition $\ref{locstadef}$ with 
\begin{equation}\label{localalter2}
    \Log_{\mu_T(i/T)} = e_i^\T \mathbf{E}(\tau(T),{i}/{T}).
\end{equation}
{A visual illustration of this local alternative is provided Figure \ref{tpmu}(b).}
% \begin{figure}[h!]
%     \centering
%     \includegraphics[width=0.8\textwidth]{localalter.png}
%     \caption{Illustration of the local alternative. We consider a perturbation $\tau b(t)$  on the tangent space $\cT_\mu\cM$. Let $\gamma(\tau,t) = \Exp_\mu \big(\tau b(t)\big)$. 
% w    As $T\to \infty$, $\tau(T)$ converges to $0$, and  the mean function $\mu_T(t) =\gamma(\tau(T),t)$ converges to $\mu$.    }
%      \label{la_viz}
% \end{figure}
This local alternative scheme  possesses two properties.  First, for each $T$, the data is locally stationary associated with the mean curve $\mu_T(\cdot)$ and parallel orthonormal frame $\mathbf{E}(\tau,\cdot)$. Second, as $T\to\infty$, the time series  smoothly changes and uniformly converges to a first-order stationary time series at a rate $\tau(T)$. In Euclidean space, this local alternative scheme is identical to the case where $X_i = \mu_T(i/T)+ e_i$ with  $\mu_T(t)=\mu+\tau(T) b(t)$ for a smooth function $b(t)$ and $\{e_i\}_{i=1}^T$ is a zero-mean  locally stationary time series. 
The following theorems present the asymptotic results for the local alternative.
\begin{theorem}\label{thm3}
    Assume   \ref{LipHess}-\ref{G:subexp} and the local alternative hypothesis given by Eq.\eqref{localalter} and Eq.\eqref{localalter2}.
    \begin{enumerate}
        \item If $\lim_{T\to\infty} T^{1/2}\tau({T})\to \infty  $, then $Q_T \to \infty$ almost surely.
        \item  If $\tau(T)= T^{-1/2}$, then,      with $\cH(t)$ defined in Lemma \ref{lem1}, we have 
        \begin{equation}\label{la2}
        \begin{aligned}
             Q_T &\overset{\mathcal{D}}{\to }\sup_{0\leq t\leq 1} \|U(t)-\cH(t)\circ \cH^{-1}(1)\circ U(1) \\&+ \cH(t)\circ\cH^{-1}(1)\circ\int_0^1 \frac{\partial}{\partial \xi}\cH(\xi)\circ b(\xi) d\xi-
             \int_0^t \frac{\partial}{\partial \xi}\cH(\xi) \circ b(\xi)d\xi\|_{\mu}.
        \end{aligned}
        \end{equation}
   
    \end{enumerate}
\end{theorem}

\begin{theorem}\label{thm4}
Under the conditions of Theorem \ref{thm3}, if we further assume that $\lim_{T\to\infty}n(T) =\infty$ and \\$\lim_{T\to\infty}{n(T)}^{1/2}\tau(T)=0$, then the bootstrap procedure in Algorithm \ref{boot1} is consistent in the sense that, conditioning on $\{X_i\}_{i=1}^T$,  
$Q_T^{(b)} \overset{\mathcal{D}}{\to} \sup_{0\leq t\leq 1} \|U(t)-\cH(t)\circ \cH^{-1}(1)\circ U(1)\|_{\mu}.$    
\end{theorem}
Theorem \ref{thm4} suggests that, even under the local alternative, the bootstrap samples $Q_T^{(b)}$ are asymp-totically drawn from the limiting null distribution, with some suitable block size $n$ that meets a stronger condition  $n(T)^{1/2}\tau(T)\to 0$ 
compared with those in Theorem \ref{thm2}. 
Theorems \ref{thm3} and \ref{thm4} together show that  our method can detect the first-order non-stationarity with rate $T^{-1/2}$ and has asymptotic power $1$ whenever $\lim_{T\to\infty}{n(T)}^{1/2}\tau(T) =0$ and $\lim_{T\to\infty}{n(T)}=\infty$.
    Note that, in Theorem \ref{thm3}, 
    if $b(\cdot)\equiv 0$, i.e., under the null hypothesis, the asymptotic distribution of $Q_T$ given by Eq.\eqref{la2} is identical to  the one in Theorem \ref{thm2}.
\begin{remark}
    Our test for first-order stationarity differs from previous change point detection methods \citep{dubey2020frechet,wang2023online,jiang2024two} by examining whether the mean is constant or varies (continuously or discontinuously) over time, allowing gradual changes. In contrast, their methods detect abrupt changes and assume the time series can be segmented into blocks of constant mean and variance, a condition not required in our test.
\end{remark}

\subsection{Second-order stationarity test}
If a manifold time series is first-order stationary, it is natural to further test the second-order stationarity.
% \kong{Any application in real world?} 
Below we propose a second-order stationarity test for  first-order stationary manifold time series using  local spectral density \citep{dal1997,holger2011,anne2021}.

Let $\{X_i\}_{i=1}^T$  be a first-order stationary manifold time series with constant intrinsic mean $\mu$. 
% , and $\{e_i\}_{i=1}^T$ be the corresponding $\mathbb{R}^D$-valued process of $\{\Log_\mu X_i\}_{i=1}^T$ .
The local spectral density of the coordinate representation of  $\{\Log_\mu X_i\}_{i=1}^T$ under a given orthonormal frame $\mathbf{E}$, i.e., the time series $\{e_i \}_{i=1}^T$,  is 
  $ F_\mathbf{E}(\omega,t) = ({2\pi})^{-1}\sum_{h\in\mathbb{Z}}\mathbb{E}\{G_\mathbf{E}(t,\mathcal{F}_0)G^\T_\mathbf{E}(t,\mathcal{F}_h)\}e^{-\bi\omega h},~\lambda\in[-\pi,\pi]$,
where we define $\bi=\sqrt{-1}$  throughout this paper. Under some technical assumptions introduced later, the local spectral density is well-defined.
% Note that $F(\omega,t)$ is independent of the choice of $\mathbf{E}$.
The second-order stationarity of $\{X_i\}_{i=1}^T$  is equivalent to 
$ F_\mathbf{E}(\omega,t)\equiv F_\mathbf{E}(\omega)$ a.e. on $ [-\pi,\pi]\times [0,1],~\text{for some function}~F_\mathbf{E}(\omega)$.
Thus, testing the second-order stationarity is  equivalent to testing the following hypothesis:
\begin{equation}\label{secondhypo}
\begin{aligned}
    \boldsymbol{H_0}: &~~ F_\mathbf{E}(\omega,t)\equiv F_\mathbf{E}(\omega), ~\text{a.e. on}~ [-\pi,\pi]\times [0,1];\\
    \boldsymbol{H_1}: &~~ F_\mathbf{E}(\omega,t)\neq F_\mathbf{E}(\omega) ~\text{for all}~ F_\mathbf{E}(\omega) ~\text{on a subset of} ~[-\pi,\pi]\times [0,1] \\
    & \quad \text{with positive Lebesgue measure}.
\end{aligned}   
\end{equation}
We then define  squared variation  of $F_\mathbf{E}(\omega,t)$ by
\begin{equation*}
V^2_F= \int_{-\pi}^\pi\int_0^1\|F_\mathbf{E}(\omega,u)\|_{\text{HS}}^2 dud\omega -\int_{-\pi}^\pi \|\bar{F}_\mathbf{E}(\omega)\|_{\text{HS}}^2 d\omega ,
\end{equation*}
where $\bar{F}_\mathbf{E}(\omega) = \int_0^1 F_\mathbf{E}(\omega,t)dt$ and $\|\cdot\|_{\text{HS}}$ is the Hilbert-Schmidt norm of complex matrices. Since Hilbert-Schmit norm is invariant under unitary transformation,  $V_F^2$ is independent of choices of $\mathbf{E}$.  Note that $V^2_F = 0$ if and only if $F_\mathbf{E}(\omega,t)\equiv \bar{F}_\mathbf{E}(\omega)$, $\text{a.e. on}~ [-\pi,\pi]\times [0,1]$. Thus, the testing the hypothesis in \eqref{secondhypo} is equivalent to testing 
\begin{equation}\label{secondhypo2}
    \boldsymbol{H_0}:~ V^2_F =0,~\text{versus}~~\boldsymbol{H_1}:~ V^2_F>0.
\end{equation}

We adapt the technique from \cite{anne2021}, initially created for assessing second-order stationarity in functional time series, into our context of manifold time series. Let $m$ and $n$ be two positive integers such that $mn=T$ and $n$  is  even. The intuition is to split the time series into $m$ blocks of size $n$, and then estimate the local spectral density and the squared variation $V^2_F$. Let $\hat{\mu}$ be the empirical intrinsic mean and 
$I_n(\omega,t) = J_n(\omega,t)\otimes J_n(\omega,t)$, where { $\otimes$ is the complex tensor product (i.e., conjugation included) } and 
$$ J_n(\omega,t)= {({2\pi n})^{-1/2} } \sum_{h=0}^{n-1} \Log_{\hat{\mu}} X_{\floor{tT}-n/2+1+h} \cdot e^{-\bi  h \omega}.
 $$
 Then the coordinate representation of $I_n(\lambda,t)$ under any orthonormal frame at $\hat{\mu}$ can serve as an estimator of $F(\lambda,t)$, and the test statistic is given by
\begin{equation*}
\begin{aligned}
    {V}^2_{\hat{F}}&={4\pi}{T}^{-1}\sum_{k=1}^{n/2}\sum_{j=1}^m \langle I_n(\omega_k,t_j),I_n(\omega_{k-1},t_j)\rangle_{\text{HS}}+\widehat{W} - \\
    &\quad {4\pi}{n}^{-1} \sum_{k=1}^{n/2} \langle {m}^{-1}\sum_{j=1}^mI_n(\omega_k,t_j),{m}^{-1}\sum_{j=1}^mI_n(\omega_{k},t_j)\rangle_{\text{HS}} ,
\end{aligned}
\end{equation*}
where $\omega_k = {2 k\pi}/{n}$, $t_j = {n (j-0.5)}/{T}$, and 
$\widehat{W} ={T}^{-1} \sum_{k=1}^{n/2}\sum_{j=1}^m\|J_n(\omega_k,t_j)\|_{\hat\mu}^2 \|J_n(\omega_{k-1},t_j)\|_{\hat\mu}^2$.    

To develop a test based on the statistic $V_{\hat F}^2$,  we proceed with studying its  asymptotic distribution. Let 
$Y_i(t) =(Y_{i,1}(t), \cdots,Y_{i,d+d(d+1)/2}(t))$ be a vector  such that $\left(Y_{i,1}(t), \cdots, Y_{i,d}(t)\right) = G_\mathbf{E}(t,\mathcal{F}_i)$ and 
\begin{align*}
   Y_{i,d+(d-j/2)(j-1)+k}(t) = \langle E_j(t),H(\mu,\Exp_\mu\{G_\mathbf{E}(t,\mathcal{F}_i)^\T \mathbf{E}(t) \})\circ E_{j+k}(t)\rangle_\mu, ~1\leq j \leq d,~j\leq k \leq d .
\end{align*}
% Y_{i}(t) =\textbf{Vec}\Bigg(\{ G_\mathbf{E}(t,\mathcal{F}_i)_j,~1\leq j\leq d\}\cup\{\langle E_j(0), H(\mu,\Exp_\mu(G_\mathbf{E}(t,\mathcal{F}_i)^\T \mathbf{E}(0) ))\circ E_k(0)\rangle_\mu,~1\leq j\leq k\leq d\}  \Bigg),$$
% where $\textbf{Vec}(\cdot)$.
% $$Y_{i,j} = \langle\Log_\mu X_i, E_j \rangle_\mu,~1\leq j\leq d,$$
% and 
% $$Y_{i,(d+1):} = \langle , H(\mu,X_i) E_j \rangle_\mu,~1\leq j\leq d $$
Given $k$ random variables $Z_1,\cdots,Z_k$, we  denote  the $k^{th}$ order joint cumulant of  $k$ random variables $\{Z_1,\cdots,Z_k\}$ by  $\cum_k(Z_1,\cdots,Z_k)$. 
We assume that $\{X_i\}_{i=1}^T$  satisfies the following conditions.  For every even number $k\in\mathbb{N}$,
there exists a positive sequence $\alpha_{k;i_1,\cdots,i_{k-1}}$ such that, for all $j=0,\cdots,k-1$ and for some $\ell\in\mathbb{N} $,  we have 
$\sum_{i_1,\cdots,i_{k-1}}(1+|i_j|^\ell ) \alpha_{k;i_1,\cdots,i_k}<\infty$, and 
\begin{enumerate}[label=(\textbf{C\arabic*})]

\item \label{cum1}  $\sup_{0\leq t\leq1}|\cum_{k}\{Y_{i_1,l_1}(t),\cdots,Y_{i_k,l_k}(t)\} |\leq \alpha_{k;i_1-i_k,\cdots,i_{k-1}-i_k}$, for all $(l_1,\cdots,l_k)\in\{1,\cdots,d+d(d+1)/2\}^k$,
  \item \label{cum2} $\sup_{0\leq t\leq1}|\frac{\partial}{\partial t^\ell}\cum_{k}\{Y_{i_1,l_1}(t),\cdots,Y_{i_k,l_k}(t)\} |\leq \alpha_{k;i_1-i_k,\cdots,i_{k-1}-i_k}$, for all $(l_1,\cdots,l_k)\in\{1,\cdots,d+d(d+1)/2\}^k$.
\end{enumerate}
{The conditions \ref{cum1} and \ref{cum2} are proposed to guarantee the  existence of the local spectral density and   the  weak convergence of test statistics. Similar conditions are used in previous works \citep{anne2021}.}

\begin{theorem}\label{thm5}
    If $\{X_i\}_{i=1}^T$ is a first-order stationary time series, the  conditions \ref{LipHess}-\ref{G:subexp} and \ref{cum1}-\ref{cum2} hold, and  $T^{1/2}\ll n\ll T^{2/3}$, then under both null and fixed alternative hypotheses, we have 
    \begin{equation*}
    T^{1/2} ({V}^2_{\hat{F}}- V^2_F)\overset{\mathcal{D}}{\to} N(0,\sigma^2_V),~\text{as}~T\to\infty, 
    ~~~\text{where } 0<\sigma^2_V<\infty.
    \end{equation*}
    
    % where under $H_0$ of \eqref{secondhypo2} the variance $\sigma_V^2$ can be consistently  estimated by \lin{what if under $H_1$?}
    % \begin{equation}
    %     \hat{\sigma}_V^2 = \frac{16\pi^2}{n}\sum_{k=1}^{n/2}\left( \frac{1}{m}\sum_{j=1}^m\langle I_n(\omega_{k-1},t_j),I_n(\omega_k,t_j) \rangle_{\text{HS}}\right)^2.
    % \end{equation}
\end{theorem}

\begin{theorem}\label{thm6}
    Suppose conditions \ref{LipHess}-\ref{G:subexp} and \ref{cum1}-\ref{cum2} hold. Then under $\boldsymbol{H}_0$ of \eqref{secondhypo2}, we have 
     $\sigma_V^2=4\pi\int_{-\pi}^\pi \|\bar{F}_\mathbf{E}(\omega)\|^4_{\text{HS}}d\omega.$ In addition, the estimator 
    \begin{equation}\label{sigmahat}
        \hat{\sigma}_{V }^2 = {16\pi^2}{n}^{-1}\sum_{k=1}^{n/2}\left( {m}^{-1}\sum_{j=1}^m\langle I_n(\omega_{k-1},t_j),I_n(\omega_k,t_j) \rangle_{\text{HS}}\right)^2
    \end{equation}
    converges to $4\pi\int_{-\pi}^\pi \|\bar{F}_\mathbf{E}(\omega)\|^4_{\text{HS}}d\omega$ in probability under both $\boldsymbol{H_0}$ and $\boldsymbol{H_1}$.
\end{theorem}
The above theorem implies that  $\hat\sigma_{V}^2$    consistently estimates $\sigma_V^2$ under $\boldsymbol H_0$. It also suggests that, for a significant level $\alpha$, we can conduct the test by rejecting the null hypothesis $\boldsymbol H_0$ if  $T^{1/2}V^2_{\hat{F}}/\hat{\sigma}_V \geq z_{1-\alpha}$, where $z_{1-\alpha}$ is the $1-\alpha$ quantile of standard normal distribution. Under $\boldsymbol{H_1}$,  the quantity  $\hat{\sigma}^2_{V}$ given by \eqref{sigmahat} also  converges to $4\pi\int_{-\pi}^\pi \|\bar{F}_\mathbf{E}(\omega)\|^4_{\text{HS}}d\omega$ in probability. Thus, 
when $0<4\pi\int_{-\pi}^\pi \|\bar{F}_\mathbf{E}(\omega)\|^4_{\text{HS}}d\omega<\infty$, the probability of rejecting the null hypothesis $\boldsymbol{H_0}$ is approximately $\Phi\{({{4\pi\int_{-\pi}^\pi \|\bar{F}_\mathbf{E}(\omega)\|^4_{\text{HS}}d\omega}})^{1/2} z_{1-\alpha}/{\sigma_V} +{T}^{1/2}{V_F^2}/{\sigma_V}\}$, where $\Phi(\cdot)$ is the CDF of the standard normal distribution.
This result implies that the test has asymptotic power $1$ as $T\to \infty$ under any fixed alternative $\boldsymbol{H_1}$. 

Interestingly, Theorem \ref{thm5} implies that under certain regularity conditions,  the null distribution of our test statistic for the second-order stationarity is not affected by the curvature. This is in contrast to the first-order stationarity test, where curvature does have an impact on the null distribution.
 The reason is that the curvature effect in  ${4\pi}{T}^{-1}\sum_{k=1}^{n/2}\sum_{j=1}^m \langle I_n(\omega_k,t_j), I_n(\omega_{k-1},t_j)\rangle_{\text{HS}} $ is  asymptotically neutralized by the curvature effect in $({4\pi}{n}^{-1} )\sum_{k=1}^{n/2} \langle {m}^{-1}\sum_{j=1}^mI_n(\omega_k,t_j),{m}^{-1}\cdot\sum_{j=1}^mI_n(\omega_{k},t_j)\rangle_{\text{HS}}$ under the null hypothesis. Under the alternative  hypothesis, the curvature effect exists and  asymptotically has a form of $\langle U,\Log_\mu\hat{\mu}\rangle_\mu$, where $U$ is an unknown deterministic vector in $\cT_\mu\cM$ and  vanishes when $\cM$ is Euclidean space.  Thus, the  asymptotic alternative  distribution of ${T}^{1/2}(V^2_{\hat{F}}-V^2_{{F}})$  is  a Gaussian distribution, with a variance  different from the one in Euclidean or Hilbert space. The asymptotic variance under the alternative is complex, and therefore not included here; details on the asymptotic behavior related to the curvature impact on the second-order stationarity test can be found in Section \ref{s-proof:thm5} of the supplementary materials. 

\begin{remark}
    Our second-order stationarity test addresses a different setting from the variance change detection in non-Euclidean data proposed by \cite{dubey2020frechet} and \cite{jiang2024two}. Their focus is on detecting changes in $\mathbb{E}d^2_\mathcal{M}(\mu,X_i)$, which, in our context, corresponds to testing whether the trace of the time-varying matrix $F_\mathbf{E}(0,t)$ remains independent of $t$. In contrast, our test examines the entire variance structure, not just its trace, and also accommodates continuously varying variance over time.
\end{remark}

\begin{remark} Although the null distribution of the test statistic  for the second-order stationarity on the manifold is the same as Hilbert space or Euclidean space, the block size  $n$ is more restricted. For the test in  Hilbert space, the upper bound of $n$ is of order $T^{3/4} $ \citep{anne2021}, while for the general manifold it is of order  $T^{2/3}$, since the curvature effect   introduces  a bias of order $T^{-1}{n^{3/2}}$  in non-trivial manifolds, as discussed in the supplementary materials. 
\end{remark}

\begin{remark}
The constant mean assumption is common in the literature on second-order stationarity tests \citep{holger2011,Philip2013,anne2021}, and when $\mu$ is non-constant but smooth, it can be estimated \citep{anne2021}. For instance, we can estimate $\mu$ using methods from \cite{alex2019,AOS2095}, and then estimate $J_n(\lambda, t)$ by parallel transporting $\Log_{\hat{\mu}(i/T)} X_i$ to $\hat{\mu}(1/T)$ along the estimated  curve $\hat{\mu}(\cdot)$ as  a detrending procedure.
    \vspace{-0.2cm}
\end{remark}

\section{Simulations}\label{sim}
In this section, we conduct Monte Carlo simulation experiments to study the finite sample performance of our proposed testing procedures in two cases: (i) hypersphere, a positively-curved manifold example; (ii) SPD-matrices endowed with negatively curved manifold structure. 
\subsection{Simulations  for first-order stationarity test on spherical time series}\label{first_simulation}
In the numerical study of first-order stationarity test, we consider the following two settings, and report results for Type-I error rates under null and power under alternative hypothesis.

Setting (i): We simulate locally stationary time series on  $\mathbb{S}^6=\{x\in\mathbb{R}^7: \|x\|_2^2=1\}$, as follows. Let $t_i=i/ T$ for $1\leq i\leq T$. Take $\mu(t)$ be the  geodesic such that  $\mu(0)=(0,0,0,0,0,0,1)$ and  $\mu(1)=(1,0,0,0,0,0,0)$, and $\mu_\tau(t)=\mu(\tau t)$ be a re-scaled version of $\mu(t)$ for $\tau\in[0,1]$. We also denote $\cP_{\mu_\tau( s)}^{\mu_\tau (t)}(\cdot)$ the parallel transport map from $\mu_\tau( s)$ to $\mu_\tau( t)$ along the geodesic $\mu_\tau(\cdot)$, which is equivalent to the parallel transport map from $\mu( \tau s)$ to $\mu(  \tau t)$ along the geodesic $\mu(\cdot)$ in this simulation setting.
For $j=1,\ldots,6$, let $E_j(0)$ be the vector with $1$ at the $j^{th}$ and with $0$ at the other entries; we view $\{E_j(0)\}_{j=1}^6$ as an orthonormal basis of $\mathcal{T}_{\mu(0)}\mathbb{S}^6$.
Then we consider the following time-varying auto-regressive models 
% \kong{use bracket order $ [\{(\cdot)\}]$ when there are multiple brackets throughout the paper, see the following as an example:}
\begin{equation}\label{arsim1}
           %  inten = 1.1*(1+u)*sigma/(1+tau)     
           %  noise = generate_sample_tangent_space(base=base,n_samples=1,sigma=sigma/(1+tau))
           % # data_tv[i,:]  = rhotmp*delta + noise
 M_1(\tau):~   \rlog_{\mu_\tau( t_{i+1})}X_{i+1} = \left\{0.05+0.5 t_i\cdot(1-t_i)\right\}\cP_{\mu_\tau( t_{i})}^{\mu_\tau ( t_{i+1})} \rlog_{\mu_\tau( t_{i})} X_{i} +({1+\tau})^{-1}{\ve_i}, 
\end{equation}
where $\ve_i = \sum_{j=1}^6 \sigma_j(\tau,t_i) Z_{i,j} E_j(\tau t)$, $\sigma_j(\tau,t_i) = (1.1+1.1 t_i)/(1+\tau)$ if $j=1,2,3$ and $\sigma_j(\tau,t_i) =  1/(1+\tau)$ if $j=4,5,6$, $Z_{i,j}\overset{i.i.d}{\sim} \text{Unif}(-0.5,0.5)$, and $\{E_j(t),1\leq j\leq 6,~0\leq t\leq 1\}$ is a parallel orthonormal frame along $\mu(t)$, {i.e., $E_j(t)=\cP_{\mu(0)}^{\mu( t)}E_j(0) $}. The parameter $\tau$ in Eq.\eqref{arsim1} determines the deviation of the time series from first-order stationarity.
When $\tau=0$, the time series is first-order stationary.  
As $ \tau $ increases, the model will deviate from the null and we use it to evaluate the performance of the test statistic under the alternative. In this setting, we consider $\tau=0.125,~0.25,~ 0.375,~0.75,1.0$.

Setting (ii): We next consider $\text{Sym}^+_3$, the space of $3\times 3$ SPD matrices endowed with the affine-invariant metric \citep{moakher2005differential}, which is a  six-dimensional negatively-curved Riemannian manifold. Let $\mu(t)$ be a geodesic joining $I_3$ and $2I_3$   such that $\mu(0)=I_3$ and $\mu(1)=2I_3$, and define $\mu_\tau(t)=\mu(\tau t)$.
Let  $\{E_{j,k}(0)\}_{1\leq j\leq k\leq 3}\subset \text{Sym}_3$ be  the set of $3\times 3$ symmetric matrices  with $1$ at the  $(j,k)$ and $(k,j)$ entries   and $0$ at the remaining entries. Note that $\{E_{j,k}(0)\}_{1\leq j\leq k\leq 3}$ form an orthogonal basis of $\mathcal{T}_{\mu(0)}\text{Sym}_3^+$, with $\|E_{j,k}(0)\|_{\mu(0)} = 1 $ for $j=k$, and $\|E_{j,k}(0)\|_{\mu(0)} = \sqrt{2} $ for $j\neq k$. Let $\{E_{j,k}(t):1\leq j\leq k\leq 3,~0\leq t\leq1 \}$ be the parallel orthogonal  frame along $\mu(t)$ with initial value $E_{j,k}(0)$. 
We simulate the following time-varying auto-regressive process:
\begin{equation*}
   % rhotmp = rho+0.25*u
   %          inten = ( (2.5*(u-0.25)) **2+0.2 )*sigma
   %          noise = generate_sample_tangent_space(base=gs.eye(d), n_samples=1, sigma=(inten/(1+2*tau)))
 M_2(\tau):~    \rlog_{\mu_\tau(t_{i+1})}  X_{i+1} = \left(0.05+0.25t_i \right) \cP_{\mu_\tau( t_{i})}^{\mu_\tau ( t_{i+1})} \rlog_{\mu_\tau(t_{i})} X_{i} +{(1+2\tau)^{-1}}\{6.25 (t_i-0.25)^2+0.2 \}{\ve_i}, 
\end{equation*}
where $\ve_i = \sum_{1\leq j\leq k\leq 3}Z_{i,j,k} E_{j,k}(\tau t_i)$, and the collection of $Z_{i,j,k}$ are independent Gaussian random variables such that $Z_{i,j,k}\sim \mathcal{N}(0,1)$ if $j=k$ and $Z_{i,j,k}\sim \mathcal{N}(0,1/4)$. Similarly,   $\cP_{\mu_\tau( s)}^{\mu_\tau (t)}(\cdot)$ is the parallel transport map from $\mu_\tau( s)$ to $\mu_\tau( t)$ along the geodesic $\mu_\tau(\cdot)$.
When $\tau=0$, the manifold time series is first-order stationary. For power study under alternative, we consider $\tau = 0.25,~0.5,~0.75,~1,~1.25,~1.5,~1.75$.
%The Type-I error rates are also reported in Table. \ref{table1}. We observe that the bootstrap for residuals ignoring the curvature effect is conservative in this example, with empirical rejection probability  around $0.009$, which will lose power under the alternative hypothesis. We observe that when treating data as  Euclidean multivariate time series,   eigenvalues of the sample mean and variance of the data explode as sample size $T$ increases. The Type-I error rates of bootstrap are not reliable in this case. 
%In the right panel of Figure \ref{power_first}, we show the power curve with $\tau$ ranging between $0$  and $1.75$. As expected, when $\tau$ increases, the probability of rejecting null hypothesis grows to $1$. 

% We conduct $ 5000 $ replications of Monte Carlo runs. We compare with direct application of the  bootstrap in \cite{2013zhou} to the Riemannian gradients of data in the tangent space of the empirical intrinsic mean. 
% \kong{Is this equivalent to treating the data in Euclidean space without the composition constraint?} 

For the first-order stationary test in both scenarios, we consider $T=50,100,500$. The number of Monte Carlo runs is $5000$.
The null distribution of the test statistic $Q_T$ is estimated by the  curvature adjusted multiplier bootstrap (CAMB) method in Algorithm \ref{boot1}. We compare the proposed test against two approaches: (B1) a method that bootstrap $\sup_t\|U(t)-tU(1)\|_\mu$, neglecting the curvature effect $\cH(t)$, and (B2) an approach that considers manifold time series as Euclidean multivariate time series, utilizing the multiplier bootstrap technique suggested by \citep{2013zhou}, thereby overlooking the manifold structure. The bootstrap sample size is set as $B=2000$ for all  methods in this benchmark study.

The Type-I error rates of our method and the two comparison methods are reported in Table \ref{table1}. 
In the sphere scenario, we find that  the Type-I error rates of the two comparison methods are inflated, while our method controls the Type-I error well. In the context of SPD matrices, the first comparison method, B1, which overlooks the curvature, tends to be overly conservative in this instance, exhibiting an empirical rejection probability of approximately $0.009$. This conservative approach may result in diminished power under the alternative hypothesis. Results of the comparison method B1 in both scenarios numerically support  that the curvature term $\cH(t)$ plays an important role in manifold time series. On the other hand, the second method, B2, which treats the data as Euclidean multivariate time series, leads to an escalation in the eigenvalues of the sample mean and variance. Consequently, the variance associated with the multiplier bootstrap also surges, rendering the Type-I error rates for this approach unreliable in this scenario.  

We also evaluate the power of the first-order stationary test by varying $\tau$.  For each fixed $\tau$, the power is calculated based on $5000$ repetitions of Monte Carlo runs. We plot the power curve in 
Figure \ref{power_first}, and as expected,  one can observe that the test becomes more powerful as $ T $ increases, and the power will ultimately reach $ 1$ as $ \tau $ continues to increase. 

% \begin{figure}[h]
% \centering 
% \subfigure{
% \begin{overpic}[width=0.48\textwidth]{Power_sphere.png}
% \put(3,75){(A)} % Adjust the position as needed
% \end{overpic}
% }
% \subfigure{
% \begin{overpic}[width=0.48\textwidth]{Power_spd.png}
% \put(3,75){(B)} % Adjust the position as needed
% \end{overpic}
% }
% \vspace{-0.5cm}
% \caption{Simulated power curves for first-order stationarity test. Panel (A): power curve for first-order stationarity test of spherical time series. Right Panel: power curve for first-order stationarity test of SPD-matrix-valued time series.}
% \label{power_first}
% \end{figure}

\begin{figure}[t]
\centering 
\subfigure{
\includegraphics[width=0.4\textwidth]{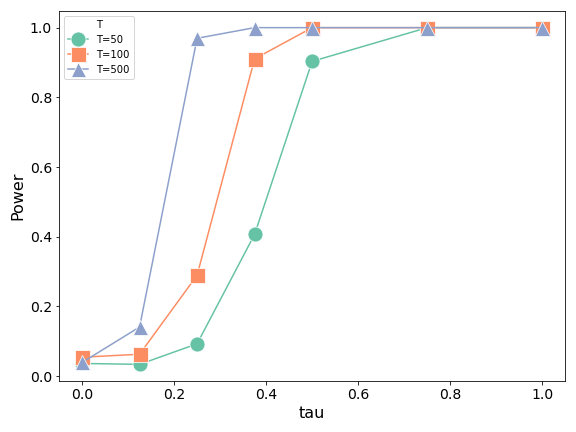}}~~~~~
\subfigure{
\includegraphics[width=0.4\textwidth]{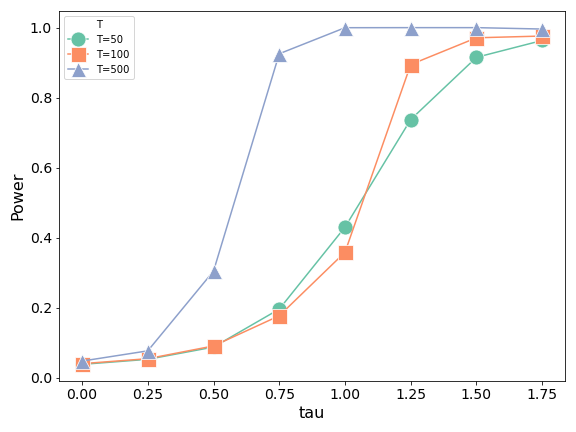}
}
\vspace{-0.1cm}
\caption{Simulated power curves for the first-order stationarity test. Left Panel: power curve for first-order stationarity test of spherical time series. Right Panel: power curve for first-order stationarity test of SPD-matrix-valued time series. The significant level is $0.05$. }
\label{power_first}
\vspace{-0.3cm}
\end{figure}

\subsection{Simulations  for second-order stationarity test}
To study  the second-order stationarity test, 
for both  $\mathbb{S}^6$ and $\text{Sym}_+^3$ settings,  we simulate locally  stationary manifold time series which are first-order stationary from the model
\begin{equation*}\label{AR1_second}
  % u =i/n_samples
  %           rhotmp = rho+ tau*(0.2*np.cos(2*u*np.pi)+u*(1-u)) 
  %           #rhotmp =  rho
  %           inten = sigma*(0.5 + tau*u**2)
  %           delta = data_tv[i - 1, :]
 M_3(\tau):~   \Log_{\mu}X_{i+1} =\left[0.1+\tau \{0.2\cos(2\pi t_i)+t_i*(1-t_i)\}\right]\cdot\Log_{\mu}X_{i} +\epsilon_i.
\end{equation*}
In the $6$-dimensional hypersphere  $\mathbb{S}^6$ case, we set $\mu = (0,0,0,0,0,0,1)$ and take $\{E_j\}_{j=1}^6$ be an orthonormal basis of $\cT_\mu\cM$, as the $E_j(0)$ in Setting (i) of Section \ref{first_simulation}.
We set $\ve_i = \sum_{j=1}^6 Z_{i,j} E_j$, and $Z_{i,j}\overset{i.i.d}{\sim} \text{Unif}(-0.75,0.75)$. 
For the $\text{Sym}_+^3$-valued time series, $\mu$ is set to be $I_3$, and $\epsilon_i$ is generated in the same way as Section \ref{first_simulation}.   When $\tau=0$, the manifold time series under both settings is second-order stationary, and when $\tau>0$, the simulated time series  is  non-stationary in terms of the second order. We consider  $\tau=0.25,~0.5,~0.75,~1.0, 1.5$  in the power study.
% For both settings and all $\tau$,  we report the results for sample sizes $T= 256, ~512$ and $1024$, with number of Monte Carlo  runs as $5000$.   The  window size is set to be  $n=8$ as the same setting in \cite{anne2021}. 
 We implement  $5000$ Monte Carlo replications with $T=256,~512,~1024$, respectively.   The  block size is set to be  $n=8$ as  suggested in \cite{anne2021}.

 Type-I error rates  are reported
in Table \ref{table2}. We observe that, under both  $\mathbb{S}^6$ and $\text{Sym}_+^3$ settings, at the significant level $\alpha=0.05$,  the Type-I error rates decrease  as $T$ increases, but are slightly inflated for relatively small $T$.  This slight inflation is  due to an intrinsic limitation of the method we adapted \citep{anne2021}, which also applies to Euclidean time series. Specifically, we show in Table \ref{table2} that for an AR$(0.1)$ process in 
$\mathbb{R}^6$, the Type-I error rates of this testing procedure are also slightly  larger than $0.05$. 
% \lin{are there other methods in Euclidean space with better type I error rate?}\zhu{Most of second-order stationarity test are based on testing whether the spectral density is constant. Most of them need to be fine tuned to get a better type-I error rate...} 
In our power study,  we observe that the testing power for both  $\mathbb{S}^6$ and $\text{Sym}_+^3$ settings increases to $1$ as $\tau$ grows; see Figure \ref{power_second}. %Regarding the second-order stationarity test for time series valued in SPD matrices, the power also increases towards $1$ with a larger sample size, specifically when  $T=512$ or  $T=1024$. 

\begin{figure}[t]
\centering 
\subfigure{
\includegraphics[width=0.4\textwidth]{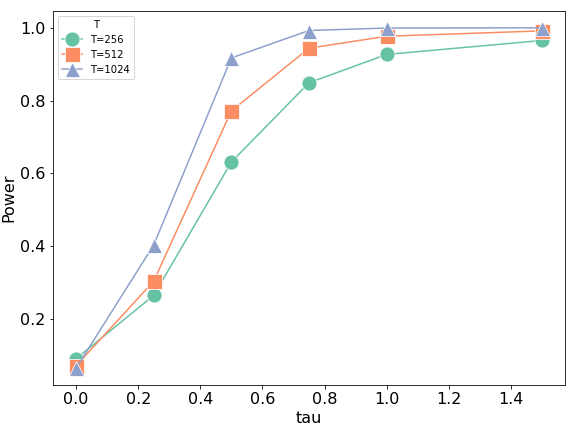}}~~~~~
\subfigure{
\includegraphics[width=0.4\textwidth]{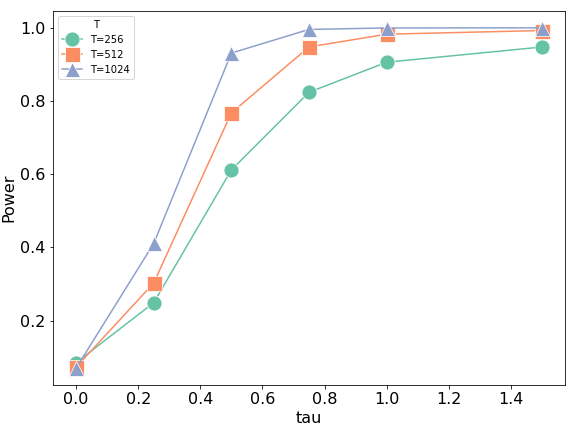}
}
\caption{Simulated power curves for the second-order stationarity test. Left Panel: power curve for the second-order stationarity test of spherical time series. Right Panel: power curve for the second-order stationarity test of SPD-matrices-valued time series.  The significant level is $0.05$.}
\label{power_second}
 \vspace{-0.2cm}
\end{figure}

\section{Application to Real Data}\label{realdata1}
In this section, we apply our stationarity test to a single-cell RNA sequencing data generated by \cite{schiebinger2019optimal}. The raw data is available at NCBI Gene Expression Omnibus %with identification number GSE122662 
(\url{https://www.ncbi.nlm.nih.gov/geo/query/acc.cgi?acc=GSE122662}).
The goal is to understand the developmental process of mouse embryonic cells  and model the change of cell-type proportion at each stage.
To achieve this goal, scientists first obtained  mouse embryonic cells  from a single female embryo, plated cells for $18$ days, measured the gene expression profiles of cells collected across $18$ days, and 
finally profiled $251,203$ high-quality cells 
with $1,479$ variable genes  after pre-processing. A nonlinear dimensionality reduction method called force-directed layout embedding \citep{fleviz} was used  to visualize the temporal change of cellular populations in 2D in the original work, as shown in Figure \ref{fig:fle}. 
These cells were  then assigned to seven  major cell types by clustering and annotation with gene signature scores provided by prior biological knowledge. 
The annotated seven cell types are Mouse Embryonic Fibroblasts (MEFs), Mesenchymal-Epithelial Transition (MET) Cells, Induced Pluripotent Stem (IPS) Cells, Stromal Cells, Epithelial Cells, Neural Cells and Trophoblasts; each cell type has their own morphological features and functions.  In this study, the proportions of these cell types are observed at each time point, with data collected at 37 time points over the course of 18 days (at 12-hour intervals).

A common approach to model the compositional data is the square-root transformation, which maps the data to a hypersphere.  
This transformation has an advantage that the composition constraint and zero components are naturally incorporated \citep{micheal1983,scealy2010}. Applying square-root transformation to our data, we finally obtain a time series in hypersphere  $\mathbb{S}^6$ with length $T=37$, with visualization provided in Figure \ref{fig:CTP}.
We aim to answer a biological question: does the  cell-type proportion have systematic change over time, or equivalently, does the cell-type transition achieve dynamic equilibrium? 
% \zhu{A recently published work \cite{deepvelo} are also interested in similar questions. They defined  cell criticality index (CCI) to quantify the unstability of cells,   and  studied whether some particular cell types achieve a stationary and stable state.    } 
% \kong{Any references that are also interested in this type of question?}
 This question is closely related to the discussion regarding validity of adopting a dynamic equilibrium assumption for modeling cellular dynamics without prior knowledge in cell biology \citep{schiebinger2019optimal,zhou2021dissecting,sha2024reconstructing}. Statistically, the question is equivalent to testing the constancy of the mean of this hyperspherical time series , i.e., the first-order stationarity, and our proposed test serves as a tool to assess the feasibility of such an assumption when applied to real data. 
 
Specifically, we apply the proposed first-order stationarity test to the data, with   bootstrap sample size $B =2000$  and block size selected by the minimum volatility method \citep{politis2012subsampling}. The corresponding p-value is $0.0005$,  providing a strong evidence to reject the null hypothesis. Thus, the cell-type proportions in this  cell population undergo  systematic temporal change, and cell-type transitions are still out of dynamic equilibrium. 
This result is consistent with the findings in \cite{schiebinger2019optimal}, as they discovered that the extracted mouse embryonic cells have a strong ability of differentiation, and 
gradually  moves to   a terminal stromal state or a MET state, where the latter  further generates pluripotent, extra-embryonic, and neural cells. 

We then use the same dataset as an illustrative example to evaluate the proposed second-order stationarity test.  In particular, we first estimate the mean curve $\hat{\mu}(\cdot)$ using  the total-variation regression  with regularization parameters selected by leave-one-out cross validation \citep{AOS2095}.  Then we parallelly  transport
$\Log_{\hat{\mu}(i/T)} X_i$ from $\hat{\mu}(i/T)$ to $\hat{\mu}(1/T)$  along the $\hat{\mu}(\cdot)$ as a detrend procedure, and apply the second-order test to the detrend version time-series  $\{\mathcal{P}^{\hat{\mu}(i/T)}_{\hat{\mu}(1/T)}(\Log_{\hat{\mu}(i/T)} X_i )\}_{i=1}^T$. Since the sample size is small, we divide the data into $5$ overlapped blocks of size $n=8$, which are  $[1,8],[8,15],[15,22],[22,29]$ and $[30,37]$. 
The p-value associated to the second-order stationarity test is $0.223$. The result shows that there is no significant evidence suggesting the uncertainty caused by the rate of random proliferation and apoptosis or noises due to technical issues in the sequencing platform varies over time. The constant uncertainty was implicitly  made as an assumption of the biological model in \cite{schiebinger2019optimal} since the uncertainty parameter was shared by all time points  in their models and numerical analysis, and our testing result provides a numerical support for the assumption in this dataset. 

\begin{figure}[h]
    \centering
   \subfigure{\label{fig:fle}
\begin{overpic}[width=0.48\textwidth]{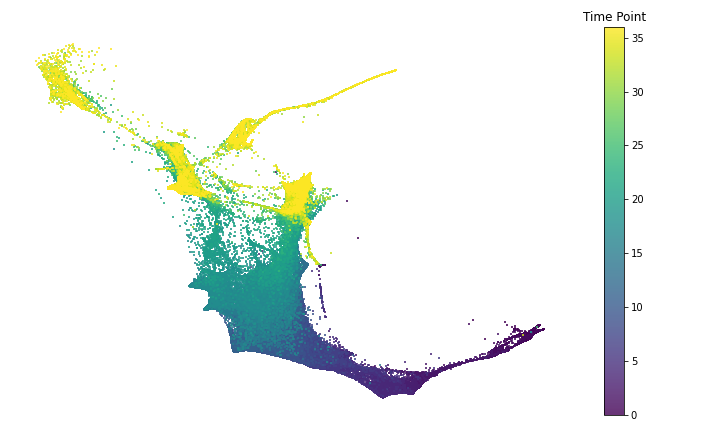}
\put(3,60){(A)} % Adjust the position as needed
\end{overpic}
}\hfill
%     \subfigure{%
%         \label{fig:fle}
% \includegraphics[width=0.52\textwidth]{fle_viz.png}}\hfill
    \subfigure{%
        \label{fig:CTP}
    \begin{overpic}[width=0.48\textwidth]{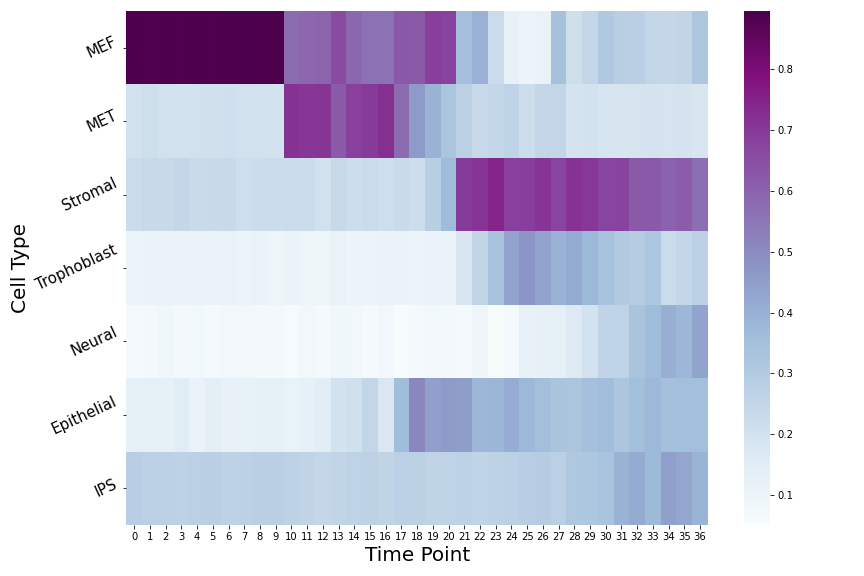}
\put(1,60){(B)} % Adjust the position as needed
\end{overpic}
    }
    \vspace{-0.3cm}
    \caption{\textbf{(A)}:Visualization of gene expression profiles of cells using force directed layout embedding ( a type of nonlinear dimension reduction).  {This figure is adopted from \cite{schiebinger2019optimal}, which was originally used to visualize the  temporal change of cell populations.} In this visualization, each cell is depicted as a dot. The coloring of these dots corresponds to the time point at which each cell was sequenced, with darker shades indicating later time points. 
    This visualization illustrates that the temporal dynamics of the cell population {varies continuously over time, hence 
    is locally stationary.} \textbf{(B)}: A heatmap to visualize the square-root transformed cell-type proportion time series data of seven cell types at 37 time points across 18 days. The square-root transform outputs a spherical time series of length $T=37$ in the manifold $\mathbb{S}^6$. Each row corresponds to the square-root of the time-varying proportion of a pxarticular cell type within the cellular population, while each column denotes the observed value in the manifold time series at a particular time point. A darker hue signifies a larger proportion. }
    \vspace{-0.15cm}
\end{figure}

% \begin{figure}[htbp]
%     \centering
%     \includegraphics[width=0.6\textwidth]{fle_viz.png}
%     \caption{  Visualization of gene expression profiles of cells from \cite{schiebinger2019optimal} using force directed layout embedding \citep{fleviz}.  In this visualization, each cell is depicted as a dot. The coloring of these dots corresponds to the time point at which each cell was sequenced, with darker shades indicating later time points. This visualization illustrates the temporal dynamics of cell population is a locally stationary. 
%     }
%      \label{fle}
% \end{figure}
% \begin{figure}[htbp]
% \centering
% \includegraphics[width=\textwidth]{Cell_type_prop_SR.png}
% \caption{\zhu{We use a heatmap to visualize the square-root transformed cell-type proportion time series data of seven cell types at 37 time point across 18 days. The square-root transform outputs a spherical time series of length $T=37$ in the manifold $\mathbb{S}^6$.   In this visualization, each row corresponds  to  square-root of the time-varying proportion of a particular cell type within the cellular population, while each column denotes the observed value in the manifold time series at a particular time point.
% A darker hue signifies a larger proportion.} 
% }
% \label{CTP}
% \end{figure}

% \section{Extension to Wasserstein space}\label{wass}

\section{Discussion}\label{discuss}
In this paper, we  introduce the definition of first-order and second-order stationarity of manifold-valued time series. We propose testing methods to test both  first-order and second-order  stationarity. Our methods can account for the curved nature of general manifolds. We derive the asymptotic consistency  and asymptotic local powers of the tests. Numerical simulation studies and real data analysis are provided to illustrate the efficiency of our methods.

One limitation of our work is the dependency of our method for spectral density-based testing second-order stationarity on the choice of  block size, a process that lacks a universally accepted benchmark and requires further improvement. This issue is not exclusive to our approach but is a widespread concern in the context of second-order stationarity assessments for time series within linear spaces  \citep{holger2011,anne2021}.

There are a few interesting future directions of our work. For example, in neuroscience study, an interesting question is how to detect structural break of dynamic functional connectivity \cite{hutchison2013dynamic} when the state change. This issue can be approached as a problem of identifying breakpoints in manifold time series, which can be potentially  solved by an extension of our framework to detect abrupt change in a block-wise locally stationary manifold time series.  Another interesting extension is to generalize our framework and methods to the Wasserstein space $\mathcal{W}_1([0,1])$, since  $\mathcal{W}_1([0,1])$  can be viewed as an infinite-dimensional  Hilbert manifold \citep{chen2023} by proper definition.  However, an extension to general metric spaces is challenging and is still an open question, and we leave it for future research.

\begin{table}[t]
\centering
\begin{tabular}{ccccccc}
\toprule
& \multicolumn{3}{c}{\(\mathbb{S}^6\)} & \multicolumn{3}{c}{\(\text{Sym}^+_3\)} \\
\cmidrule(lr){2-4} \cmidrule(lr){5-7}
T & CAMB & B1 & B2 & CAMB & B1 & B2 \\
\midrule
50 & 0.0364 & 0.0584 & 0.3936 & 0.0384 & 0.0082 & 0.0098  \\
100 & 0.0544 & 0.1070 & 0.9806 & 0.0404 & 0.0086 & 0.0254\\
500 & 0.0392 & 0.1008 & 1.0000 & 0.0478 & 0.0098 & 0.1116 \\
\bottomrule
\end{tabular}
\caption{Type-I error rates of the first-order stationarity test of three benchmarked methods under \(\mathbb{S}^6\) and \(\text{Sym}^+_3\) scenarios. CAMB represents our method, B1 represents the first comparison method and B2 represents the second comparison method. The bootstrap size is \(B=2000\) for all methods in this study. The results are based on \(5000\) repetitions of Monte Carlo runs. The significant level is set to be $0.05$.}
\label{table1}
\end{table}

\begin{table}[th]
% \floatbox[{\capbeside\thisfloatsetup{capbesideposition={right,top},capbesidewidth=10cm}}]{table}[\FBwidth]
{\caption{Type-I errors of second-order stationarity test for $T=256,512,1024$ and $n=T/8$ with values in sphere, SPD matrices, and Euclidean space, respectively. The results are based on $5000$ Monte Carlo runs. The significant level is set to be $0.05$.}\label{table2}}
{\begin{tabular}{ c cccc|} 
    \hline
     T& $\mathbb{S}^6$ & $\text{Sym}_+^3$ & $\mathbb{R}^6$ \\
     \hline
    256 & 0.090 & 0.085 & 0.106 \\
    512 & 0.071 & 0.073 & 0.090 \\
    1024 & 0.064 & 0.070 & 0.074 \\
    \hline
  \end{tabular}}
\end{table}

\section*{Acknowledgment}
We thank Robert J. McCann and Zhou Zhou for their insightful conversations and suggestions.

\section*{Supplementary Materials}
The supplementary file contains technical proofs for the theorems in this article.

\bibliographystyle{agsm}

\bibliography{ref}



\section*{Supplementary material}

# Supplementary Material for "Stationarity of Manifold Time Series"

Junhao Zhu, Dehan Kong, Zhaolei Zhang and Zhenhua Lin

## S1 Technical Proofs

### S1.1 Lemma 1

*Proof.* We begin by showing that $\hat{\mu}$ is a $\sqrt{T}$-consistent estimator of $\hat{\mu}$. Let $F_T(p) = \frac{1}{T}\sum_{i=1}^{T} d_{\mathcal{M}}^2(p, X_i)$. As $\mathcal{M}$ satisfies the condition **(M1)** or **(M2)**, $F_T$ is strongly convex in the sense

$$F_T(p) \geq F_T(q) + \langle \frac{1}{T}\sum_{j=1}^{T} \mathrm{Log}_q X_i, \mathrm{Log}_q p \rangle_q + \lambda d_{\mathcal{M}}^2(p, q) \tag{S1}$$

for some constant $\lambda > 0$ depending on $\mathcal{M}$ only. Let $\mathbf{E}$ be an orthonormal frame at $\mathcal{T}_\mu \mathcal{M}$, and $e_i$ be the coordinate representation of $\mathrm{Log}_\mu X_i$. When conditions **(A1)**-**(A4)** hold, by Proposition 5 in Zhou (2013), then on a richer probability space, there exists i.i.d standard normal random vectors $\{V_j\}_{j \in \mathbb{N}}$ such that

$$\sup_{1 \leq i \leq T} \left| \sum_{j=1}^{i} \frac{1}{\sqrt{T}} e_i - \frac{1}{\sqrt{T}} \Sigma_{\mathbf{E}}^{1/2}(\frac{j}{T}) V_j \right| = o_p(T^{-1/2} \log^2 T),$$

1

which implies that $\frac{1}{T}\sum_{j=1}^{T} \mathrm{Log}_q X_i = O_p(\frac{1}{\sqrt{T}})$. With $p = \hat{\mu}$ and $q = \mu$ in Eq.(S1), we have

$$0 \geq F_T(\hat{\mu}) - F_T(\mu) \geq \{O_p(\frac{1}{\sqrt{T}}) + d_{\mathcal{M}}(\mu, \hat{\mu})\} \cdot d_{\mathcal{M}}(\mu, \hat{\mu}),$$

whic implies $d_{\mathcal{M}}(\hat{\mu}, \mu) = O_p(\frac{1}{\sqrt{T}})$.

We next show that

$$\sup_{1 \leq k \leq T} \left\| \mathcal{H}(\frac{k}{T}) - \frac{1}{T}\sum_{i=1}^{k} H_i \right\|_\mu = O_p(\frac{1}{T^{1/2}}).$$

Since $H_i = H(\mu, X_i) = H(\mu, \mathrm{Exp}_\mu(G_{\mathbf{E}}(t_i, \mathcal{F})_i^\top \mathbf{E}))$, where $\mathbf{E} = \{E_j\}_{j=1}^d$ is an orthonormal frame on $\mathcal{T}_\mu\mathcal{M}$. By the assumption (**A1**), there exists a constant $L_H$ such that :

$$\mathbb{E}\left\| H(\mu, \mathrm{Exp}_\mu(G_{\mathbf{E}}(s, \mathcal{F}_0)^\top E)) - H(\mu, \mathrm{Exp}_\mu(G_{\mathbf{E}}(t, \mathcal{F}_0)^\top E)) \right\|_{\mathrm{HS}} \leq L_H C |t - s| \tag{S2}$$

By conditions (**A3**) and the (**A5**), there exists a constant $\tilde{\alpha} < 1$ which only depends on $C_1$ and $\alpha$ such that $\langle H(\mu, \mathrm{Exp}_\mu(G_{\mathbf{E}}(s, \mathcal{F}_0)^\top E))E_j, E_k\rangle$ also satisfies (**A3**) for all $1 \leq j \leq k \leq d$. Let $\{H_{i,jk}\}_{1 \leq j \leq k \leq d}$ be the coordinate representation of $H(\mu, X)$ under the basis $\{E_1, \cdots, E_d\}$, i.e. $A_{i,jk} = \langle H_i E_j, E_k \rangle_\mu$. It suffices to show that

$$A_{T,jk} := \max_{1 \leq l \leq T} \left| \frac{1}{T}\sum_{i=1}^{l} H_{i,jk} - \frac{1}{T}\sum_{i=1}^{l} \mathbb{E} H_{i,jk} \right| = O_P(\frac{1}{T^{1/2}}).$$

Let $\Xi_{l,jk} = \sum_{i=1}^{l}(H_{i,jk} - \mathbb{E}[H_{i,jk}\|\mathcal{F}_{i-1}])$ and $\Lambda_{l,jk} = \sum_{i=1}^{l}(\mathbb{E}[H_{i,jk}|\mathcal{F}_{i-1}] - \mathbb{E}[H_{i,jk}])$. Then $A_{T,jk}$ can be bounded by

$$A_{T,jk} = \max_{1 \leq l \leq T} \frac{1}{T}|\Xi_{l,jk} + \Lambda_{l,jk}| \leq \max_{1 \leq l \leq T} \frac{1}{T}|\Xi_{l,jk}| + \max_{1 \leq l \leq T} \frac{1}{T}|\Lambda_{l,jk}|.$$

Noting that $\Xi_{l,jk}$ is a bounded martingale, hence by Doob's $L^p$ inequality (Durrett 2019) we have

$$\mathbb{E}[\frac{1}{T}\max_{1\leq l\leq T}|\Xi_{l,jk}|^2] \leq \frac{C}{T^2}\mathbb{E}|\Xi_{T,jk}|^2 = O(\frac{1}{T}). \tag{S3}$$

Now We start to bound the quantity $\max_{1\leq l\leq T}\frac{1}{T}|\Lambda_{i,jk}|$. Define $\Pi_i Y = \mathbb{E}[Y|\mathcal{F}_i] - \mathbb{E}[Y|\mathcal{F}_{i-1}]$. Then

$$\max_{1\leq l\leq T}|\Lambda_{l,jk}| = \max_{1\leq l\leq T}|\sum_{i=1}^{l}\sum_{a=0}^{\infty}\Pi_{i-a}H_{i,jk}| = \max_{1\leq l\leq T}\left|\sum_{a=0}^{\infty}\left(\sum_{i=1}^{l}\Pi_{i-a}H_{i,jk}\right)\right|$$
$$\leq \sum_{a=0}^{\infty}\max_{1\leq l\leq T}\left|\left(\sum_{i=1}^{l}\Pi_{i-a}H_{i,jk}\right)\right|.$$

The triangle inequality implies that

$$\sqrt{\mathbb{E}(\max_{1\leq l\leq T}|\Lambda_{l,jk}|)^2} \leq \sum_{a=0}^{\infty}\sqrt{\mathbb{E}\max_{1\leq l\leq T}\left|\left(\sum_{i=1}^{l}\Pi_{i-a}H_{i,jk}\right)\right|^2}.$$

Note that $\left(\sum_{i=1}^{l}\Pi_{i-a}H_{i,jk}\right)$ is a martingale. By Doob's $L^p$ inequality again, there exists a constant $C > 0$ such that

$$\sqrt{\mathbb{E}(\max_{1\leq l\leq T}|\Lambda_{l,jk}|)^2} \leq C\sum_{a=0}^{\infty}\sqrt{\mathbb{E}\left|\left(\sum_{i=1}^{T}\Pi_{i-a}H_{i,jk}\right)\right|^2}.$$

By Theorem 1 in Wu (2005), $\mathbb{E}\left|\left(\sum_{i=1}^{T}\Pi_{i-a}H_{i,jk}\right)\right|^2 = O(T\tilde{\alpha}^a)$ with $\tilde{\alpha} < 1$, which implies that

$$\sqrt{\mathbb{E}(\max_{1\leq l\leq T}|\Lambda_{l,jk}|)^2} \leq C\sqrt{T}\sum_{a=0}^{\infty}\tilde{\alpha}^a = O(\sqrt{T}). \tag{S4}$$

By the upper bounds provided by Eq.(S3) and Eq.(S4) we can deduce that

$$A_{T,jk} \leq \max_{1\leq l\leq T}\frac{1}{T}|\Xi_{l,jk}| + \max_{1\leq l\leq T}\frac{1}{T}|\Lambda_{l,jk}| = O_p(\frac{1}{\sqrt{T}}).$$

3

Let $\mathcal{H}(t) = \int_0^t \mathbb{E}H(\mu, \mathrm{Exp}_\mu(G_{\mathbf{E}}(s, \mathcal{F}_0)^\top E))ds$. $\mathbb{E}H(\mu, \mathrm{Exp}_\mu(G_{\mathbf{E}}(s, \mathcal{F}_0)^\top E))$ is uniformly bounded and Lipschitz continuous in $s$, so the integral is well-defined. By the property of Riemann sum, we have:

$$\sup_{0 \leq t \leq 1} \left\| \mathcal{H}(t) - \frac{1}{T} \sum_{i/T \leq t} \mathbb{E}H\big(\mu, \mathrm{Exp}_\mu(G_{\mathbf{E}}(i/T, \mathcal{F}_0)^\top E)\big) \right\|_{HS} = O(\frac{1}{T}).$$

Hence, we conclude that:

$$\sup_{1 \leq k \leq T} \left\| \mathcal{H}(\frac{k}{T}) - \frac{1}{T} \sum_{i=1}^{k} H_i \right\|_{HS}$$
$$\leq \sup_{1 \leq k \leq T} \left\| \mathcal{H}(\frac{k}{T}) - \frac{1}{T} \sum_{i=1}^{k} \mathbb{E}H_i \right\|_\mu + \sup_{1 \leq k \leq T} \left\| \frac{1}{T} \sum_{i=1}^{k} H_i - \frac{1}{T} \sum_{i=1}^{k} \mathbb{E}H_i \right\|_{HS}$$
$$= O_P(\frac{1}{T^{1/2}}).$$

□

## S1.2 Theorem 1

*Proof.* For simplicity, throughout this proof we use $\tilde{e}_i$ to represent $\mathrm{Log}_\mu X_i$. The proof is based on the Taylor expansion of Riemannian log map on manifold (Kendall & Le 2011). Let $\gamma : [0, 1]$ be the geodesic from $\hat{\mu} = \gamma(0)$ to $\mu = \gamma(1)$. By Taylor expansion, we have

$$\sup_{1 \leq k \leq T} \left\| \mathcal{P}_{\gamma(0)}^{\gamma(1)}(\frac{1}{\sqrt{T}} S_k) - \{ \frac{1}{\sqrt{T}} \sum_{j=1}^{k} \tilde{e}_j - \frac{1}{\sqrt{T}} \sum_{j=1}^{k} H(\mu, X_j) \circ \mathrm{Log}_\mu \hat{\mu} \} \right\|_\mu = O_p\left(\frac{1}{\sqrt{T}}\right).$$

Since parallel transport is a linear map and $S_T = 0$, the above equation implies that $\sqrt{T} \mathrm{Log}_\mu \hat{\mu} = (\frac{1}{T} \sum_{j=1}^{T} H(\mu, X_j))^{-1} \frac{1}{\sqrt{T}} \sum_{j=1}^{T} \tilde{e}_j + O_p(\frac{1}{T})$. By Proposition 5 in Zhou (2013),

4

on a richer probability space, there exists a Gaussian process $U(t)$ such that

$$\sup_{1\leq k\leq T}\left\|\frac{1}{\sqrt{T}}\sum_{j=1}^{k}\tilde{e}_j - U(\frac{k}{T})\right\|_\mu = o_p(T^{-1/4}\log^2 T).$$

The above equation combined with Lemma 1 yields that

$$\sup_{1\leq k\leq T}\left\|\mathcal{P}_{\hat{\mu}}^{\mu}(\frac{1}{\sqrt{T}}S_k) - \left(U(\frac{k}{T}) - \mathcal{H}(\frac{k}{T})\circ\mathcal{H}^{-1}\circ(1)U(1)\right)\right\|_\mu = o_p(T^{-1/4}\log^2 T),$$

which implies that

$$\sup_{1\leq k\leq T}\left\|\mathcal{P}_{\hat{\mu}}^{\mu}(\frac{1}{\sqrt{T}}S_k)\right\|_\mu = \sup_{0\leq k\leq T}\left\|U(\frac{k}{T}) - \mathcal{H}(\frac{k}{T})\circ\mathcal{H}^{-1}(1)\circ U(1)\right\|_\mu + o_p(T^{-1/4}\log^2 T).$$

Since parallel transport $\mathcal{P}_{\gamma(0)}^{\gamma(1)}$ is an isometry from $\mathcal{T}_{\gamma(0)}\mathcal{M}$ to $\mathcal{T}_{\gamma(1)}\mathcal{M}$, we have

$$\sup_{1\leq k\leq T}\left\|\frac{1}{\sqrt{T}}S_k\right\|_{\hat{\mu}} = \sup_{0\leq k\leq T}\left\|U(\frac{k}{T}) - \mathcal{H}(\frac{k}{T})\circ\mathcal{H}^{-1}(1)\circ U(1)\right\|_\mu + o_p(T^{-1/4}\log^2 T),$$

which completes the proof. □

## S1.3 Theorem 2

*Proof.* To establish the consistency of the debiased bootstrap procedure, we first show that $\hat{\mathcal{H}}_k$ is a consistent estimate of $\mathcal{H}(k/T)$. Let $\gamma : [0,1] \to \mathcal{M}$ be the geodesic from $\hat{\mu} = \gamma(0)$ to $\mu = \gamma(1)$. Note that $H(\mu, X)$ is Lipschitz continuous w.r.t $\mu$ by (**A1**), i.e.,

$$|\langle\mathcal{P}_{\gamma(1)}^{\gamma(0)}(H(\mu, X)\circ E_i), \mathcal{P}_{\gamma(1)}^{\gamma(0)}(E_j)\rangle_\mu - \langle H(\hat{\mu}, X)\circ\mathcal{P}_{\gamma(1)}^{\gamma(0)}(E_i), \mathcal{P}_{\gamma(1)}^{\gamma(0)}(E_j)\rangle_{\hat{\mu}}| \leq L_H d_\mathcal{M}(\mu, \hat{\mu}).$$

The above inequality implies

$$|\langle \mathcal{P}_{\gamma(1)}^{\gamma(0)}(\frac{1}{T}\sum_{j=1}^{k} H(\mu, X_j) \circ E_i), \mathcal{P}_{\gamma(1)}^{\gamma(0)}(E_j)\rangle_\mu - \langle \hat{\mathcal{H}}_k \circ \mathcal{P}_{\gamma(1)}^{\gamma(0)}(E_i), \mathcal{P}_{\gamma(1)}^{\gamma(0)}(E_j)\rangle_{\hat{\mu}}| \tag{S5}$$

$$\leq \frac{k}{T} L_H d_\mathcal{M}(\mu, \hat{\mu}),$$

and

$$\sup_{1\leq k\leq T} |\langle \mathcal{P}_{\gamma(1)}^{\gamma(0)}(\frac{1}{T}\sum_{j=1}^{k} H(\mu, X_j) \circ E_i), \mathcal{P}_{\gamma(1)}^{\gamma(0)}(E_j)\rangle_\mu - \langle \hat{\mathcal{H}}_k \circ \mathcal{P}_{\gamma(1)}^{\gamma(0)}(E_i), \mathcal{P}_{\gamma(1)}^{\gamma(0)}(E_j)\rangle_{\hat{\mu}}|$$

$$\leq L_H d_\mathcal{M}(\mu, \hat{\mu}) = O_p(\frac{1}{T^{1/2}}).$$

By the bound in Lemma 1 that $\sup_{1\leq k\leq T} \left\|\mathcal{H}(\frac{k}{T}) - \frac{1}{T}\sum_{i=1}^{k} H_i\right\|_{HS} = O_P(\frac{\log^2 T}{T^{1/2}})$, we have

$$\sup_{1\leq k\leq T} |\langle \mathcal{P}_{\gamma(1)}^{\gamma(0)}(\mathcal{H}(k/T) \circ E_i), \mathcal{P}_{\gamma(1)}^{\gamma(0)}(E_j)\rangle_\mu - \langle \hat{\mathcal{H}}_k \circ \mathcal{P}_{\gamma(1)}^{\gamma(0)}(E_i), \mathcal{P}_{\gamma(1)}^{\gamma(0)}(E_j)\rangle_{\hat{\mu}}|$$

$$= O_P(\frac{1}{T^{1/2}}). \tag{S6}$$

We next show that $\{V_{k,n}\}$ consistently mimics the Gaussian process $U(t): 0 \leq t \leq 1$ up to an isometry, where $V_{k,n}$ is a generic version of $V_{k,n}^{(b)}$ defined in Algorithm 1. Recall $v_i = \mathrm{Log}_{\hat{\mu}} X_i$ be the residuals at the tangent space of $\hat{\mu}$, $S_{j,n} = \sum_{i=j}^{j+n-1} v_i$ for $1 \leq j \leq T - n + 1$, and $V_{k,n}^{(b)} = \sum_{j=1}^{k} \{n(T-n+1)\}^{-1/2} S_{j,n} R_j$ for $k = n, \cdots, T-n+1$, where $\{R_j\}_{j\in\mathbb{N}}$ are i.i.d standard normal random variables. Define

$$B_{i,j} = \frac{1}{n(T-n+1)} \sum_{i\leq l\leq j} S_{l,n} \otimes S_{l,n}$$

and

$$B_{i,j,a,b} = \frac{1}{n(T-n+1)} \sum_{i\leq l\leq j} \langle S_{l,n}, \mathcal{P}_{\gamma(1)}^{\gamma(0)} E_a\rangle_{\hat{\mu}} \langle S_{l,n}, \mathcal{P}_{\gamma(1)}^{\gamma(0)} E_b\rangle_{\hat{\mu}}. \tag{S7}$$

6

We first show that

$$\max_{n+1\leq i\leq j\leq T-n+1}\left|B_{i,j,a,b} - \int_{i/T}^{j/T}\{\Sigma_{\mathbf{E}}(\xi)\}_{a,b}d\xi\right| = o_p(1),$$

where $\{\Sigma_{\mathbf{E}}(\xi)\}_{a,b}$ is the $(a,b)$-entry of the matrix $\Sigma_{\mathbf{E}}(\xi)$. To this end, applying Taylor expansion of Riemannian log map on $\mathcal{T}_\mu \mathcal{M}$ yields

$$\begin{aligned}
\frac{1}{\sqrt{n}}\mathcal{P}_{\gamma(0)}^{\gamma(1)} S_{l,n} &= \frac{1}{\sqrt{n}}\mathcal{P}_{\gamma(0)}^{\gamma(1)}\sum_{k=l}^{i+n-1} v_k = \sum_{k=l}^{i+n-1}\frac{1}{\sqrt{n}}\mathcal{P}_{\gamma(0)}^{\gamma(1)} v_k \\
&= \sum_{k=l}^{l+n-1}\frac{1}{\sqrt{n}}\tilde{e}_k - \frac{1}{\sqrt{n}}\sum_{k=l}^{l+n-1} H(\mu, X_k) \circ \mathrm{Log}_\mu\hat{\mu} + o_p(1/T) \\
&= \frac{1}{\sqrt{n}}\sum_{k=l}^{l+n-1}\tilde{e}_k - \left(\frac{1}{n}\sum_{k=l}^{l+n-1} H(\mu, X_k)\right)\circ(\sqrt{n}\mathrm{Log}_\mu\hat{\mu}) + o_p(1/T).
\end{aligned}$$

By the bounded curvature condition of the manifold, $H(\mu, X)$ is uniformly bounded, and thus,

$$\begin{aligned}
\langle \frac{1}{\sqrt{n}}\mathcal{P}_{\gamma(0)}^{\gamma(1)} S_{l,n}, E_a\rangle_\mu &= \langle \frac{1}{\sqrt{n}}\mathcal{P}_{\gamma(0)}^{\gamma(1)}\sum_{k=l}^{i+n-1} v_k, E_a\rangle_\mu = \langle \sum_{k=l}^{i+n-1}\frac{1}{\sqrt{n}}\mathcal{P}_{\gamma(0)}^{\gamma(1)} v_k, E_a\rangle_\mu \\
&= \langle \frac{1}{\sqrt{n}}\sum_{k=l}^{l+n-1} e_k, E_a\rangle_\mu + O_p(n^{1/2}/T^{3/2}) \\
&\quad - \langle \left(\frac{1}{n}\sum_{k=l}^{l+n-1} H(\mu, X_k)\right)\circ(\sqrt{n}\mathrm{Log}_\mu\hat{\mu}), E_a\rangle_\mu.
\end{aligned}$$

We apply the above expansion of $S_{l,n}$ to $B_{i,j,ab}$ and deduce

$$\max_{1\leq i\leq j\leq T-n+1}\left|B_{i,j,a,b} - \frac{1}{n(T-n+1)}\sum_{i\leq l\leq j}\langle\frac{1}{\sqrt{n}}\sum_{k=l}^{l+n-1} e_k, E_a\rangle_\mu\langle\frac{1}{\sqrt{n}}\sum_{k=l}^{l+n-1} e_k, E_b\rangle_\mu\right| = O_p(\sqrt{\frac{n}{T}})$$

for any $1\leq a\leq b\leq d$. When the conditions (**A2**)-(**A4**) hold, by Theorem 4 in Zhou (2013),

7

we have

$$
\max_{n+1\leq i\leq j\leq T-n+1}\left|B_{i,j,a,b} - \int_{i/T}^{j/T}\{\Sigma_{\mathbf{E}}(\xi)\}_{a,b}d\xi\right|
$$

$$
\leq \max_{1\leq i\leq j\leq T-n+1}\left|B_{i,j,a,b} - \frac{1}{n(T-n+1)}\sum_{i\leq l\leq j}\langle\frac{1}{\sqrt{n}}\sum_{k=l}^{l+n-1}e_k, E_a\rangle_\mu\langle\frac{1}{\sqrt{n}}\sum_{k=l}^{l+n-1}e_k, E_b\rangle_\mu\right| \quad \text{(S8)}
$$

$$
+ \max_{n+1\leq i\leq j\leq T-n+1}\left|\frac{1}{n(T-n+1)}\sum_{i\leq l\leq j}\langle\frac{1}{\sqrt{n}}\sum_{k=l}^{l+n-1}e_k, E_a\rangle_\mu\langle\frac{1}{\sqrt{n}}\sum_{k=l}^{l+n-1}e_k, E_b\rangle_\mu\right.
$$

$$
\left. - \int_{i/T}^{j/T}\{\Sigma_{\mathbf{E}}(\xi)\}_{a,b}d\xi\right| = O_p(\sqrt{\frac{n}{T}} + \frac{1}{n}).
$$

Now we are ready to prove the statement of Theorem 2. Let $U_n(t) : [\frac{n}{T}, 1-\frac{n}{T}] \to \mathcal{T}_{\hat{\mu}}\mathcal{M}$ be the linear interpolation of $V_{k,n}$, i.e., $U_n(t) = V_{\lfloor tT \rfloor,n} + (tT - \lfloor tT \rfloor)(V_{\lfloor tT \rfloor+1,n} - V_{\lfloor tT \rfloor,n})$. By (S8) and Theorem 3 in Zhou (2013), whenever $n(T) \to \infty$ and $T \to \infty$, $\mathcal{P}_{\gamma(0)}^{\gamma(1)} U_n(t)$ converges to $U(t)$ under the uniform topology of $\mathcal{C}([0,1], (\mathcal{T}_\mu\mathcal{M}, \|\cdot\|_\mu))$. By (S6) and the uniform convergence of $U_n(t)$, we conclude

$$
\max_{n\leq k\leq T-n+1}\left\|V_{k,n} - \hat{\mathcal{H}}_k \circ \hat{\mathcal{H}}_T^{-1} \circ V_{T-n+1,n}\right\|_{\hat{\mu}} \xrightarrow{\mathcal{D}} \sup_{0\leq t\leq 1}\left\|U(t) - \mathcal{H}(t) \circ \mathcal{H}^{-1}(1) \circ U(1)\right\|_\mu. \quad \text{(S9)}
$$

□

## S1.4 Theorem 3

*Proof.* For simplicity, we write $\tau(T)$ as $\tau$. Let $e_i$ be the coordinate-representation of $\text{Log}_{\mu_T(i/T)}$ under the frame $\mathbf{E}(\tau, i/T)$, i.e., $\text{Log}_{\mu_T(i/T)} = e_i^\top \mathbf{E}(\tau, i/T)$; see also (6) in the main text. To begin with, for a fixed $p \in \mathcal{M}$, when conditions (M1) or (M2) hold, there exists a constant

$\lambda > 0$ such that

$$\frac{1}{2T}\sum_{i=1}^{T}\left(d_{\mathcal{M}}^2(p, X_i) - d_{\mathcal{M}}^2(\mu, X_i)\right) \geq \langle\frac{1}{T}\sum_{i=1}^{T}\mathrm{Log}_\mu X_i, \mathrm{Log}_\mu p\rangle_\mu + \lambda d_{\mathcal{M}}^2(p, \mu).$$

Taylor expansion of vector fields on manifold yields

$$\mathrm{Log}_\mu X_i = \mathcal{P}_{\mu_T(i/T)}^{\mu}\left(e_i^\top E(\tau, \frac{i}{T})\right) - \tau H(\mu, X_i) \circ b(\frac{i}{T}) + O_p(\tau^2),$$

and

$$\frac{1}{2T}\sum_{i=1}^{T}\left(d_{\mathcal{M}}^2(p, X_i) - d_{\mathcal{M}}^2(\mu, X_i)\right) \geq \langle\frac{1}{T}\sum_{i=1}^{T}\mathrm{Log}_\mu X_i, \mathrm{Log}_\mu p\rangle_\mu + \lambda d_{\mathcal{M}}^2(p, \mu)$$

$$= \frac{1}{2T}\sum_{i=1}^{T}\langle e_i^\top E(\tau, \frac{i}{T}), \mathcal{P}_{\mu}^{\mu_T(i/T)}\mathrm{Log}_\mu p\rangle_\mu - \tau\langle\frac{1}{T}\sum_{i=1}^{T}H(\mu, X_i)\circ b(\frac{i}{T}), \mathrm{Log}_\mu p\rangle_\mu$$

$$+ \lambda d_{\mathcal{M}}^2(p, \mu) + O_p(\tau^2).$$

By the bounded conditions on the curvature and the curve $b(\cdot)$, there exists a constant $C_b$ such that $\left\|\tau\langle\frac{1}{T}\sum_{i=1}^{T}H(\mu, X_i)\circ b(\frac{i}{T})\right\|_\mu \leq \tau C_b$. Let $\overline{\mu p}$ be the coordinate representation of $\mathrm{Log}_\mu p$ under the orthonormal basis $\{\tilde{E}_j, 1 \leq j \leq d\}$ and $\{R_{\tau,T,i} : 1 \leq i \leq T\}$ be a collection of $d \times d$ matrices such that

$$(R_{\tau,T,i})_{j,k} = \langle E_j(\tau, \frac{i}{T}), \mathcal{P}_{\mu}^{\mu_T(i/T)} E_k\rangle_{\mu_T(i/T)}.$$

Then $\langle e_i^\top E(\tau, \frac{i}{T}), \mathcal{P}_{\mu}^{\mu_T(i/T)}\mathrm{Log}_\mu p\rangle_{\mu_T(i/T)}$ can be rewritten as $e_i^\top R_{\tau,T,i}\overline{\mu p}$. Moreover, the local isometry property of parallel transport on smooth manifold implies $R_{\tau,T,i} = \mathrm{I}_d + O(\frac{\tau}{T})$, and hence

$$\frac{1}{2T}\sum_{i=1}^{T}\left(d_{\mathcal{M}}^2(p, X_i) - d_{\mathcal{M}}^2(\mu, X_i)\right) \geq \frac{1}{2T}\sum_{i=1}^{T}e_i^\top\overline{\mu p} + \lambda d_{\mathcal{M}}^2(p, \mu) + O_p(\tau). \quad (S10)$$

9

Since $\left|\frac{1}{2T}\sum_{i=1}^{T} e_i^\top \overline{\mu p}\right| \leq \|\frac{1}{2T}\sum_{i=1}^{T} e_i\| \cdot d(\mu, p)$, by taking $p = \hat{\mu}$, we deduce from (S10) that $d_{\mathcal{M}}(\hat{\mu}, \mu) = O_p(\max\{\tau, \sqrt{1/T}\})$.

Now we show that when $\lim_{T\to\infty} \frac{\tau}{T^{-1/2}} = \infty$, the CUSUM statistic $Q_T \to \infty$ almost surely. Let $\gamma : [0,1] \to \mathcal{M}$ be the geodesic from $\hat{\mu} = \gamma(0)$ to $\mu = \gamma(1)$. Similarly to the proof of Theorem 1, we can show that

$$\sup_{1\leq k\leq T} \left\|\mathcal{P}_{\gamma(0)}^{\gamma(1)}(\frac{1}{\sqrt{T}}S_k) - \left(\frac{1}{\sqrt{T}}\sum_{j=1}^{k}\tilde{e}_j - \frac{1}{\sqrt{T}}\sum_{j=1}^{k} H(\mu, X_j) \circ \mathrm{Log}_\mu\hat{\mu}\right)\right\|_\mu = O_p(\tau^2),$$

and that

$$\sup_{1\leq k\leq T} \left\|\mathcal{P}_{\gamma(0)}^{\gamma(1)}(\frac{1}{\sqrt{T}}S_k)\right\|_\mu$$
$$= \sup_{1\leq k\leq T} \left\|\frac{1}{\sqrt{T}}\sum_{j=1}^{k}\tilde{e}_j - \frac{1}{\sqrt{T}}\sum_{j=1}^{k} H(\mu, X_j) \circ \mathrm{Log}_\mu\hat{\mu}\right\|_\mu + O_p(\tau^2)$$
$$\geq -\sup_{1\leq k\leq T} \left\|\frac{1}{\sqrt{T}}\sum_{j=1}^{k}\tilde{e}_j\right\|_\mu + \sqrt{T}\left\|\left(\frac{1}{T}\sum_{j=1}^{T} H(\mu, X_j)\right) \circ \mathrm{Log}_\mu\hat{\mu}\right\|_\mu + O_p(\tau^2).$$

Note that $\sup_{1\leq k\leq T}\left\|\frac{1}{\sqrt{T}}\sum_{j=1}^{k}\tilde{e}_j\right\|_\mu = O_p(1)$ and that $\left\|\left(\frac{1}{T}\sum_{j=1}^{T} H(\mu, X_j)\right) \circ \mathrm{Log}_\mu\hat{\mu}\right\|_\mu \geq \lambda d_{\mathcal{M}}(\hat{\mu}, \mu) \asymp O_p(\tau)$ when condtions (M1) and (M2) hold. Thus, we have

$$Q_T = \sup_{1\leq k\leq T} \left\|\mathcal{P}_{\gamma(0)}^{\gamma(1)}(\frac{1}{\sqrt{T}}S_k)\right\|_\mu$$
$$\geq -\sup_{1\leq k\leq T}\left\|\frac{1}{\sqrt{T}}\sum_{j=1}^{k}\tilde{e}_j\right\|_\mu + \lambda\sqrt{T}d_{\mathcal{M}}(\mu, \hat{\mu}) + O_p(\tau^2) \to +\infty, \text{ a.e.},$$

whenever $\lim_{T\to\infty} \frac{\tau}{\sqrt{T}} \to \infty$.

We next show that, if $\tau(T) = T^{-1/2}$, then

$$Q_T \xrightarrow{\mathcal{D}} \sup_{0 \leq t \leq 1} \left\| U(t) - \mathcal{H}(t) \circ \mathcal{H}^{-1}(1) \circ U(1) \right. \\ \left. + \mathcal{H}(t) \circ \mathcal{H}^{-1}(1) \circ \int_0^1 \frac{\partial}{\partial \xi} \mathcal{H}(\xi) \circ v(\xi) d\xi - \int_0^t \frac{\partial}{\partial \xi} \mathcal{H}(\xi) \circ v(\xi) d\xi \right\|_\mu, \quad \text{(S11)}$$

where $\mathcal{H}(t)$ is defined in Lemma 1. For convenience, we define $\Pi(s,t)$ be the parallel transport map from $\gamma(s,t)$ to $\mu$ along the geodesic $s \to \gamma(s,t)$. Notice that $\sum_i \text{Log}_{\hat{\mu}} X_i = 0$, and that $d_{\mathcal{M}}(\hat{\mu}, \mu) = O_p(T^{-1/2})$, we can apply Taylor expansion at $\hat{\mu}$ similarly to (S10) for $\tau = T^{-1/2}$ and have

$$0 = \sum_{i=1}^T \Pi(\tau, \frac{i}{T}) \left( e_i^\top E(\tau, \frac{i}{T}) \right) - \frac{\tau}{T} \sum_{i=1}^T H(\mu, X_i) \circ b(\frac{i}{T}) - \frac{1}{T} \sum_{i=1}^T H(\mu, X_i) \text{Log}_\mu \hat{\mu} + O_p(\frac{1}{T}).$$

The above identity implies that

$$\text{Log}_\mu \hat{\mu} = \left( \frac{1}{T} \sum_{i=1}^T H(\mu, X_i) \right)^{-1} \circ \left( \sum_{i=1}^T \Pi(\tau, \frac{i}{T}) \left( e_i^\top E(\tau, \frac{i}{T}) \right) \right. \\ \left. - \frac{\tau}{T} \sum_{i=1}^T H(\mu, X_i) \circ b(\frac{i}{T}) \right) + O_p(\frac{1}{T}). \quad \text{(S12)}$$

Substituting $\text{Log}_\mu \hat{\mu}$ by (S12) in the following equation

$$\sup_{1 \leq k \leq T} \left\| \mathcal{P}_{\gamma(0)}^{\gamma(1)}(\frac{1}{\sqrt{T}} S_k) - \left( \frac{1}{\sqrt{T}} \sum_{j=1}^k \tilde{e}_j - \frac{1}{\sqrt{T}} \sum_{j=1}^k H(\mu, X_j) \circ \text{Log}_\mu \hat{\mu} \right) \right\|_\mu = O_p(\frac{1}{\sqrt{T}}),$$

11

we have

$$\sup_{1\leq k\leq T}\left\|\mathcal{P}^{\gamma(1)}_{\gamma(0)}(\frac{1}{\sqrt{T}}S_k)\right\|_\mu = \frac{1}{\sqrt{T}}\sup_{1\leq k\leq T}\left\|\sum_{j=1}^k \tilde{e}_j - \sum_{j=1}^k H(\mu, X_j)\circ \mathrm{Log}_\mu \hat{\mu}\right\|_\mu + O_p(\frac{1}{T})$$

$$= \frac{1}{\sqrt{T}}\sup_{1\leq k\leq T}\left\|(\sum_{j=1}^k \tilde{e}_j - \sum_{j=1}^k H(\mu, X_j)\circ \left(\frac{1}{T}\sum_{i=1}^T H(\mu, X_i)\right)^{-1}\right.$$

$$\left.\circ \left(\sum_{i=1}^T \Pi(\tau, \frac{i}{T})\left(e_i^\top E(\tau, \frac{i}{T})\right) - \frac{\tau}{T}\sum_{i=1}^T H(\mu, X_i)\circ b(\frac{i}{T})\right)\right\|_\mu + O_p(\frac{1}{\sqrt{T}}).$$

By Taylor expansion of $\tilde{e}_j$, i.e.,

$$\tilde{e}_j = \Pi(\tau, \frac{j}{T})\left(e_j^\top E(\tau, \frac{j}{T})\right) - \tau H(\mu, X_j)\circ b(\frac{i}{T}) + O_p(\frac{1}{\sqrt{T}}),$$

the test statistic $Q_T = \sup_{1\leq k\leq T}\left\|\mathcal{P}^{\gamma(1)}_{\gamma(0)}(\frac{1}{\sqrt{T}}S_k)\right\|_\mu$ could be further represented by

$$\sup_{1\leq k\leq T}\left\|\mathcal{P}^{\gamma(1)}_{\gamma(0)}(\frac{1}{\sqrt{T}}S_k)\right\|_\mu = \frac{1}{\sqrt{T}}\sup_{1\leq k\leq T}\left\|\sum_{j=1}^k \tilde{e}_j - \sum_{j=1}^k H(\mu, X_j)\circ \mathrm{Log}_\mu \hat{\mu}\right\|_\mu + O_p(\frac{1}{T})$$

$$= \frac{1}{\sqrt{T}}\sup_{1\leq k\leq T}\left\|\sum_{j=1}^k \left(\Pi(\tau, \frac{j}{T})\left(e_j^\top E(\tau, \frac{j}{T})\right) - \tau H(\mu, X_j)\circ b(\frac{i}{T})\right)\right.$$

$$- \sum_{j=1}^k H(\mu, X_j)\circ \left(\frac{1}{T}\sum_{i=1}^T H(\mu, X_i)\right)^{-1} \quad\quad\quad\text{(S13)}$$

$$\left.\circ \left(\sum_{i=1}^T \Pi(\tau, \frac{i}{T})\left(e_i^\top E(\tau, \frac{i}{T})\right) - \frac{\tau}{T}\sum_{i=1}^T H(\mu, X_i)\circ b(\frac{i}{T})\right)\right\|_\mu + O_p(\frac{1}{\sqrt{T}}).$$

To show our desired asymptotic results, it suffices to prove the following claims:

(**C1**) $\frac{1}{\sqrt{T}}\sum_{j=1}^k \Pi(\tau, \frac{j}{T})\left(e_j^\top E(\tau, \frac{j}{T})\right) \xrightarrow{\mathcal{D}} U(\frac{k}{T})$ as $T\to\infty$;

12

**(C2)** $\max_{1\leq k\leq T} \left\| \frac{1}{T} \sum_{i=1}^{k} H(\mu, X_i) - \mathcal{H}(\frac{k}{T}) \right\| = O_p(1/\sqrt{T})$;

**(C3)** $\max_{1\leq k\leq T} \left\| \frac{1}{T} \sum_{i=1}^{k} H(\mu, X_i) \circ b(\frac{i}{T}) - \int_0^{k/T} \frac{\partial}{\partial \xi} \mathcal{H}(\xi) \circ b(\xi) d\xi \right\| = O_p(1/\sqrt{T})$.

Now we prove the claim **(C1)**. Given $\tau = 1/\sqrt{T}$, Taylor expansion of parallel transport yields that

$$\Pi(\tau, \frac{j}{T}) \left( e_j^\top E(\tau, \frac{j}{T}) \right) = \sum_{l=1}^{d} e_{j,l} \Pi(\tau, \frac{j}{T}) \left( E_l(\tau, \frac{j}{T}) \right)$$

$$= e_j^\top E + \frac{1}{\sqrt{T}} \sum_{l=1}^{d} e_{j,l} \left( \nabla_{b(j/T)} E_l(s, \frac{j}{T}) \right) \Big|_{s=0} + O_p(\frac{1}{T}),$$

which implies that

$$\max_{1\leq k\leq T} \left\| \frac{1}{\sqrt{T}} \sum_{j=1}^{k} \left[ \Pi(\tau, \frac{j}{T}) \left( e_j^\top E(\tau, \frac{j}{T}) \right) - e_j^\top E - \frac{1}{\sqrt{T}} \sum_{l=1}^{d} e_{j,l} \left( \nabla_{b(j/T)} E_l(s, \frac{j}{T}) \right) \Big|_{s=0} \right] \right\|_\mu = O_p(\frac{1}{\sqrt{T}}).$$

For fixed $l$ and $T$, define

$$\Xi_{k,l} = \frac{1}{T} \sum_{j=1}^{k} e_{j,l} \left( \nabla_{b(j/T)} E_l(s, \frac{j}{T}) \right) \Big|_{s=0}, 1 \leq k \leq T.$$

Then $\{\Xi_{k,l}\}_{k=1}^T$ is a $L^2$-martingale since $\nabla_{b(j/T)} E_l(s, \frac{j}{T})$ is uniformly bounded whenever $b(\cdot)$ is bounded continuous. By the Doob's inequality, we have $\mathbb{E} \max_{1\leq k\leq T} \|\Xi_{k,l}\|_\mu^2 \leq C \mathbb{E} \|\Xi_{k,l}\|_\mu^2$ for some constant $C$. Similar to proof of Lemma 1, $\mathbb{E} \|\Xi_{k,l}\|_\mu^2 = O_p(\frac{1}{T})$ and thus ensuring

13

that $\mathbb{E}\max_{1\leq k\leq T}\|\Xi_{k,l}\|_\mu = O(\frac{1}{\sqrt{T}})$ and that

$$\max_{1\leq k\leq T}\left\|\frac{1}{\sqrt{T}}\sum_{j=1}^{k}\left[\Pi(\tau,\frac{j}{T})\left(e_j^\top E(\tau,\frac{j}{T})\right) - e_j^\top E\right]\right\|_\mu = O_p(\frac{1}{\sqrt{T}}).$$

Applying the weak convergence of $\frac{1}{\sqrt{T}}\sum_{j=1}^{k}e_j^\top E$ to $U(\cdot)$ provided by Lemma 1, we conclude that $\frac{1}{\sqrt{T}}\sum_{j=1}^{k}\Pi(\tau,\frac{j}{T})\left(e_j^\top E(\tau,\frac{j}{T})\right)$ converges to $U(\cdot)$, as claimed.

To prove the Claim (**C2**), it suffices to show that

$$\left\|\mathbb{E}H(\mu,X_i) - \mathbb{E}H\left(\mu,\operatorname{Exp}_\mu(G_{\mathbf{E}}(\frac{i}{T},\mathcal{F}_0)^\top E)\right)\right\|_{HS} \leq C/\sqrt{T},$$

for some constant $C$, since the physical dependency meansure is the same as Theorem 1. Let $\mathcal{J}(t,e_i)$ be the Jacobi field along the geodesic $t\to\operatorname{Exp}_\mu(t\tau b(\frac{i}{T}))$ such that $\mathcal{J}(0,e_i) = \tau b(\frac{i}{T})$ and

$$\left.\frac{\partial}{\partial t}\mathcal{J}(t,e_i)\right|_{t=0} = \tau\sum_{l=1}^{d}e_{i,l}\nabla_{b(\frac{i}{T})}E_l(s,\frac{i}{T})\bigg|_{s=0}.$$

By the Lipschitz condition of the Hessian tensor and the sub-exponential condition of $G_{\mathbf{E}}(\frac{i}{T},\mathcal{F}_0)$, we have

$$\begin{aligned}
\left\|\mathbb{E}H(\mu,X_i) - \mathbb{E}H\left(\mu,\operatorname{Exp}_\mu(G_{\mathbf{E}}(\frac{i}{T},\mathcal{F}_0)^\top E)\right)\right\|_{HS} \\
\leq L_H\mathbb{E}d_\mathcal{M}\left(\operatorname{Exp}_\mu\{G_{\mathbf{E}}(\frac{i}{T},\mathcal{F}_0)^\top E\}, \operatorname{Exp}_{\mu_T(i/T)}\{G_{\mathbf{E}}(\frac{i}{T},\mathcal{F}_0)^\top E(\tau,\frac{i}{T})\}\right) \\
\leq L_H\mathbb{E}\left\|\mathcal{J}(1,e_i)\right\|_{\operatorname{Exp}_\mu(\tau b(\frac{i}{T}))} \\
\leq \tau CL_H, \text{ (by Gronwall's inequality)(Taylor 2010)}
\end{aligned} \quad (S14)$$

for some constant $C<\infty$ and $\tau = 1/\sqrt{T}$.

Claim (**C3**) holds by applying similar argument to $H(\mu,X_i)\circ b(\frac{i}{T})$ instead of $H(\mu,X_i)$. By

14

applying results of Claims (**C1**)-(**C3**) to (S13), we obtain our desired asymptotic convergence in (S11).

□

## S1.5 Theorem 4

*Proof.* To prove the consistency of the bootstrap procedure, it suffices to show that the estimated covariance function is consistent. The argument is similar to the proof of Theorem 2. Applying Taylor expansion twice, we have

$$\frac{1}{\sqrt{n}}\mathcal{P}_{\gamma(0)}^{\gamma(1)}S_{l,n} = \frac{1}{\sqrt{n}}\sum_{j=l+1}^{l+n}\Pi(\tau,\frac{j}{T})\left(e_j^\top E(\tau,\frac{j}{T})\right) + \frac{\tau}{\sqrt{n}}\sum_{j=l+1}^{l+n} H(\mu,X_j)\circ b(\frac{j}{T})$$
$$- \frac{1}{\sqrt{n}}\sum_{j=l+1}^{l+n} H(\mu,X_j)\circ \mathrm{Log}_\mu\hat\mu + o_p(\tau\sqrt{n}).$$

Moreover, Taylor expansion of parallel transport yields that

$$\Pi(\tau,\frac{j}{T})\left(e_j^\top E(\tau,\frac{j}{T})\right) = \sum_{l=1}^{d} e_{j,k}\Pi(\tau,\frac{j}{T})\left(E_k(\tau,\frac{j}{T})\right)$$
$$= e_j^\top E + \tau \sum_{l=1}^{d} e_{j,k}\left(\nabla_{b(j/T)}E_l(s,\frac{j}{T})\right)\bigg|_{s=0} + O_p(\tau^2),$$

15

and that

$$\frac{1}{\sqrt{n}}\mathcal{P}_{\gamma(0)}^{\gamma(1)}S_{l,n} = \frac{1}{\sqrt{n}}\sum_{j=l+1}^{l+n}\left(e_j^\top E + \tau\sum_{l=1}^{d}e_{j,k}\left(\nabla_{b(j/T)}E_l(s,\frac{j}{T})\right)\bigg|_{s=0}\right)$$

$$+ \frac{\tau}{\sqrt{n}}\sum_{j=l+1}^{l+n}H(\mu,X_j)\circ b(\frac{j}{T})$$

$$- \frac{1}{\sqrt{n}}\sum_{j=l+1}^{l+n}H(\mu,X_j)\circ \mathrm{Log}_\mu\hat{\mu} + o_p(\tau\sqrt{n}).$$

If $\lim_{T\to\infty}\sqrt{n}\tau \to 0$ and $n \to \infty$, we have

$$\frac{1}{\sqrt{n}}\mathcal{P}_{\gamma(0)}^{\gamma(1)}S_{l,n} = \frac{1}{\sqrt{n}}\sum_{j=l+1}^{l+n}\left(e_j^\top E\right) + O_p(\tau\sqrt{n}),$$

which implies that

$$\max_{n+1\leq i\leq j\leq T-n+1}\left|B_{i,j,a,b} - \int_{i/T}^{j/T}\{\Sigma_\mathbf{E}(\xi)\}_{a,b}d\xi\right| = o_p(1), \tag{S15}$$

where $B_{i,j,a,b}$ is defined in (S7).

We then show that $\{\frac{1}{T}\sum_{i=1}^{k}H(\hat{\mu},X_i)\}_{k=1}^{T}$ consistently estimates $\mathcal{H}(\cdot)$. Since convergence of $\{\frac{1}{T}\sum_{i=1}^{k}H(\hat{\mu},X_i)\}_{k=1}^{T}$ to $\{\frac{1}{T}\sum_{i=1}^{k}H(\mu,X_i)\}_{k=1}^{T}$ with rate $O_p(\max\{\tau,1/\sqrt{T}\})$ is guaranteed by (S5), it suffices to show that $\{\frac{1}{T}\sum_{i=1}^{k}H(\mu,X_i)\}_{k=1}^{T}$ converges to $\mathcal{H}(\cdot)$. $H(\mu,X_i) - \mathbb{E}H(\mu,\mathrm{Exp}_\mu(G_\mathbf{E}(\frac{i}{T},\mathcal{F}_0)^\top E)$ can be rewritten as

$$H(\mu,X_i) - \mathbb{E}H\left(\mu,\mathrm{Exp}_\mu(G_\mathbf{E}(\frac{i}{T},\mathcal{F}_0)^\top E\right) = H(\mu,X_i) - \mathbb{E}H(\mu,X_i)$$

$$+ \mathbb{E}H(\mu,X_i) - \mathbb{E}H\left(\mu,\mathrm{Exp}_\mu(G_\mathbf{E}(\frac{i}{T},\mathcal{F}_0)^\top E\right).$$

16

By (S14), the bias term $\mathbb{E}H(\mu, X_i) + \mathbb{E}H(\mu, X_i) - \mathbb{E}H\left(\mu, \mathrm{Exp}_\mu(G_{\mathbf{E}}(\frac{i}{T}, \mathcal{F}_0)^\top E\right)$ is bounded by $O(\tau)$. Similarly to the argument in the proof for Lemma 1, we have

$$\max_{1\leq k\leq T}\left\|\frac{1}{T}\sum_{i=1}^{k} H(\mu, X_i) - \mathbb{E}H(\mu, X_i)\right\|_{HS}^2 = O_p(\frac{1}{T}).$$

Hence, we can deduce that

$$\sup_{1\leq k\leq T} |\langle \mathcal{P}_{\gamma(1)}^{\gamma(0)}(\mathcal{H}(k/T)\circ E_i), \mathcal{P}_{\gamma(1)}^{\gamma(0)}(E_j)\rangle_\mu - \langle \hat{\mathcal{H}}_k \circ \mathcal{P}_{\gamma(1)}^{\gamma(0)}(E_i), \mathcal{P}_{\gamma(1)}^{\gamma(0)}(E_j)\rangle_{\hat\mu}| \tag{S16}$$
$$= O_P(\max\{\tau, \frac{1}{T^{1/2}}\}).$$

Let $U_n(t) : [\frac{n}{T}, 1-\frac{n}{T}] \to \mathcal{T}_{\hat\mu}\mathcal{M}$ be the piece-wise linear interpolation of $V_{k,n}$, i.e., $U_n(t) = V_{\lfloor tT\rfloor,n} + (tT - \lfloor tT\rfloor)(V_{\lfloor tT\rfloor+1,n} - V_{\lfloor tT\rfloor,n})$. Combining the results given by (S15) and (S16), conditioned on $\{X_i\}_{i=1}^T$, $U_n(t)$ converges to $U(t)$ on $\mathcal{C}([0,1], (\mathcal{T}_\mu\mathcal{M}, \|\cdot\|_\mu))$ with the uniform topology. □

## S1.6 Theorem 5

To establish the asymptotic result, we follow the arguments in Aue & van Delft (2020) and van Delft et al. (2021). Different from the Hilbert space, we need to investigate the curvature effect in the general Riemannian manifold. When $\mathcal{M}$ is a Riemannian manifold, we have to account for the curvature effect in $J_n$:

$$\mathcal{P}_{\gamma(0)}^{\gamma(1)} J_n(\omega, t) = \mathcal{P}_{\gamma(0)}^{\gamma(1)} \frac{1}{\sqrt{2\pi n}} \sum_{s=0}^{n-1} \mathrm{Log}_{\hat\mu} X_{\lfloor tT\rfloor-n/2+1+s,T} \cdot e^{-\mathbf{i}s\omega}$$
$$= \frac{1}{\sqrt{2\pi n}} \sum_{s=0}^{n-1} \mathrm{Log}_\mu X_{\lfloor tT\rfloor-n/2+1+s,T} \cdot e^{-\mathbf{i}s\omega}$$

$$-\frac{1}{\sqrt{2\pi n}}\sum_{s=0}^{n-1}H(\mu, X_{\lfloor tT\rfloor-n/2+1+s,T})\circ \mathrm{Log}_\mu\hat{\mu}\cdot e^{-\mathbf{i}s\omega}+O_P(\frac{\sqrt{n}}{T}).$$

Define $\tilde{\mathcal{P}}_{\gamma(0)}^{\gamma(1)}(U\otimes V) = \mathcal{P}_{\gamma(0)}^{\gamma(1)}(U)\otimes \mathcal{P}_{\gamma(0)}^{\gamma(1)}(V)$, then we have

$$\tilde{\mathcal{P}}_{\gamma(0)}^{\gamma(1)}I_n(\omega,u) = (\frac{1}{2\pi n}\sum_{s=0}^{n-1}\mathrm{Log}_\mu X_{\lfloor tT\rfloor-n/2+1+s,T}\cdot e^{-\mathbf{i}s\omega})\otimes(\sum_{r=0}^{n-1}\mathrm{Log}_\mu X_{\lfloor tT\rfloor-n/2+1+r,T}\cdot e^{-\mathbf{i}r\omega})$$
$$+\left(\frac{1}{2\pi n}\sum_{s=0}^{n-1}H(\mu,X_{\lfloor tT\rfloor-n/2+1+s,T})\circ\mathrm{Log}_\mu\hat{\mu}\cdot e^{-\mathbf{i}s\omega}\right)\otimes(\sum_{r=0}^{N-1}\mathrm{Log}_\mu X_{\lfloor tT\rfloor-n/2+1+r,T}\cdot e^{-\mathbf{i}r\omega})$$
$$+(\frac{1}{2\pi n}\sum_{r=0}^{n-1}\mathrm{Log}_\mu X_{\lfloor tT\rfloor-n/2+1+r,T}\cdot e^{-\mathbf{i}r\omega})\otimes\left(\sum_{s=0}^{n-1}H(\mu,X_{\lfloor tT\rfloor-n/2+1+s,T})\circ\mathrm{Log}_\mu\hat{\mu}\cdot e^{-\mathbf{i}s\omega}\right)$$
$$+\left(\frac{1}{2\pi n}\sum_{s=0}^{n-1}H(\mu,X_{\lfloor tT\rfloor-n/2+1+s,T})\circ\mathrm{Log}_\mu\hat{\mu}\cdot e^{-\mathbf{i}s\omega}\right)$$
$$\otimes\left(\sum_{r=0}^{n-1}H(\mu,X_{\lfloor tT\rfloor-n/2+1+r,T})\circ\mathrm{Log}_\mu\hat{\mu}\cdot e^{-\mathbf{i}r\omega}\right)$$
$$+O_P(\frac{\sqrt{n}}{T}).$$

Let

$$A_n(\omega,t) = (\frac{1}{2\pi n}\sum_{s=0}^{n-1}\mathrm{Log}_\mu X_{\lfloor tT\rfloor-n/2+1+s,T}\cdot e^{-\mathbf{i}s\omega})\otimes(\sum_{r=0}^{n-1}\mathrm{Log}_\mu X_{\lfloor tT\rfloor-n/2+1+r,T}\cdot e^{-\mathbf{i}r\omega}),$$
$$B_n(\omega,t) = \left(\frac{1}{2\pi n}\sum_{s=0}^{n-1}H(\mu,X_{\lfloor tT\rfloor-n/2+1+s,T})\circ\mathrm{Log}_\mu\hat{\mu}\cdot e^{-\mathbf{i}s\omega}\right)\otimes(\sum_{r=0}^{n-1}\mathrm{Log}_\mu X_{\lfloor tT\rfloor-n/2+1+r,T}\cdot e^{-\mathbf{i}r\omega}),$$
$$C_n(\omega,t) = \left(\frac{1}{2\pi n}\sum_{s=0}^{n-1}H(\mu,X_{\lfloor tT\rfloor-n/2+1+s,T})\circ\mathrm{Log}_\mu\hat{\mu}\cdot e^{-\mathbf{i}s\omega}\right)$$
$$\otimes\left(\sum_{r=0}^{n-1}H(\mu,X_{\lfloor tT\rfloor-n/2+1+r,T})\circ\mathrm{Log}_\mu\hat{\mu}\cdot e^{-\mathbf{i}r\omega}\right).$$

The asymptotic behaviour of $A_n(\omega, t)$ is well-studied in the work by Aue & van Delft (2020) and van Delft et al. (2021). We then investigate the property of $B_n(\omega, t)$. Notice that $B_n(\omega, t)$ is multi-linear in $\sum_{s=0}^{n-1} H(\mu, X_{\lfloor tT \rfloor - n/2 + 1 + s, T}) \cdot e^{-\mathbf{i} s\omega}$, $\text{Log}_\mu \hat{\mu}$ and $(\sum_{r=0}^{n-1} \text{Log}_\mu X_{\lfloor tT \rfloor - n/2 + 1 + r, T} \cdot e^{-\mathbf{i} r\omega})$, where the asymptotic distribution of the latter two terms are known. Thus, we only need to identify the asymptotic behaviour of $\sum_{s=0}^{n-1} H(\mu, X_{\lfloor tT \rfloor - n/2 + 1 + s, T}) \cdot e^{-\mathbf{i} s\omega}$. We rewrite $\sum_{s=0}^{n-1} H(\mu, X_{\lfloor tT \rfloor - n/2 + 1 + s, T}) \cdot e^{-\mathbf{i} s\omega}$ as

$$\sum_{s=0}^{n-1} H(\mu, X_{\lfloor tT \rfloor - n/2 + 1 + s, T}) \cdot e^{-\mathbf{i} s\omega}$$
$$= \sum_{s=0}^{n-1} \left\{ H(\mu, X_{\lfloor tT \rfloor - n/2 + 1 + s, T}) - \mathbb{E} H(\mu, X_{\lfloor tT \rfloor - n/2 + 1 + s, T}) \right\} \cdot e^{-\mathbf{i} s\omega}$$
$$+ \sum_{s=0}^{n-1} \mathbb{E} H(\mu, X_{\lfloor tT \rfloor - n/2 + 1 + s, T}) \cdot e^{-\mathbf{i} s\omega}.$$

We claim:

**(C4)** $\sum_{s=0}^{n-1} \mathbb{E} H(\mu, X_{\lfloor tT \rfloor - n/2 + 1 + s, T}) \cdot e^{-\mathbf{i} s\omega} = O(n^2/T)$, for $\omega = 2\pi k/n$, $k = 1, \cdots, n$;

**(C5)** $\sum_{s=0}^{n-1} \left( H(\mu, X_{\lfloor tT \rfloor - n/2 + 1 + s, T}) - \mathbb{E} H(\mu, X_{\lfloor tT \rfloor - n/2 + 1 + s, T}) \right) \cdot e^{-\mathbf{i} s\omega} = O_p(\sqrt{n})$.

To prove the claim **(C4)**, without loss of generality, we consider $t = \frac{n}{2T}$, and rewrite $\sum_{s=1}^{n} \mathbb{E} H(\mu, X_{\lfloor tT \rfloor - n/2 + 1 + s, T}) \cdot e^{-\mathbf{i} s\omega}$ as follows:

$$\sum_{s=0}^{n-1} \mathbb{E} H(\mu, X_{\lfloor tT \rfloor - n/2 + 1 + s, T}) \cdot e^{-\mathbf{i} s\omega} = \sum_{s=1}^{n} \mathbb{E} H(\mu, \text{Exp}_\mu(G_{\mathbf{E}}(\frac{s+1}{T}, \mathcal{F}_0)^\top E)) \cdot e^{-\mathbf{i} s\omega}$$
$$= \sum_{s=0}^{n-1} \left( \mathbb{E} H(\mu, \text{Exp}_\mu(G_{\mathbf{E}}(\frac{1+s}{T}, \mathcal{F}_0)^\top E)) - \mathbb{E} H(\mu, \text{Exp}_\mu(G_{\mathbf{E}}(\frac{n}{2T}, \mathcal{F}_0)^\top E)) \right) \cdot e^{-\mathbf{i} s\omega}$$
$$+ \mathbb{E} H(\mu, \text{Exp}_\mu(G_{\mathbf{E}}(\frac{n}{2T}, \mathcal{F}_0)^\top E)) \cdot \sum_{s=0}^{n-1} e^{-\mathbf{i} s\omega}.$$

19

Notice that when $\omega = 2\pi k/n$, $k = 1, \cdots, n$, $\sum_{s=0}^{n-1} e^{-is\omega} = 0$ by the property of discrete Fourier transform. By the Lipschitz continuity provided by Eq.(S2), we have

$$\sum_{s=0}^{n-1} \left( \mathbb{E} H\big(\mu, \mathrm{Exp}_\mu(G_{\mathbf{E}}(\frac{1+s}{T}, \mathcal{F}_0)^\top E)\big) - \mathbb{E} H\big(\mu, \mathrm{Exp}_\mu(G_{\mathbf{E}}(\frac{n}{2T}, \mathcal{F}_0)^\top E)\big) \right) \cdot e^{-is\omega}$$

$$\leq \sum_{s=0}^{n-1} CL_H |\frac{s+1-n/2}{T}| = O(\frac{n^2}{T}).$$

We then apply the martingale techniques to prove the Claim (**C5**). Let $\{H_{i,jk}\}_{1 \leq j \leq k \leq d}$ be the coordinate representation of $H(\mu, X)$ under the basis $\{E_1, \cdots, E_d\}$, i.e. $H_{i,jk} = \langle H_i E_j, E_k \rangle_\mu$. It suffices to show that

$$\sum_{s=0}^{n-1} (H_{s+1,jk} - \mathbb{E} H_{s+1,jk}) \cdot e^{-is\omega} = O_p(\sqrt{n}).$$

We consider the real part of $\sum_{s=0}^{n-1} (H_{s+1,jk} - \mathbb{E} H_{s+1,jk}) \cdot e^{-is\omega}$. Let $\Xi_{l,jk} = \sum_{s=0}^{l-1} (H_{s+1,jk} - \mathbb{E}[H_{s+1,jk}|\mathcal{F}_s]) \cos(s\omega)$ and $V_{l,jk} = \sum_{s=0}^{l-1} (\mathbb{E}[H_{s+1,jk}|\mathcal{F}_s] - \mathbb{E}[H_{s+1,jk}]) \cos(s\omega)$. $\{\Xi_{l,jk}\}_{l=1}^{n}$ is a bounded martingale, and by elementary calculation we have $\mathbb{E}|\Xi_{n,jk}|^2 = O(n)$. Similar to argument in proof of Lemma 1, we apply Theorem 1 in Wu (2005) and have $\mathbb{E}|V_{n,jk}|^2 = O(n)$. The imaginary part is also bounded by $O(n)$ by similar argument, and thus (S1.6) is proved. Claims (**C4**) and (**C5**) indicate that $\frac{1}{n} \sum_{s=0}^{n-1} H(\mu, X_{\lfloor tT \rfloor - n/2 + 1 + s, T}) \cdot e^{-is\omega} = O_p(\frac{1}{\sqrt{n}} + \frac{n}{T})$, and that $B_n(\omega, u) = O_p(\frac{n^{3/2}}{T^{3/2}} + \frac{1}{T^{1/2}})$, which may not converge to 0 in the test statistic scaled by $\sqrt{T}$. Similarly, we have $C_n(\omega, t) = O_p(\frac{1}{T} + \frac{n^3}{T^3} + \frac{n^{3/2}}{T^2}) = o_p(\frac{1}{\sqrt{T}})$, which is negligible in the test statistic as $T \to \infty$. Denote $\bar{H}_i = H_i - \mathbb{E} H_i$ the centered version of $H_i$. Then

$$B_n(\omega, t) = O_p(\frac{n^{3/2}}{T^{3/2}}) +$$

$$(\frac{1}{2\pi n} \sum_{s=0}^{n-1} \bar{H}_{\lfloor tT \rfloor - n/2 + 1 + s, T} \circ \mathrm{Log}_\mu \hat{\mu} \cdot e^{-is\omega}) \otimes (\sum_{r=0}^{n-1} \mathrm{Log}_\mu X_{\lfloor tT \rfloor - n/2 + 1 + r, T} \cdot e^{-ir\omega}).$$

Define $F_{H,G}(\lambda, t)$ the cross-spectral density function of $G_{\mathbf{E}}(t, \mathcal{F}_0)^\top E$ and $H(\mu, \text{Exp}_\mu(G_{\mathbf{E}}(t, \mathcal{F}_0)^\top E)) - \mathbb{E}H(\mu, \text{Exp}_\mu(G_{\mathbf{E}}(t, \mathcal{F}_0)^\top E))$, i.e., $\forall v \in \mathcal{T}_\mu \mathcal{M}$, we have $F_{H,G}(\omega, t)(v) = \frac{1}{2\pi} \sum_{h \in \mathbb{Z}} \mathbb{E}[\bar{H}(t, \mathcal{F}_h) \circ v \otimes \{G_{\mathbf{E}}(t, \mathcal{F}_0)^\top E\}] e^{i\omega h}$. Then $B_n(\omega, t)$ intuitively serves as an estimate of $F_{H,G}(\omega, t)(\text{Log}_\mu \hat{\mu})$.

We observe that

$$\frac{\sqrt{T}}{T} \sum_{j=1}^{m} \sum_{k=1}^{n/2} (\langle I_n(\omega_k, t_j), I_n(\omega_{k-1}, t_j) \rangle - \langle A_n(\omega_k, t_j), A_n(\omega_{k-1}, t_j) \rangle)$$

$$= \frac{\sqrt{T}}{T} \sum_{j=1}^{m} \sum_{k=1}^{n/2} \langle A_n(\omega_k, t_j), B_n(\omega_{k-1}, t_j) \rangle$$

$$+ \frac{\sqrt{T}}{T} \sum_{j=1}^{m} \sum_{k=1}^{n/2} \langle B_n(\omega_k, t_j), A_n(\omega_{k-1}, t_j) \rangle$$

$$+ \frac{\sqrt{T}}{T} \sum_{j=1}^{m} \sum_{k=1}^{n/2} \langle B_n(\omega_k, t_j), B_n(\omega_{k-1}, t_j) \rangle + o_p(1/\sqrt{T}).$$

Since $B_n(\omega_k, t_j) = O_p(1/\sqrt{T})$, we have $\frac{\sqrt{T}}{T} \sum_{j=1}^{m} \sum_{k=1}^{n/2} \langle B_n(\omega_k, t_j), B_n(\omega_{k-1}, t_j) \rangle = O_p(1/\sqrt{T})$. For the quantity $\sqrt{T} \langle A_n(\omega_k, t_j), B_n(\omega_{k-1}, t_j) \rangle$, Taylor expansion yields that

$$\sqrt{T} \langle A_n(\omega_k, t_j), B_n(\omega_{k-1}, t_j) \rangle = \langle J_n(-\omega_k, t_j), J_n(\omega_{k-1}, t_j) \rangle_\mu$$

$$\times \langle J_n(\omega_k, t_j), J_n^{\bar{H}}(-\omega_{k-1}, t_j) \circ U_T) \rangle_\mu + O_p(\frac{1}{T}),$$

where

$$J_n^{\bar{H}}(-\omega, t_j) = \frac{1}{\sqrt{2\pi n}} \sum_{s=0}^{n-1} e^{-is\omega} \bar{H}_{\lfloor tT \rfloor - n/2 + 1 + s, T} \circ \mathcal{H}^{-1}(1),$$

and

$$U_T = \frac{1}{\sqrt{T}} \sum_{j'=1}^{T} e_{j'}^\top E$$

21

is the local DFT of the tensor $\bar{H}_s \circ \mathcal{H}^{-1}(1)$. Under the orthonormal basis $\mathbf{E} = \{E_1, \cdots, E_d\}$ on $\mathcal{T}_\mu \mathcal{M}$, we can rewrite $\langle J_n(-\omega_k, t_j), J_n(\omega_{k-1}, t_j)\rangle_\mu \times \langle J_n(\omega_k, t_j), J_n^{\bar{H}}(-\omega_{k-1}, t_j) \circ U_T)\rangle_\mu$ as

$$\langle J_n(-\omega_k, t_j), J_n(\omega_{k-1}, t_j)\rangle_\mu \times \langle J_n(\omega_k, t_j), J_n^{\bar{H}}(-\omega_{k-1}, t_j) \circ U_T)\rangle_\mu$$

$$= \sum_{l_3=1}^d \langle U_T, E_{l_3}\rangle_\mu \sum_{l_1=1}^d \sum_{l_2=1}^d \langle J_n(-\omega_k, t_j), E_{l_1}\rangle_\mu \langle J_n(\omega_{k-1}, t_j), E_{l_1}\rangle_\mu$$

$$\times \langle J_n(\omega_k, t_j), E_{l_2}\rangle_\mu \langle E_{l_2}, J_n^{\bar{H}}(-\omega_{k-1}, t_j) \circ E_{l_3}\rangle_\mu.$$

We apply the similar expansion to $\frac{\sqrt{T}}{n} \sum_{k=1}^{n/2} \langle \frac{1}{m}\sum_{j=1}^m I_n(\omega_k, t_j), \frac{1}{m}\sum_{j=1}^m I_n(\omega_k, t_j)\rangle$. Similarly, we have

$$\frac{\sqrt{T}}{n}\sum_{k=1}^{n/2}\langle \frac{1}{m}\sum_{j=1}^m I_n(\omega_k, t_j), \frac{1}{m}\sum_{j=1}^m I_n(\omega_k, t_j)\rangle - \frac{\sqrt{T}}{n}\sum_{k=1}^{n/2}\langle \frac{1}{m}\sum_{j=1}^m A_n(\omega_k, t_j), \frac{1}{m}\sum_{j=1}^m A_n(\omega_k, t_j)\rangle,$$

$$= \frac{\sqrt{T}}{n}\sum_{k=1}^{n/2}\langle \frac{1}{m}\sum_{j=1}^m B_n(\omega_k, t_j), \frac{1}{m}\sum_{j=1}^m A_n(\omega_k, t_j)\rangle$$

$$+ \frac{\sqrt{T}}{n}\sum_{k=1}^{n/2}\langle \frac{1}{m}\sum_{j=1}^m A_n(\omega_k, t_j), \frac{1}{m}\sum_{j=1}^m B_n(\omega_k, t_j)\rangle + O_p(1/\sqrt{T}).$$

(S17)

Note that $\frac{\sqrt{T}}{n}\sum_{k=1}^{n/2}\langle \frac{1}{m}\sum_{j=1}^m B_n(\omega_k, t_j), \frac{1}{m}\sum_{j=1}^m A_n(\omega_k, t_j)\rangle$ is asymptotically equivalent to

$$\frac{\sqrt{T}}{n}\sum_{k=1}^{n/2}\langle \frac{1}{m}\sum_{j=1}^m A_n(\omega_k, t_j), \frac{1}{m}\sum_{j=1}^m B_n(\omega_k, t_j)\rangle$$

$$= \frac{1}{n}\sum_{k=1}^{n/2}\langle \frac{1}{m}\sum_{j=1}^m A_n(\omega_k, t_j), \frac{1}{m}\sum_{j=1}^m J_n^{\bar{H}}(-\omega_k, t_j) \circ U_T \otimes J_n(\omega_k, t_j)\rangle + O_p(1/\sqrt{T}).$$

22

Extracting the common linear factor $U_T$ above, we have

$$\langle \frac{1}{m}\sum_{j=1}^{m} A_n(\omega_k, t_j), \frac{1}{m}\sum_{j=1}^{m} J_n^{\bar{H}}(-\omega_k, t_j) \circ U_T \otimes J_n(-\omega_k, t_j)\rangle$$

$$= \sum_{l_3=1}^{d}\langle U_T, E_{l_3}\rangle_\mu \frac{1}{m^2}\sum_{l_1=1}^{d}\sum_{l_2=1}^{d}\sum_{j_1=1}^{m}\sum_{j_2=1}^{m}\langle J_n(-\omega_k, t_{j_1}), E_{l_1}\rangle_\mu \langle J_n(\omega_k, t_{j_2}), E_{l_1}\rangle_\mu$$

$$\times \langle J_n(\omega_k, t_{j_1}), E_{l_2}\rangle_\mu \langle E_{l_2}, J_n^{\bar{H}}(-\omega_k, t_{j_2}) \circ E_{l_3}\rangle_\mu$$

$$= \sum_{l_3=1}^{d}\sum_{l_1=1}^{d}\sum_{l_2=1}^{d}\langle U_T, E_{l_3}\rangle_\mu \left(\frac{1}{m}\sum_{j_1=1}^{m}\langle J_n(-\omega_k, t_{j_1}), E_{l_1}\rangle_\mu \langle J_n(\omega_k, t_{j_1}), E_{l_2}\rangle_\mu\right)$$

$$\times \left(\frac{1}{m}\sum_{j_2=1}^{m}\langle J_n(\omega_k, t_{j_2}), E_{l_1}\rangle_\mu J_n^{\bar{H}}(-\omega_k, t_{j_2}) \circ E_{l_3}\rangle_\mu\right)$$

Hence, the additional term induced by the curvatures of manifold compared to Euclidean space is asymptotically equivalent to $\Delta + \bar{\Delta}$, where $\Delta$ is given by

$$\Delta = \sum_{l_3=1}^{d}\sum_{l_1=1}^{d}\sum_{l_2=1}^{d}\langle U_T, E_{l_3}\rangle_\mu \left\{\frac{1}{T}\sum_{k=1}^{n/2}\sum_{j=1}^{m}\langle J_n(-\omega_k, t_j), E_{l_1}\rangle_\mu \langle J_n(\omega_k, t_j), E_{l_2}\rangle_\mu\right.$$

$$\times \langle J_n(\omega_{k-1}, t_j), E_{l_1}\rangle_\mu \langle E_{l_2}, J_n^{\bar{H}}(-\omega_{k-1}, t_j) \circ E_{l_3}\rangle_\mu$$

$$- \sum_{k=1}^{n/2}\left(\frac{1}{m}\sum_{j_1=1}^{m}\langle J_n(-\omega_k, t_{j_1}), E_{l_1}\rangle_\mu \langle J_n(\omega_k, t_{j_1}), E_{l_2}\rangle_\mu\right)$$

$$\left.\times \left(\frac{1}{m}\sum_{j_2=1}^{m}\langle J_n(\omega_k, t_{j_2}), E_{l_1}\rangle_\mu J_n^{\bar{H}}(-\omega_k, t_{j_2}) \circ E_{l_3}\rangle_\mu\right)\right\},$$

and $\bar{\Delta}$ is the complex conjugate of $\Delta$. Denote $F_{l_1,l_2}^{GG}(u,w)$ the cross spectral density of $\langle G_{\mathbf{E}}(u, \mathcal{F}), E_{l_1}\rangle_\mu$ and $\langle G_{\mathbf{E}}(u, \mathcal{F}), E_{l_2}\rangle_\mu$, and $F_{l_1,l_2,l_3}^{GH}(u,w)$ the cross spectral density of $\langle G_{\mathbf{E}}(u, \mathcal{F}), E_{l_1}\rangle_\mu$

and $\langle E_{l_2}, H(\mu, \mathrm{Exp}_\mu(G_\mathbf{E}(u, \mathcal{F})) \circ E_{l_3}\rangle_\mu$. We claim that:

$$\begin{aligned}
\Delta_{l_1,l_2,l_3} &= \Bigg\{ \frac{1}{T} \sum_{k=1}^{n/2} \sum_{j=1}^{m} \langle J_n(-\omega_k, t_j), E_{l_1}\rangle_\mu \langle J_n(\omega_k, t_j), E_{l_2}\rangle_\mu \\
&\qquad \times \langle J_n(\omega_{k-1}, t_j), E_{l_1}\rangle_\mu \langle E_{l_2}, J_n^{\bar{H}}(-\omega_{k-1}, t_j) \circ E_{l_3}\rangle_\mu \\
&\quad - \frac{1}{n} \sum_{k=1}^{n/2} \left( \frac{1}{m} \sum_{j_1=1}^{m} \langle J_n(-\omega_k, t_{j_1}), E_{l_1}\rangle_\mu \langle J_n(\omega_k, t_{j_1}), E_{l_2}\rangle_\mu \right) \\
&\qquad \times \left( \frac{1}{m} \sum_{j_2=1}^{m} \langle J_n(\omega_k, t_{j_2}), E_{l_1}\rangle_\mu J_n^{\bar{H}}(-\omega_k, t_{j_2}) \circ E_{l_3}\rangle_\mu \right) \Bigg\} \\
&= \frac{1}{4\pi} \int_{-\pi}^{\pi} \int_0^1 F^{GG}_{l_1,l_2}(u, w) F^{GH}_{l_1,l_2,l_3}(u, w) du d\omega \\
&\quad - \frac{1}{4\pi} \int_{-\pi}^{\pi} \left( \int_0^1 F^{GG}_{l_1,l_2}(u, w) du \right) \left( \int_0^1 F^{GH}_{l_1,l_2,l_3}(u, w) du \right) d\omega + O_p(\frac{1}{m}),
\end{aligned} \quad (\mathrm{S}18)$$

which indicates that whenever $G_\mathbf{E}(u, \mathcal{F})$ is independent of $u$, $\Delta + \bar{\Delta} = O_p(\frac{1}{m})$, and the asymptotic distribution of the test statistics $V^2_{\hat{F}}$ only depends on $A_n(u, \omega)$, as per the distribution in Euclidean space or Hilbert space; and under $\boldsymbol{H_1}$ the asymptotic distribution will differ from the counterpart in Euclidean or Hilbert space in the asymptotic variance. In the remainder of this section we apply the cumulant techniques to show the claim given by (S18) holds.

For a complex random variable $X$, we define $\bar{X}$ as the complex conjugate of $X$. Denote $\mathbf{cum}_k(\cdot)$ the $k^{th}$-order joint cumulant of random variables $X_1, \cdots, X_k$, which is given by the coefficient of $\mathbf{i}^k(t_1 \cdots t_k)$ in the complex Taylor expansions of $\log \mathbb{E}[e^{\mathbf{i}\sum_j^k t_j X_j}]$ at the origin (Brillinger 2001). Particularly, $\mathbf{cum}_1(X) = \mathbb{E}X$ and $\mathbf{cum}_2(\{X, \bar{Y}\}) = \mathbb{E}(XY) - \mathbb{E}X\mathbb{E}Y$. For comprehensive details of cumulants, we refer to the book by Brillinger (2001).

We mainly follow the discussions by van Delft et al. (2021). For simplicity, we use $D_{n,l}(\omega_k, t_j)$ to represent $\langle J_n(\omega_k, t_j), E_l\rangle_\mu$ and $D^H_{n,l,l'}(\omega_k, t_j)$ to represent $\langle E_l, J_n^{\bar{H}}(-\omega_{k-1}, t_j) \circ$

24

$E_{l'}\rangle_\mu$. Then by the expansion of cumulant (Brillinger 2001), Lemma S4.1 and Corollary S4.1 in van Delft et al. (2021), we have

$$\mathbb{E}\frac{1}{T}\sum_{k=1}^{n/2}\sum_{j=1}^{m}\langle J_n(-\omega_k, t_j), E_{l_1}\rangle_\mu \langle J_n(\omega_k, t_j), E_{l_2}\rangle_\mu$$

$$\times \langle J_n(\omega_{k-1}, t_j), E_{l_1}\rangle_\mu \langle E_{l_2}, J_n^{\bar{H}}(-\omega_{k-1}, t_j) \circ E_{l_3}\rangle_\mu$$

$$= \frac{1}{T}\sum_{k=1}^{n/2}\sum_{j=1}^{m}\mathbf{cum}_1(D_{n,l_1}(-\omega_k,t_j)D_{n,l_2}(\omega_k,t_j)D_{n,l_1}(\omega_{k-1},t_j)D_{n,l_2,l_3}^H(\omega_{k-1},t_j))$$

$$= \frac{1}{T}\sum_{k=1}^{n/2}\sum_{j=1}^{m}\mathbf{cum}_4(D_{n,l_1}(-\omega_k,t_j),\ D_{n,l_2}(\omega_k,t_j),\ D_{n,l_1}(\omega_{k-1},t_j),\ D_{n,l_2,l_3}^H(-\omega_{k-1},t_j))$$

$$+ \frac{1}{T}\sum_{k=1}^{n/2}\sum_{j=1}^{m}\mathbf{cum}_2(D_{n,l_1}(-\omega_k,t_j),\ D_{n,l_2}(\omega_k,t_j))\mathbf{cum}_2(D_{n,l_1}(\omega_{k-1},t_j),\ D_{n,l_2,l_3}^H(-\omega_{k-1},t_j))$$

$$+ \frac{1}{T}\sum_{k=1}^{n/2}\sum_{j=1}^{m}\mathbf{cum}_2(D_{n,l_1}(-\omega_k,t_j),\ D_{n,l_1}(-\omega_{k-1},t_j))\mathbf{cum}_2(D_{n,l_2}(\omega_k,t_j),\ D_{n,l_2,l_3}^H(\omega_{k-1},t_j))$$

$$+ \frac{1}{T}\sum_{k=1}^{n/2}\sum_{j=1}^{m}\mathbf{cum}_2(D_{n,l_1}(-\omega_k,t_j),\ D_{n,l_2,l_3}^H(-\omega_{k-1},t_j))\mathbf{cum}_2(D_{n,l_2}(\omega_k,t_j),\ D_{n,l_1}(-\omega_{k-1},t_j))$$

$$= \frac{1}{T}\sum_{k=1}^{n/2}\sum_{j=1}^{m}\mathbf{cum}_2(D_{n,l_1}(-\omega_k,t_j),\ D_{n,l_2}(\omega_k,t_j))\mathbf{cum}_2(D_{n,l_1}(\omega_{k-1},t_j),\ D_{n,l_2,l_3}^H(-\omega_{k-1},t_j))$$

$$+ O(\frac{1}{T}) + O(\frac{1}{m^2}) + O(\frac{1}{n})$$

$$= \frac{1}{4\pi}\int_{-\pi}^{\pi}\int_0^1 F_{l_1,l_2}^{GG}(u,w)F_{l_1,l_2,l_3}^{GH}(u,w)dud\omega + O(\frac{1}{m^2}) + O(\frac{1}{n}) + O(\frac{1}{T}).$$

Similarly, we have

$$\mathbb{E}\frac{1}{n}\sum_{k=1}^{n/2}\left(\frac{1}{m}\sum_{j_1=1}^{m}\langle J_n(-\omega_k,t_{j_1}),E_{l_1}\rangle_\mu \langle J_n(\omega_k,t_{j_1}),E_{l_2}\rangle_\mu\right)$$
$$\times\left(\frac{1}{m}\sum_{j_2=1}^{m}\langle J_n(\omega_k,t_{j_2}),E_{l_1}\rangle_\mu J_n^{\bar{H}}(-\omega_k,t_{j_2})\circ E_{l_3}\rangle_\mu\right)$$

$$=\frac{1}{mT}\sum_{k=1}^{n/2}\sum_{j_1=1}^{m}\sum_{j_2=1}^{m}\mathbf{cum}_1(D_{n,l_1}(-\omega_k,t_{j_1})D_{n,l_2}(\omega_k,t_{j_1})D_{n,l_1}(\omega_k,t_{j_2})D_{n,l_2,l_3}^H(-\omega_k,t_{j_2}))$$

$$=\frac{1}{mT}\sum_{k=1}^{n/2}\sum_{j_1=1}^{m}\sum_{j_2=1}^{m}\mathbf{cum}_4(D_{n,l_1}(-\omega_k,t_{j_1}),\ D_{n,l_2}(\omega_k,t_{j_1}),\ D_{n,l_1}(\omega_k,t_{j_2}),\ D_{n,l_2,l_3}^H(-\omega_k,t_{j_2}))$$

$$+\frac{1}{mT}\sum_{k=1}^{n/2}\sum_{j_1=1}^{m}\sum_{j_2=1}^{m}\mathbf{cum}_2(D_{n,l_1}(-\omega_k,t_{j_1}),\ D_{n,l_2}(\omega_k,t_{j_1}))\mathbf{cum}_2(D_{n,l_1}(\omega_k,t_{j_2}),\ D_{n,l_2,l_3}^H(-\omega_k,t_{j_2}))$$

$$+\frac{1}{mT}\sum_{k=1}^{n/2}\sum_{j_1=1}^{m}\sum_{j_2=1}^{m}\mathbf{cum}_2(D_{n,l_1}(-\omega_k,t_{j_1}),\ D_{n,l_1}(\omega_k,t_{j_2}))\mathbf{cum}_2(D_{n,l_2}(\omega_k,t_{j_1}),\ D_{n,l_2,l_3}^H(-\omega_k,t_{j_2}))$$

$$+\frac{1}{mT}\sum_{k=1}^{n/2}\sum_{j_1=1}^{m}\sum_{j_2=1}^{m}\mathbf{cum}_2(D_{n,l_1}(-\omega_k,t_j),\ D_{n,l_2,l_3}^H(-\omega_k,t_{j_2}))\mathbf{cum}_2(D_{n,l_1}(\omega_k,t_{j_2}),\ D_{n,l_2}(\omega_k,t_{j_1}))$$

$$=\frac{1}{mT}\sum_{k=1}^{n/2}\sum_{j_1=1}^{m}\sum_{j_2=1}^{m}\mathbf{cum}_2(D_{n,l_1}(-\omega_k,t_{j_1}),\ D_{n,l_2}(\omega_k,t_{j_1}))\mathbf{cum}_2(D_{n,l_1}(\omega_k,t_{j_2}),\ D_{n,l_2,l_3}^H(-\omega_k,t_{j_2}))$$

$$+\frac{1}{mT}\sum_{k=1}^{n/2}\sum_{j_1=1}^{m}\sum_{j_2=1}^{m}\mathbf{cum}_2(D_{n,l_1}(-\omega_k,t_{j_1}),\ D_{n,l_1}(\omega_k,t_{j_2}))\mathbf{cum}_2(D_{n,l_2}(\omega_k,t_{j_1}),\ D_{n,l_2,l_3}^H(-\omega_k,t_{j_2}))$$

$$+O(\frac{1}{T})+O(\frac{1}{m^2})$$

$$=\frac{1}{4\pi}\int_{-\pi}^{\pi}\left(\int_0^1 F_{l_1,l_2}^{GG}(u,w)du\right)\left(\int_0^1 F_{l_1,l_2,l_3}^{GH}(u,w)du\right)+$$

$$+\frac{1}{4\pi m}\int_{-\pi}^{\pi}\int_0^1 F_{l_1,l_1}^{GG}(u,w)F_{l_2,l_2,l_3}^{GH}(u,w)dud\omega+O(\frac{1}{T})+O(\frac{1}{m^2}).$$

We then compute $\mathbf{cum}_2(\Delta_{l_1,l_2,l_3},\Delta_{l_1,l_2,l_3})$, the second order cumulant of $\Delta_{l_1,l_2,l_3}$. We begin

26

by considering the quantity:

$$V_1(\Delta_{l_1,l_2,l_3}) = \frac{1}{T^2} \sum_{k_1,k_2} \sum_{j_1,j_2} \mathbf{cum}_2(D_{n,l_1}(-\omega_{k_1},t_{j_1})D_{n,l_2}(\omega_{k_1},t_{j_1})D_{n,l_1}(\omega_{k_1-1},t_{j_1})D^H_{n,l_2,l_3}(-\omega_{k_1-1},t_{j_1}),$$
$$D_{n,l_1}(-\omega_{k_2},t_{j_2})D_{n,l_2}(\omega_{k_2},t_{j_2})D_{n,l_1}(\omega_{k_2-1},t_{j_2})D^H_{n,l_2,l_3}(-\omega_{k_2-1},t_{j_2})$$

It suffices to compute the multiplication of cumulant induced by all indecomposable partitions with size larger than 3 of the following array

$$\overbrace{D_{n,l_1}(-\omega_{k_1},t_{j_1})}^{1}, \overbrace{D_{n,l_2}(\omega_{k_1},t_{j_1})}^{2}, \overbrace{D^H_{n,l_2,l_3}(-\omega_{k_1-1},t_{j_1})}^{3}, \overbrace{D_{n,l_1}(\omega_{k_1-1},t_{j_1})}^{4},$$
$$\overbrace{D_{n,l_1}(-\omega_{k_2},t_{j_2})}^{5}, \overbrace{D_{n,l_2}(\omega_{k_2},t_{j_2})}^{6}, \overbrace{D^H_{n,l_2,l_3}(-\omega_{k_2-1},t_{j_2})}^{7}, \overbrace{D_{n,l_1}(\omega_{k_2-1},t_{j_2})}^{8}.$$

Particularly, the significant terms of cumulant have structures $\mathbf{cum}_4(\cdot)\mathbf{cum}_2(\cdot)\mathbf{cum}_2(\cdot)$ and $\mathbf{cum}_2(\cdot)\mathbf{cum}_2(\cdot)\mathbf{cum}_2(\cdot)\mathbf{cum}_2(\cdot)$. By Lemma S4.1 and Lemma S4.2 in van Delft et al. (2021), the significant terms with the structure $\mathbf{cum}_4(\cdot)\mathbf{cum}_2(\cdot)\mathbf{cum}_2(\cdot)$ can only be included in arrays with $j_1 = j_2$ and induced by one of the following partitions:

$$\{(1256)(34)(78)\}, \ \{(1278)(34)(56)\}, \ \{(3456)(12)(78)\}, \ \{(3478)(12)(56)\}.$$

There are $nm^2$ terms satisfying these two conditions, and each of them is at of order $\frac{1}{n}$, which means that their contribution to the quantity $V_1(\Delta_{l_1,l_2,l_3})$ is of order $O(\frac{1}{T^2} \times n^2m \times \frac{1}{n}) = O(\frac{1}{T})$.

The terms with structures $\mathbf{cum}_2(\cdot)\mathbf{cum}_2(\cdot)\mathbf{cum}_2(\cdot)\mathbf{cum}_2(\cdot)$ that is significant asymptotically can only be included in arrays with $j_1 = j_2$ and $|k_1 - k_2| \leq 1$ and induced by one of the following partitions:

$$\{12)(37)(56)(48)\}, \ \{12)(36)(78)(45)\}, \ \{(15)(26)(34)(78)\}, \ \{(18)(27)(34)(56)\}$$
$$\{(15)(26)(37)(48)\}.$$

There are at most $\frac{3}{2}(mn)$ terms satisfying these two restrictions, with each is of order $O(1)$ and uniformly bounded, hence will contribute to the quantity $V_1(\Delta_{l_1,l_2,l_3})$ with asymptotic order $O(mn \times \frac{1}{T^2}) = O(\frac{1}{T})$. It remains to show that $V_2(\delta_{l_1,l_2,l_3})$ also vanishes as $T \to \infty$, where

$$V_2(\Delta_{l_1,l_2,l_3})$$
$$= \frac{1}{T^2} \sum_{k_1,k_2} \sum_{j_1,j_2,j_3,j_4} \mathbf{cum}_2(D_{n,l_1}(-\omega_{k_1}, t_{j_1}) D_{n,l_2}(\omega_{k_1}, t_{j_1}) D_{n,l_1}(\omega_{k_1}, t_{j_2}) D^H_{n,l_2,l_3}(-\omega_{k_1}, t_{j_2}),$$
$$D_{n,l_1}(-\omega_{k_2}, t_{j_3}) D_{n,l_2}(\omega_{k_2}, t_{j_3}) D_{n,l_1}(\omega_{k_2}, t_{j_4}) D^H_{n,l_2,l_3}(-\omega_{k_2}, t_{j_4})).$$

Similar to the discussion of $V_1(\delta_{l_1,l_2,l_3})$, we calculate the order of cumulants of indecomposable partitions of arrays:

$$\overbrace{D_{n,l_1}(-\omega_{k_1}, t_{j_1})}^{1}, \overbrace{D_{n,l_2}(\omega_{k_1}, t_{j_1})}^{2}, \overbrace{D^H_{n,l_2,l_3}(-\omega_{k_1}, t_{j_2})}^{3}, \overbrace{D_{n,l_1}(\omega_{k_1}, t_{j_2})}^{4},$$
$$\overbrace{D_{n,l_1}(-\omega_{k_2}, t_{j_3})}^{5}, \overbrace{D_{n,l_2}(\omega_{k_2}, t_{j_3})}^{6}, \overbrace{D^H_{n,l_2,l_3}(-\omega_{k_2}, t_{j_4})}^{7}, \overbrace{D_{n,l_1}(\omega_{k_2}, t_{j_4})}^{8}.$$

Similar to $V_1(\Delta_{l_1,l_2,l_3})$, we only need to consider cumulants with structures $\mathbf{cum}_4(\cdot)\mathbf{cum}_2(\cdot)\mathbf{cum}_2(\cdot)$ and $\mathbf{cum}_2(\cdot)\mathbf{cum}_2(\cdot)\mathbf{cum}_2(\cdot)\mathbf{cum}_2(\cdot)$. The only significant terms with structure $\mathbf{cum}_4(\cdot)\mathbf{cum}_2(\cdot)\mathbf{cum}_2(\cdot)$ are the same as $V_1(\Delta_{l_1,l_2,l_3})$:

$$\{(1256)(34)(78)\}, \{(1278)(34)(56)\}, \{(3456)(12)(78)\}, \{(3478)(12)(56)\}.$$

which is of order $O(\frac{1}{T})$. Similarly, there are $O(nm^3)$ significant terms with structure $\mathbf{cum}_2(\cdot)\mathbf{cum}_2(\cdot)\mathbf{cum}_2(\cdot)\mathbf{cum}_2(\cdot)$ and of order $O(1)$, which will contribute to $V_2(\Delta_{l_1,l_2,l_3})$ with

asymptotic order $\frac{1}{T}$. Consequently, we have

$$\begin{aligned}\Delta_{l_1,l_2,l_3} &= \frac{1}{4\pi}\int_{-\pi}^{\pi}\int_0^1 F_{l_1,l_2}^{GG}(u,w)F_{l_1,l_2,l_3}^{GH}(u,w)dud\omega \\ &\quad - \frac{1}{4\pi}\int_{-\pi}^{\pi}\left(\int_0^1 F_{l_1,l_2}^{GG}(u,w)du\right)\left(\int_0^1 F_{l_1,l_2,l_3}^{GH}(u,w)du\right)d\omega + O_p(\frac{1}{m}) + O_p(\frac{1}{T}),\end{aligned} \quad \text{(S19)}$$

as claimed.

When $\boldsymbol{H_0}$ holds, (S19) suggests that $\Delta + \bar{\Delta} = O_p(\frac{1}{m})$, and the null distribution is the same as the null distribution given by Theorem 1 in van Delft et al. (2021). When $\boldsymbol{H_1}$ holds, $\Delta + \bar{\Delta}$ weakly converges to a normal distribution, since $U_T$ is asymptotically normally distributed as shown in the proof to Lemma 1 and $\Delta + \bar{\Delta}$ is asymptotically equivalent to a linear transform of $U_T$. By the asymptotic normality of $\frac{4\pi}{T}\sum_{k=1}^{n/2}\sum_{j=1}^m \langle A_n(\omega_k, t_j), A_n(\omega_{k-1}, t_j)\rangle + \hat{W} - \frac{4\pi}{n}\sum_{k=1}^{n/2}\langle \frac{1}{m}\sum_{j=1}^m A_n(\omega_k, t_j), \frac{1}{m}\sum_{j=1}^m A_n(\omega_k, t_j)\rangle$ provided by Theorem 1 in van Delft et al. (2021), and the asymptotic normality of $\Delta + \bar{\Delta}$, we deduce that $\sqrt{T}(V_{\hat{F}}^2 - V_F^2)$ also converges to a normal distribution under $\boldsymbol{H_1}$.

## S1.7 Theorem 6

Here we show that the estimate given by

$$\hat{\sigma}_V^2 = 16\pi^2 n^{-1}\sum_{k=1}^{n/2}\left(m^{-1}\sum_{j=1}^m \langle I_n(\omega_{k-1}, t_j), I_n(\omega_k, t_j)\rangle_{\text{HS}}\right)^2 \quad \text{(S20)}$$

converges to the asymptotic variance under $\boldsymbol{H_0}$.

Applying the similar technique to establish (S17), we have

$$\left(m^{-1}\sum_{j=1}^{m}\langle I_n(\omega_{k-1},t_j), I_n(\omega_k,t_j)\rangle_{\text{HS}}\right)^2 = \left(m^{-1}\sum_{j=1}^{m}\langle A_n(\omega_{k-1},t_j), A_n(\omega_k,t_j)\rangle_{\text{HS}}\right.$$
$$+ m^{-1}\sum_{j=1}^{m}\langle A_n(\omega_{k-1},t_j), B_n(\omega_k,t_j)\rangle_{\text{HS}} + m^{-1}\sum_{j=1}^{m}\langle B_n(\omega_{k-1},t_j), A_n(\omega_k,t_j)\rangle_{\text{HS}}$$
$$\left. + m^{-1}\sum_{j=1}^{m}\langle B_n(\omega_{k-1},t_j), B_n(\omega_k,t_j)\rangle_{\text{HS}} + O_p(T^{-1})\right)^2.$$

Noting that $B_n(\omega_k, t_j) = O_p(T^{-1/2})$, we further have

$$\left(m^{-1}\sum_{j=1}^{m}\langle I_n(\omega_{k-1},t_j), I_n(\omega_k,t_j)\rangle_{\text{HS}}\right)^2 = \left(m^{-1}\sum_{j=1}^{m}\langle A_n(\omega_{k-1},t_j), A_n(\omega_k,t_j)\rangle_{\text{HS}} + O_p(T^{-1/2})\right)^2$$
$$= \left(m^{-1}\sum_{j=1}^{m}\langle A_n(\omega_{k-1},t_j), A_n(\omega_k,t_j)\rangle_{\text{HS}}\right)^2 + O_p(T^{-1/2})$$

and

$$\hat{\sigma}_V^2 = 16\pi^2 n^{-1} \sum_{k=1}^{n/2} \left(m^{-1}\sum_{j=1}^{m}\langle I_n(\omega_{k-1},t_j), I_n(\omega_k,t_j)\rangle_{\text{HS}}\right)^2$$
$$= 16\pi^2 n^{-1} \sum_{k=1}^{n/2} \left(m^{-1}\sum_{j=1}^{m}\langle A_n(\omega_{k-1},t_j), A_n(\omega_k,t_j)\rangle_{\text{HS}}\right)^2 + O_p(T^{-1/2}).$$

By Lemma 3.1 in van Delft et al. (2021), the quantity

$$16\pi^2 n^{-1} \sum_{k=1}^{n/2} \left(m^{-1}\sum_{j=1}^{m}\langle A_n(\omega_{k-1},t_j), A_n(\omega_k,t_j)\rangle_{\text{HS}}\right)^2$$

is a consistent estimate $4\pi \int_{-\pi}^{\pi} \|\bar{F}_{\mathbf{E}}(\omega)\|_{\text{HS}}^4 d\omega$. Thus $\hat{\sigma}_V^2$ also converges to $4\pi \int_{-\pi}^{\pi} \|\bar{F}_{\mathbf{E}}(\omega)\|_{\text{HS}}^4 d\omega$,

as desired.

# References

Aue, A. & van Delft, A. (2020), 'Testing for stationarity of functional time series in the frequency domain', *The Annals of Statistics* **48**(5), 2505 – 2547.

Brillinger, D. R. (2001), *Time series: data analysis and theory*, SIAM.

Durrett, R. (2019), *Probability: Theory and Examples*, Cambridge University Press.

Kendall, W. S. & Le, H. (2011), 'Limit theorems for empirical Fréchet means of independent and non-identically distributed manifold-valued random variables', *Brazilian Journal of Probability and Statistics* **25**(3), 323 – 352.

Moakher, M. (2005), 'A differential geometric approach to the geometric mean of symmetric positive-definite matrices', *SIAM Journal on Matrix Analysis and Applications* **26**(3), 735–747.

Taylor, M. (2010), *Partial Differential Equations I: Basic Theory*, Applied Mathematical Sciences, Springer New York.

van Delft, A., Characiejus, V. & Dette, H. (2021), 'A nonparametric test for stationarity in functional time series', *Statistica Sinica* **31**(3), pp. 1375–1395.

Wu, W. B. (2005), 'Nonlinear system theory: Another look at dependence', *Proceedings of the National Academy of Sciences* **102**(40), 14150–14154.

Zhou, Z. (2013), 'Heteroscedasticity and autocorrelation robust structural change detection', *Journal of the American Statistical Association* **108**(502), 726–740.



\end{document}